\documentclass[aps,preprint,preprintnumbers,nofootinbib,superscriptaddress,10pt]{revtex4}

\include{Figures}
\usepackage{framed,multirow}
\usepackage{amssymb}
\usepackage{latexsym}
\usepackage{amsmath}
\usepackage{url}
\usepackage{xcolor}
\definecolor{newcolor}{rgb}{.8,.349,.1}
\usepackage{grffile}
\usepackage{eurosym}
\usepackage{rotating}
\usepackage{hyperref} 
\usepackage{xcolor}
\definecolor{cite}{rgb}{0.,0.,0.5}   
\hypersetup{ 
    colorlinks,
    linkcolor={cite},
    citecolor={cite},
    urlcolor={cite}
}

\begin{document}

\title{\LARGE High magnetic fields for fundamental physics}

\author{R\'{e}my Battesti}
\affiliation{Laboratoire National des Champs Magn\'etiques Intenses (UPR 3228, CNRS-UPS-UGA-INSA), F-31400 Toulouse Cedex, France}
\author{Jerome Beard}
\affiliation{Laboratoire National des Champs Magn\'etiques Intenses (UPR 3228, CNRS-UPS-UGA-INSA), F-31400 Toulouse Cedex, France}
\author{Sebastian B\"{o}ser}
\affiliation{Johannes Gutenberg University, 55099 Mainz, Germany}
\author{Nicolas Bruyant}
\affiliation{Laboratoire National des Champs Magn\'etiques Intenses (UPR 3228, CNRS-UPS-UGA-INSA), F-31400 Toulouse Cedex, France}
\author{Dmitry Budker}\email{budker@uni-mainz.de}
\affiliation{Johannes Gutenberg University, 55099 Mainz, Germany}
\affiliation{Helmholtz Institute Mainz, 55099 Mainz, Germany}
\affiliation{Department of Physics, University of California, Berkeley, Berkeley, California 94720-\affiliation{Johannes Gutenberg University, 55099 Mainz, Germany}
\affiliation{Helmholtz Institute Mainz, 55099 Mainz, Germany}
\affiliation{School of Physics, University of New South Wales, Sydney, New South Wales 2052,  Australia}7300, USA}
\affiliation{Nuclear Science Division, Lawrence Berkeley National Laboratory, Berkeley, California 94720-7300, USA}
\author{Scott A.\,Crooker}
\affiliation{National High Magnetic Field Laboratory, Los Alamos, New Mexico 87545, USA}
\author{Edward J.\,Daw}
\affiliation{Department of Physics and Astronomy, The University of Sheffield, Hicks Building, Hounsfield Road,  Sheffield S3 7RH, United Kingdom}
\author{Victor V.\,Flambaum}
\affiliation{Johannes Gutenberg University, 55099 Mainz, Germany}
\affiliation{Helmholtz Institute Mainz, 55099 Mainz, Germany}
\affiliation{School of Physics, University of New South Wales, Sydney, New South Wales 2052,  Australia}
\author{Toshiaki Inada}
\affiliation{International Center for Elementary Particle Physics, The University of Tokyo,  7-3-1 Hongo, Bunkyo, Tokyo 113-0033,  Japan}
\author{Igor G. Irastorza}
\affiliation{Grupo de Fisica Nuclear y Astroparticulas. Departamento de Fisica Teorica. Universidad de Zaragoza,  50009 Zaragoza,  Spain}
\author{Felix Karbstein}
\affiliation{Helmholtz-Institut Jena, Fr\"obelstieg 3,  07743 Jena,  Germany}
\affiliation{Theoretisch-Physikalisches Institut, Abbe Center of Photonics,  Friedrich-Schiller-Universit\"at Jena, Max-Wien-Platz 1,  07743 Jena,  Germany}
\author{Dong Lak Kim}
\affiliation{Center for Axion and Precision Physics Research, IBS,  Daejeon 34051,  Republic of Korea}
\author{Mikhail G.\,Kozlov}
\affiliation{Petersburg Nuclear Physics Institute of NRC ``Kurchatov Institute'', Gatchina 188300, Russia}
\affiliation{St. Petersburg Electrotechnical University ``LETI'', Professor Popov Street 5,   St. Petersburg 197376,  Russia}
\author{Ziad Melhem}
\affiliation{Oxford Instruments NanoScience,   Tubney Woods, Abingdon, Oxon, OX13 5QX,  United Kingdom}
\author{Arran Phipps}
\affiliation{Department of Physics, Stanford University,   Stanford, California 94305,  USA}
\author{Pierre Pugnat}
\affiliation{Laboratoire National des Champs Magn\'{e}tiques Intenses (LNCMI), European Magnetic Field Laboratory (EMFL), CNRS \& Universit\'{e} Grenoble-Alpes,  38042 Grenoble Cedex 9,  France}
\author{Geert Rikken}\email{geert.rikken@lncmi.cnrs.fr}
\affiliation{Laboratoire National des Champs Magn\'etiques Intenses (UPR 3228, CNRS-UPS-UGA-INSA), F-31400 Toulouse Cedex, France}
\author{Carlo Rizzo}\email{carlo.rizzo@lncmi.cnrs.fr}
\affiliation{Laboratoire National des Champs Magn\'etiques Intenses (UPR 3228, CNRS-UPS-UGA-INSA), F-31400 Toulouse Cedex, France}
\author{Matthias Schott}
\affiliation{Johannes Gutenberg University, 55099 Mainz, Germany}
\author{Yannis K. Semertzidis}
\affiliation{Center for Axion and Precision Physics Research, IBS,  Daejeon 34051,  Republic of Korea}
\affiliation{Department of Physics, KAIST,  Daejeon 34051,  Republic of Korea}
\author{Herman H.\,J.\,ten Kate}
\affiliation{CERN Experimental Physics Department,  POB Geneva 23, CH1211,   Switzerland}
\author{Guido Zavattini}
\affiliation{INFN-Sezione di Ferrara and Dipartimento di Fisica e Scienze della Terra, Universit\`a di Ferrara, Polo Scientifico, Via Saragat 1, Blocco C,  44100 Ferrara,   Italy}

\begin{abstract}
Various fundamental-physics experiments such as measurement of the birefringence of the vacuum, searches for ultralight dark matter (e.g., axions), and precision spectroscopy of complex systems (including exotic atoms containing antimatter constituents) are enabled by high-field magnets. We give an overview of current and future experiments and discuss the state-of-the-art DC- and pulsed-magnet technologies and prospects for future developments. 
\end{abstract}

\maketitle

\tableofcontents

\section{Introduction}

%















While production of high magnetic fields is a mature field, recent years have seen significant developments in both pulsed and continuous-operation (often referred to as ``DC'') magnets. These developments are driven by applications in materials sciences, nuclear magnetic resonance (NMR), thermonuclear-energy research, particle accelerators, and particle detectors (see, for example, a High-Magnetic-Field study by the US National Academy of Sciences  \cite{national2013high}).

In this review we discuss the state-of-the-art of magnet technologies with an eye towards their application to ``small-scale'' fundamental physics such as searches for magnetic birefringence of the vacuum and for axions/axion-like particles (ALPs) with haloscope and helioscope devices and light-through-wall experiments, etc. 

\begin{figure}[h!]
    \begin{center}
    \includegraphics[width=5.5 in]{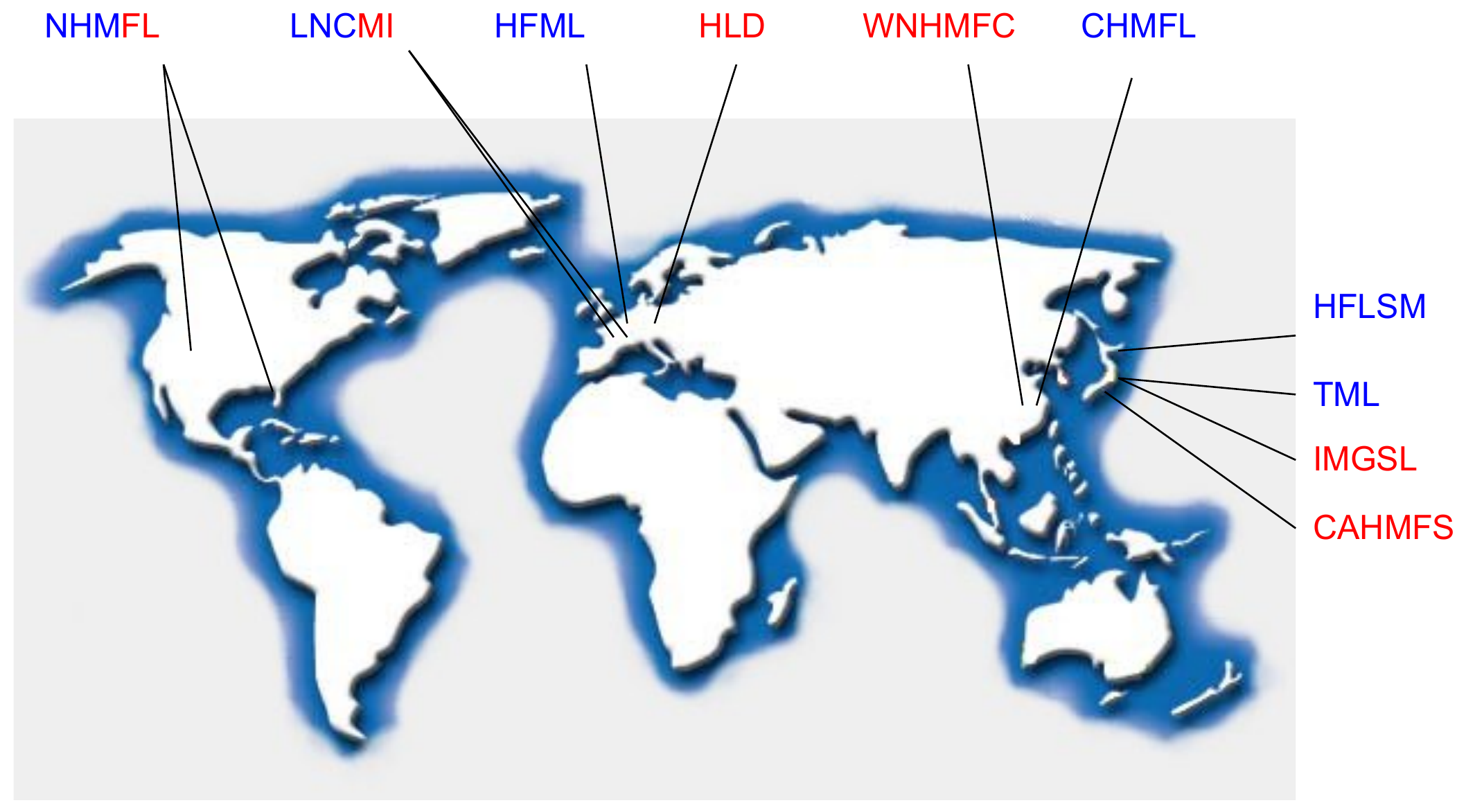}
    \caption{Laboratories around the world pursuing high pulsed (red) and DC (blue) magnetic fields.}
    \label{Fig-High Field Facilities}
    \end{center}
\end{figure}

There are several laboratories around the world (Fig.\,\ref{Fig-High Field Facilities} and Table\, \ref{Fig-High Field Facilities}) pursuing research into high magnetic fields, including their production and scientific applications. 
In Europe, there is the French Laboratoire National des Champs Magn\'{e}tiques Intenses (LNCMI) which has two sites, one in Grenoble with a specialization in DC magnets and one in Toulouse with a specialization in pulsed magnets, the German Dresden High Magnetic Field Laboratory (Hochfeld-Magnetlabor Dresden, HLD), and the High Field Magnet Laboratory (HMFL) in Nijmegen,  
the Netherlands. The European high-field laboratories operate as coordinated infrastructure within the framework of the European Magnetic Field Laboratory (EMFL). In America, there is the US National High Magnetic Field Laboratory (NHMFL) with one location at Los Alamos, New Mexico, and two locations in Florida  (Tallahassee and Gainesville). 
In Asia, there are two laboratories in China, the High Magnetic Field Laboratory of the Chinese Academy of Sciences (CHMFL) in Hefei and the Wuhan National High Magnetic Field Center (WHMFC), and four laboratories in Japan: the Tsukuba Magnet Laboratory (TML), the High Field Laboratory for Superconducting Materials in Sendai, the International Megagauss Science Laboratory (IMGSL) in Kashiwa, and the Center for Advanced High Magnetic Field Science in Osaka.  The laboratories and the parameters of some of their magnets are listed in Table\,\ref{Fig-High Field Facilities}.

\begin{table}[h]
\caption{High-magnetic-field laboratories around the globe. SD stands for ``semi-destructive, SC stands for superconducting, '' $B/T$ is magnetic field/temperature; the dimension stands for the diameter of the magnet's bore. ``Best performance'' does not include many practical factors such as whether it represents routine operation or a rare record and to what extent the facility is available to users.}
\smallskip 
\center
\begin{tabular}{llll}
\hline \hline
Name &	Location &	Type &	Best performance \\
\hline
\rule{0ex}{2.6ex} Laboratoire National des Champs Magnetiques Intenses & &	 \\
\rule{0ex}{2.6ex} - DC-field facility		& Grenoble, France 	& DC 	& 37 T in 34 mm \\
\rule{0ex}{2.6ex} - pulsed-field facility		& Toulouse, France 	& pulsed	& 98.8 T in 8 mm \\
\rule{0ex}{2.6ex}  				& 			 	& 		& 209 T in 8 mm SD \\
\rule{0ex}{2.6ex} High Field Magnet Laboratory & 	the Netherlands &	DC	 & 38 T in 32 mm \\
\rule{0ex}{2.6ex} Hochfeld Labor Dresden	& Dresden, Germany &	pulsed &	95.6 T in 16 mm \\
\rule{0ex}{2.6ex} National High Magnetic Field Laboratory	& 	 &  \\
\rule{0ex}{2.6ex} - DC-field facility	& Tallahassee, USA	& DC &	45 T in 32 mm hybrid \\
\rule{0ex}{2.6ex} - pulsed-field facility 	& Los Alamos, USA	& pulsed  & 100.7 T in 10 mm \\
\rule{0ex}{2.6ex} 			 	& 				& 	    & 240 T in 8 mm SD \\
\rule{0ex}{2.6ex} - high-$B/T$ facility 	& Gainesville, USA	& DC	& 16.5 T at 0.4 mK \\
\rule{0ex}{2.6ex} Wuhan National High Magnetic Field Center &	Wuhan, China	& pulsed	& 90.6 T in 12 mm \\
\rule{0ex}{2.6ex} Chinese High Magnetic Field Laboratory &	 Heifei, China &	DC &	40 T in 32 mm hybrid \\
\rule{0ex}{2.6ex} International Megagauss Science Laboratory &	Kashiwa, Japan &	pulsed &	75 T in 17 mm \\
\rule{0ex}{2.6ex}  								 &				&		  &	250 T in 8 mm SD \\
\rule{0ex}{2.6ex} High Field Laboratory for Superconducting Materials	& Sendai, Japan 	& DC &	30 T in 32 mm hybrid \\
\rule{0ex}{2.6ex}                    	                                              &                                            	&       &	24.6 T in 52 mm all-SC \\
\rule{0ex}{2.6ex} Center for Advanced High Magnetic Field Science &	Osaka, Japan	& pulsed	& 60 T \\
\rule{0ex}{2.6ex} Tsukuba Magnet Laboratory &	Tsukuba, Japan	& DC 	& 32 T in 52 mm hybrid\\
\rule{0ex}{2.6ex}  								 &				&		  &	35 T in 30 mm hybrid\\
\hline \hline
\end{tabular}
\label{Table:High Field Facilities}
\end{table}
\begin{table}[h]
\caption{Highest-field magnets currently in operation. Highest field does not necessarily imply high figure-of-merit (see text).}
\smallskip 
\center
\begin{tabular}{llll}
\hline \hline
Country & Field (T) & Laboratory & feature \\
\hline
\rule{0ex}{2.6ex} DC magnets &  &  &  \\

\rule{0ex}{2.6ex} USA & 45 & NHMFL & hybrid \\
\rule{0ex}{2.6ex} China & 40 & CHMFL & hybrid \\
\rule{0ex}{2.6ex} Europe & 38 & HFML &  \\
\rule{0ex}{2.6ex} Japan & 35 & TML & hybrid \\

\hline
\rule{0ex}{2.6ex} Pulsed nondestructive magnets &  &  &  \\

\rule{0ex}{2.6ex} USA & 100.7 & NHMFL &  \\
\rule{0ex}{2.6ex} Europe & 98.8 & LNCMI &  \\
\rule{0ex}{2.6ex} China & 90.6 & WHMFC &  \\
\rule{0ex}{2.6ex} Japan & 86 & IMGSL &  \\

\hline \hline
\end{tabular}
\label{Table:high-field magnets}
\end{table}

Generation of magnetic fields involves flowing an electric current through a conductor (typically, a coil), with the field intensity $B$ being proportional to the current $I$. For a coil with a fixed resistance $R$, the heat dissipation $I^2R$ is proportional to the square of  the magnetic field. At the same time, mechanical pressure also scales as $B^2$, with a proportionality coefficient of approximately 4\,atm/T$^2$. Heating and mechanical forces are the two essential challenges for the design of high-field electromagnets. The stored energy in magnetic field is proportional to the integral of $B^2$ over the volume of the magnet, so the size of the field volume is an important characteristic of the magnet along with the maximum field. Another key parameter is, of course, whether a magnet is DC or pulsed and, in the latter case, additional important parameters are the duration and temporal profile of the pulse, as well as the pulse repetition rate.

\begin{figure}[h!]
    \begin{center}
    \includegraphics[width=5.5 in]{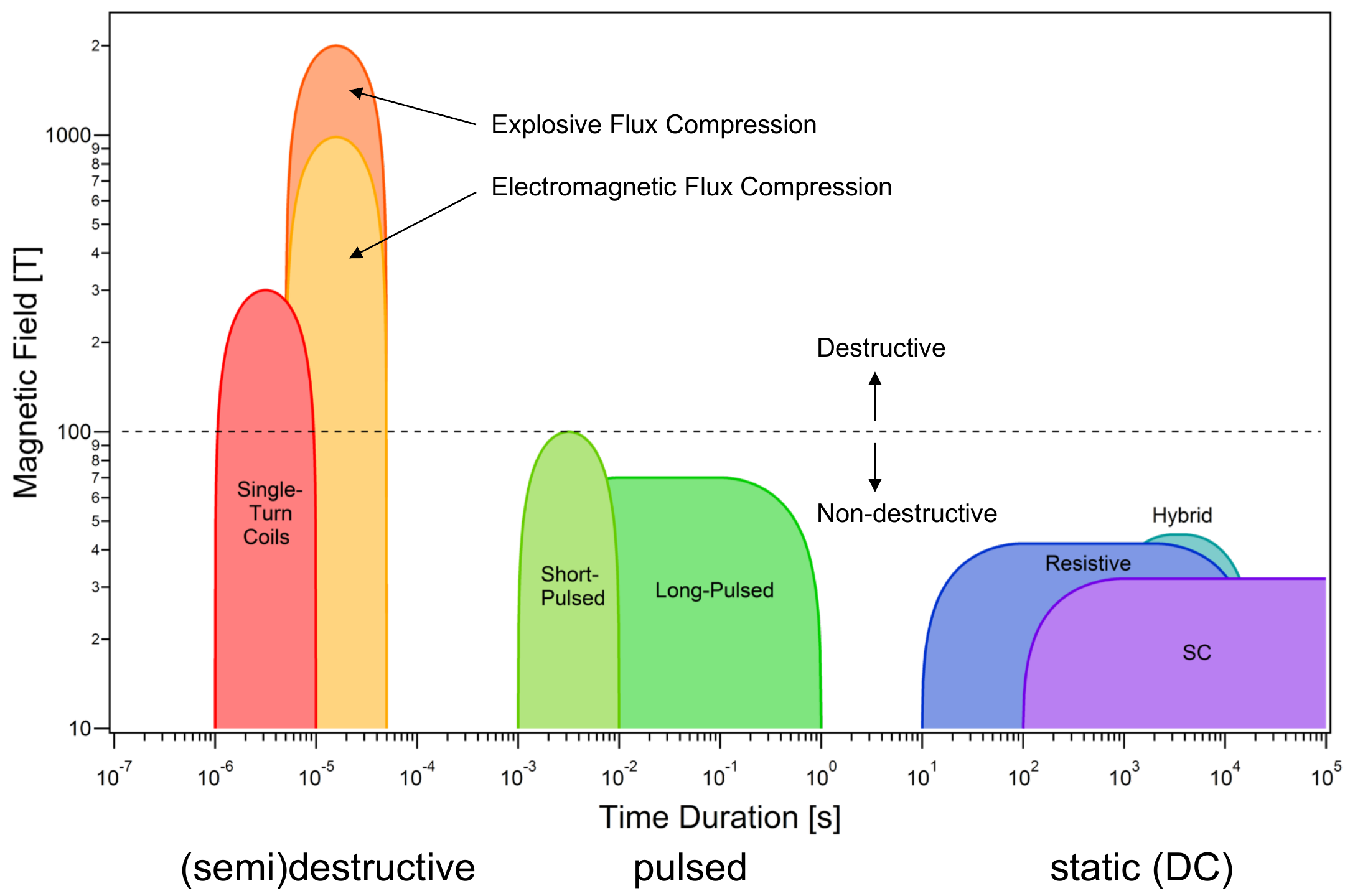}
    \caption{Overview of strong-magnet methods; see also Table \ref{Table:High Field Facilities}.}
    \label{Fig high field magnet methods}
    \end{center}
\end{figure}

The approaches to meeting the heat-dissipation and mechanical-stability challenges are different for different types of magnets as summarized in Fig.\,\ref{Fig high field magnet methods} and discussed in Sec.\,\ref{Sec: Current and forthcoming magnet technologies}. However, in all cases, materials are crucially important for the development of advanced magnets;  such development is thus a task at the intersection of physics, engineering, and materials sciences. 

Some of the world's highest-field nondestructive magnets are listed in Table\, \ref{Table:high-field magnets}. However, it is frequently the case that the highest-field magnet is not necessarily the best choice for a particular experiment. Indeed, as summarized in Table\,\ref{Table:FOMs}, the figure-of-merit (FOM), depending on the experiment, could be the stored energy $B^2V$ (where $B$ is the magnetic field and $V$ is the field volume), or the effective $B^2L$ (where $L$ is the length of the field), or something completely different such as the ability to tune the magnet to a desired field value, or the ability to tune the field over a broad range,  while maintaining spatial homogeneity and low ripple. 
\begin{table}[h]
\caption{Figure-of-merit (FOM) for various fundamental-physics experiments. $L$, $A$ and $V$ are the characteristic length, transverse area and volume of the magnetic-field region. The last column lists the sections of this paper where the corresponding experiments are discussed.}
\smallskip 
\center
\begin{tabular}{llll}
\hline \hline
Experiment & FOM & Examples & Section \\
\hline
\rule{0ex}{2.6ex}Vacuum birefringence & $B^2L$  & BMV, PVLAS, OVAL & \ref{Subsect:VB} \\
\rule{0ex}{2.6ex}Light shining through wall & $B^4L^4$  & ALPS, OSQAR, ... & \ref{Subsubsec:LSW}  \\
\rule{0ex}{2.6ex}Helioscope & $B^2L^2A=B^2VL$ & CAST, IAXO  & \ref{subsubsect:Helioscopes} \\
\rule{0ex}{2.6ex}Haloscope (Primakoff) & $B^2V$  & ADMX, HAYSTAC, ORGAN, CULTASK ... & \ref{Subsubsec: Haloscopes}  \\
\rule{0ex}{2.6ex}Haloscope (other) & None of the above  & CASPEr, QUAX, ... & \ref{Subsubsec: Haloscopes}  \\

\hline \hline
\end{tabular}
\label{Table:FOMs}
\end{table}

Maxwell's equations possess symmetry with respect to electric and magnetic fields. While this symmetry is explicitly seen in Gaussian units where electric field is measured in the same units as magnetic induction (gauss), the symmetry is somewhat obscured in the International System. To compare electric and magnetic fields, one may use the fact that $1\,$T corresponds to $3\cdot 10^8\,$V/m. This allows us to ``compare'' strong laboratory electric and magnetic fields. For example, the highest static electric fields that can be produced in macroscopic volumes are significantly below $10^8\,$V/m, which is a mere 0.3 T! Clearly, magnetic fields win at the ``strong-field game,'' particularly when one is studying the properties of the vacuum such as field-induced birefringence! 

The present review is structured as follows. We begin with a discussion of some of the current fundamental-physics experiments (Sec.\,\ref{Sec:Expts}) using strong fields, leaving aside many important topics such as magnets for accelerators, high-energy detectors, NMR, thermonuclear fusion, and neutron scattering, only mentioning some of the corresponding developments in passing. We then present several ideas for future experiments in Sec.\,\ref{Sec:Ideas}. State-of-the-art magnet technologies are discussed in Sec.\,\ref{Sec: Current and forthcoming magnet technologies}, while magnetic-field measurement techniques are discussed in Sec.\ref{Sec:B_Measurement}.  Conclusions and outlook on the future of the field are presented in Sec.\,\ref{Sec:Conclusions}.

\clearpage
\section{Experiments}
\label{Sec:Expts}

In this section, we discuss several classes of experiments that use strong magnetic field to probe the quantum vacuum close to the conditions where nonlinear-electrodynamics effects such as vacuum optical birefringence (Sec.\,\ref{Subsect:VB}) are expected to be observable and that are sensitive to various beyond-the-standard-model scenarios. One of the major goals of high-magnetic-field experiments is a search, either direct or indirect, for possible constituents of dark matter (Sec.\,\ref{Subsect: AxionSearches}).

\subsection{Vacuum birefringence}
\label{Subsect:VB}

\subsubsection{Vacuum-birefringence tests with quasi-stationary magnetic fields}
\label{subsubsection:Vacuum-birefringence tests with quasi-stationary magnetic fields}
 
In the framework of classical electrodynamics, governed by Maxwell's linear equations, vacuum isotropy and the speed of light are not affected by the application of electromagnetic fields. This property distinguishes vacuum from material medium. Going beyond classical electrodynamics, for example, by accounting for virtual particle-antiparticle fluctuations in the framework of quantum electrodynamics (QED), mediating effective interactions among electromagnetic fields \cite{Heisenberg1935,Weisskopf:1936,Schwinger1951}, one finds that nonlinear phenomena such as photon-photon scattering \cite{Euler:1935zz,Karplus:1950zz,Karplus:1950zza} and magnetic-field-induced birefringence of vacuum \cite{Toll:1952rq,Baier1967,Baier1967APA} exist, albeit in regimes that are not easy to achieve experimentally. At the same time, QED can be seen as a special case of nonlinear electrodynamics (NLED)  \cite{Fouche2016}, therefore, a more general framework exists that allows a uniform description and comparison of different experiments \cite{BattestiRPP2013} that test QED predictions and constrain alternative models that predict bigger nonlinear effects. A prominent example is Born-Infeld electrodynamics \cite{Born:1934gh}. Historically, nonlinear self-interaction of electromagnetic field as a consequence of the Dirac theory was first considered by Heisenberg and Euler \cite{Heisenberg1935}; see the pertinent historical review \cite{Scharnhorst:2017wzh}. For a modern theoretical perspective, see \cite{Dittrich:1985yb,Dittrich:2000zu,Gies:2016yaa} and references therein. While there are indications for the relevance of vacuum birefringence for explaining the observed linear polarization of the light from a neutron star \cite{Mignani:2016fwz,Capparelli:2017mlv}, unambiguous detection in the laboratory is being attempted by several groups.

\begin{figure}[h!]
    \begin{center}
    \includegraphics[width=5.5 in]{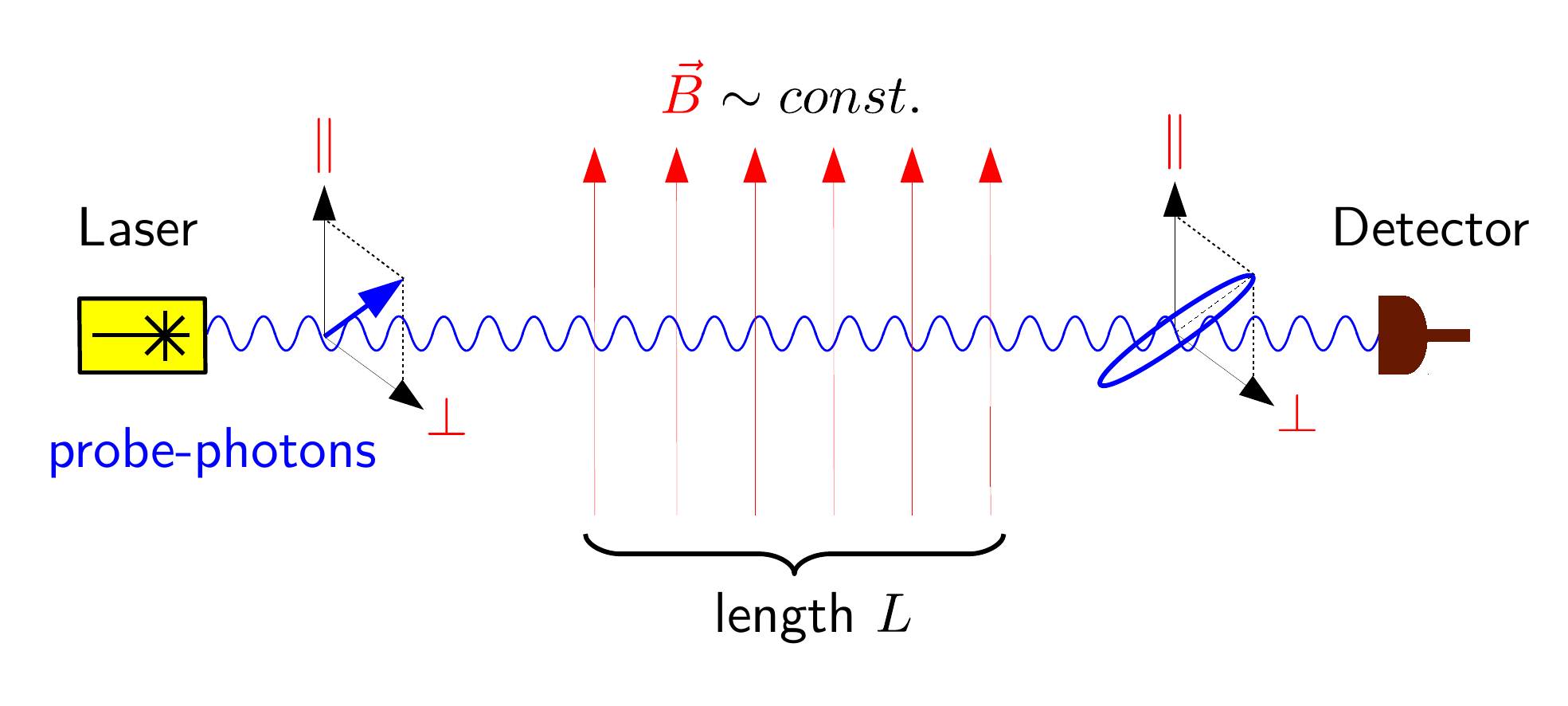}
    \caption{The concept of the vacuum-birefringence experiments.}
    \label{Fig Vacuum birefr expt principle}
    \end{center}
\end{figure}

\begin{figure}[h!]
    \begin{center}
    \includegraphics[width=5.5 in]{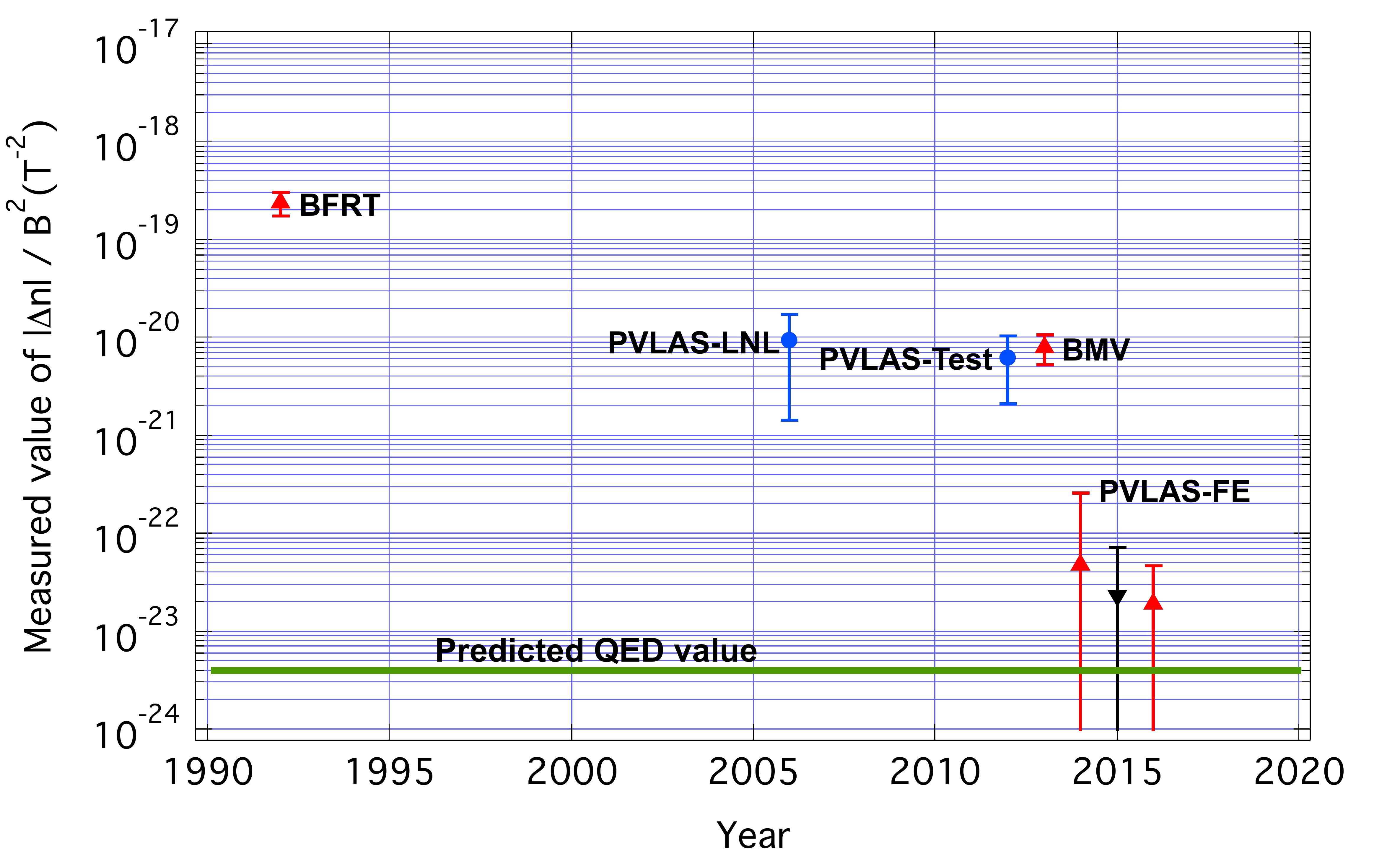}
    \caption{Time evolution of the absolute values of the measured vacuum magnetic birefringence normalized to $B^2$. Red up triangles refer to positive central values for $\Delta n/B^2$ whereas the blue circles and black down triangles refer to undetermined sign and negative central values, respectively. The error bars correspond to $1\sigma$. Several measurements correspond to central values larger than the declared errors indicating the presence of undetermined systematics. The values derive from the following references:  BFRT (the Brookhaven-Fermilab-Rutherford-Trieste experiment) \cite{Cameron:1993mr},
PVLAS-LNL (LNL: Italian National Laboratories in Legnaro) \cite{Zavattini2008}, PVLAS-Test \cite{PVLAS-TEST2013}, BMV \cite{Cadene2014}, PVLAS-FE (FE: the University of Ferrara) \cite{Ejlli2017}.}
    \label{Fig VB timeline}
    \end{center}
\end{figure}

The general idea for vacuum-birefringence experiments \cite{IACOPINI1979} is illustrated in Fig.\,\ref{Fig Vacuum birefr expt principle}. 
The initial polarization of the light is at $45^\circ$ to the magnetic field; the difference in the light speed for the light polarized along and perpendicular to the magnetic field induces light ellipticity given by
\begin{equation}
 \Phi=2\pi\frac{L}{\lambda}\Delta n\,,
\end{equation}
where $L$ is the length of the magnetic field, $\lambda$ is the wavelength of the light, and $\Delta n$ is the difference of the refractive indices for light polarized parallel and perpendicular to the magnetic field (the Cotton-Mouton effect\footnote{So named after A. Cotton and H. Mouton who studied magnetic birefringence effects in liquids in the beginning of the 20th century.} of the vacuum). More specifically, we have (in SI units)
\begin{equation}
 \Delta n=n_\parallel-n_\perp=k_\text{CMV}|\vec{B}|^2\,,\label{Eq:CMofVac Delta n}
\end{equation}
where $k_\text{CMV}$ is the Cotton-Mouton coefficient of the vacuum. According to standard QED,
\begin{equation}
k_\text{CMV}=\frac{1}{\mu_0}\frac{\alpha}{\pi}\frac{1}{30}\left ( \frac{e\hbar}{m^2c^2}\right )^2\approx4.0\cdot10^{-24}{\rm T}^{-2},\label{Eq:CMofVac}
\end{equation}
where $\mu_0$ is vacuum permeability, $\alpha\approx 1/137$ is the fine structure constant and $m$ is the electron mass. Up to a numerical constant, the difference of the refractive indices due to the vacuum Cotton-Mouton effect of Eq.\,\eqref{Eq:CMofVac Delta n} is just the square of the ratio of magnetic energy of a magnetic dipole equal to a Bohr magneton immersed in field $B$, $e\hbar B/(2m)$, and the electron rest energy, $mc^2$. The quantity $\left (\frac{e\hbar}{m^2c^2}\right )^{-1}\sim10^{10}$\,T is the critical magnetic field. Its electric-field analogue governs the onset of spontaneous electron-positron pair production.
Upon propagation of initially linearly polarized light traveling perpendicular to a quasi-static magnetic field, the magnetic field induced change in light polarization is analyzed (Fig.\ \ref{Fig Vacuum birefr expt principle}). 
With practically achievable values of $B^2L$, the expected optical phase shifts are small and challenging to detect with current polarimetry techniques, though the experiments are getting tantalizingly close to the QED predictions (Fig.\ \ref{Fig VB timeline}).

As noted above, as long as the transverse dimensions of the magnet are sufficient to accommodate the laser beam, the figure-of-merit for a vacuum-birefringence experiment is:
\begin{equation}
		\textrm{FOM}_{\textrm{Vacuum\,birefr.}}\ =B^2 L. \label{Eq:FOM for vac birefr}
\end{equation} 
\begin{table}[h]
\center
 \begin{tabular}{lccc}
 \hline\hline
 & BMV~\cite{Cadene2014} & PVLAS~\cite{PVLAS2016,Ejlli2017} & OVAL~\cite{Fan2017} \\ 
 \hline
 Magnet type & pulsed & rotating permanent & pulsed \\
 Maximum field $B_\text{max}\,[{\rm T}]$ & $6.5$ & $2.5$ & $9.0$ \\
 Field length $L\,[{\rm m}]$ & $0.14$ & $0.82+0.82$ & $0.17$ \\
 $B^2L\,[{\rm T^2m}]$ & $5.8$ & $5.06+5.06$ & $13.8$ \\
``Filtered'' $B^2L\,[{\rm T^2m}]$ & $2.7$ & $5.06+5.06$ & $4.1$ \\
 Wavelength $\lambda\,[{\rm nm}]$ & $1064$ & $1064$ & $1064$ \\
 Cavity finesse $\mathcal{F}$ & $445\,000$ & $700\,000$ & $320\,000$ \\
 Integration time $[{\rm s}]$ & $0.3$ & $2\cdot10^6$ & $0.6$ \\
 Repetition rate $[{\rm Hz}]$ & $0.0017$ & continuous & $0.17$ \\
 Uncertainty in $\Delta n\,[10^{-23}/{\rm T}^2]$ & $270$ & $2.7$ & $110\,000$ \\
 \hline\hline
 \end{tabular}
 \caption{Parameters of the three operating vacuum-birefringence experiments. Effective values are given for $L$ and $B^2L$, see text. Note that the OSQAR experiment at CERN originally aimed at measuring vacuum magnetic birefringence using an LHC dipole magnet providing $B^2L = 1150$\,T$^2$m \cite{pugnat2005}; however, it has operated as a light-shining-through-wall experiment \cite{Ballou2015}, see Sec\,\ref{Subsubsec:LSW}.} \label{Tab: VMB experiments}
\end{table}

Currently ongoing vacuum-birefringence experiments include the BMV (Bir\'{e}fringence Magn\'{e}tique du Vide) experiment in Toulouse, the
PVLAS (Polarizzazione del Vuoto con Laser) experiment in Ferrara, and the OVAL (Observing Vacuum with Laser) experiment in Tokyo. Some key parameters of these experiments are summarized in Table~\ref{Tab: VMB experiments}.
Note that it is customary to present the maximum value of the field $B_{\rm max}$, the effective value of $B^2L$ which is an integral of $B^2$ along the path of the light beam, as well as the effective value of $L$, which is the effective value of $B^2L$ divided by maximum $B_{\rm max}^2$.  All current experiments employ optical cavities in order to enhance the effect of the birefringence, by a factor of the order of the cavity finesse. For pulsed experiments, ``filtered'' values for $B^2L$ are listed in Table~\ref{Tab: VMB experiments}. What this means is the following: high-finesse optical cavities used in these experiments ``trap'' light for times that exceed the duration of the magnetic-field pulse. In this situation, the cavity acts as a low-pass filter for the signal. For instance, in BMV, a characteristic signal bandwidth  is $\approx 100\,$Hz, while the cavity cuts at $\approx 70\,$Hz. The final signal is almost half of what one would calculate without taking into account the cavity filtering. One can also see the filtering effect in the following way. Photons while trapped between the cavity mirrors ``see'' the pulsed magnetic field changing so the effect is given by an averaged value of the magnetic field,  smaller than the nominal one. The  PVLAS modulating at $5\,$Hz has almost no cavity filtering. Because of cavity filtering, there is nothing to gain by increasing the cavity finesse much further, or by shortening the magnetic-field pulse.

\begin{figure}[h!]
    \begin{center}
    \includegraphics[width=6.5 in]{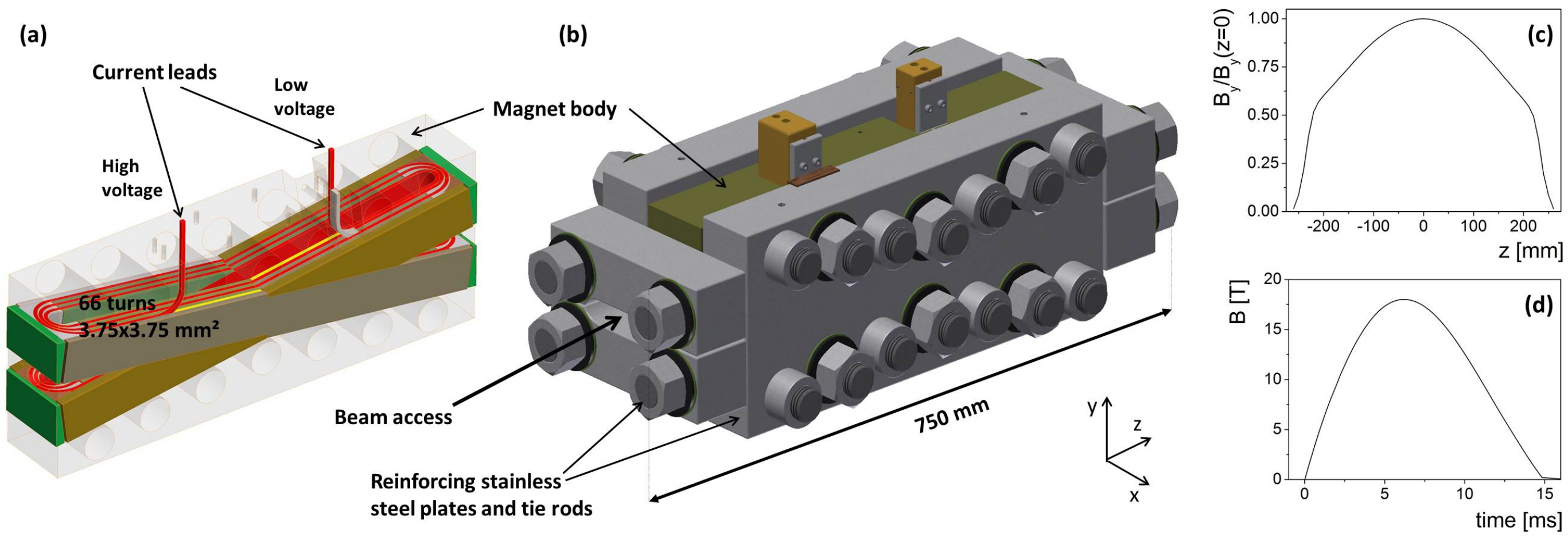}
    \caption{The LNCMI XXL-Coil (the magnet currently used in the BMV\,2 experiment). (a) Some winding details of this X-shape magnet. It consists of two interlaced coils that leave openings on two sides for the beam access. The conductor (in red) is in Hog\"{a}n\"{a}s Glidcop\textsuperscript{\textregistered} AL-15 insulated with Kapton\textsuperscript{\textregistered}. (b) Final assembly with stainless-steel reinforcement and brass connectors. The magnet is installed in a Dewar (not shown) and cooled with liquid nitrogen before a pulse. Optical-access room-temperature port is 13\,mm diameter. The symmetry in the alternating layers from one branch of the X to the other limits the axial components of the magnetic field to a few percent of the total field. Copper screens (not shown) about one centimeter thick, are added on both sides to decrease Faraday component of the magnetic field by about one order of magnitude at the positions of the cavity mirrors of the BMV experiments. (c) Spatial profile of the transverse magnetic field along the beam path. (d) Temporal profile of the magnetic field for an 18\,T pulse. Figure courtesy LNCMI-Toulouse.}
    \label{Fig BMV Magnet}
    \end{center}
\end{figure}

Since the science run with the parameters shown in Table~\ref{Tab: VMB experiments}, the BMV experiment is being significantly upgraded \cite{Hartman2017}. The so-called BMV\,2 is based on a new pulsed magnet developed at the LNCMI-Toulouse \cite{Batut2008}, the XXL-Coil (Fig.\,\ref{Fig BMV Magnet}). A prototype was tested up to 30\,T corresponding to $B^2L\approx 278\,$T$^2$m and one magnet is currently in use at LNCMI providing up to 18\,T, $B^2L\approx 100\,$T$^2$m. The rise time of the magnetic pulse is 6\,ms, corresponding to a typical frequency for $B^2$ of about 80\,Hz, which is not too far above the optical-cavity frequency cut-off, limiting the filtering effect due to the high finesse. The total pulse duration is 14\,ms with cooling duration between two pulses fixed to (conservative) one minute. The repetition rate is not limited by the Joule heating in the magnet and the cooling duration could be reduced about one order of magnitude for the final data-accumulation process. Commissioning of the apparatus is nearing completion, first data runs are expected by the end of 2018.

\begin{figure}[h!]
    \begin{center}
    \includegraphics[width=6.5 in]{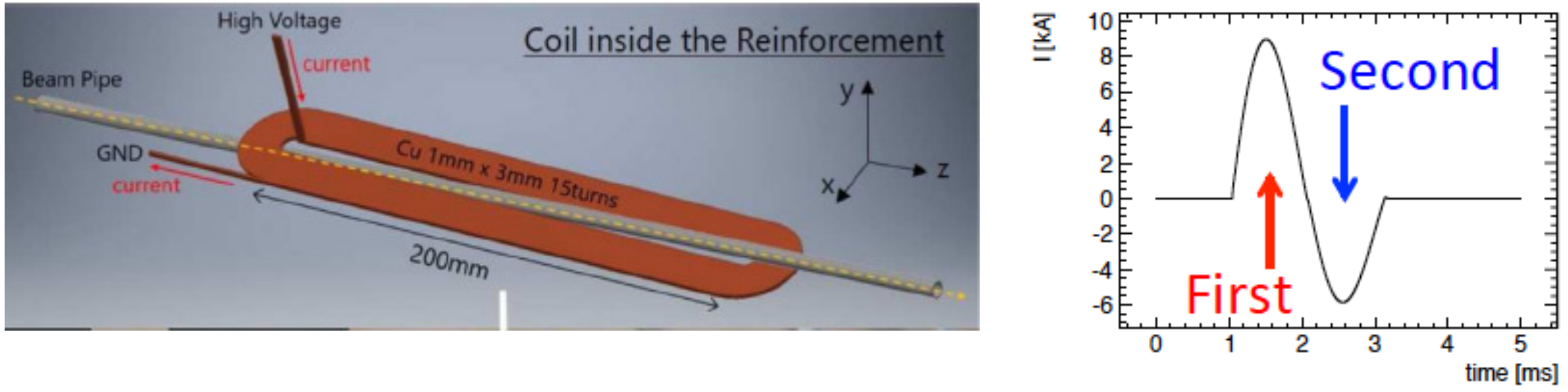}
    \caption{Left: one of the four racetrack coils used in the University of Tokyo high-repetition-rate pulsed magnet \cite{Yamazaki2016}. Right: temporal profile of the current during a single pulse. Part of the energy stored in the capacitor bank is recycled.}
    \label{Fig_TokyoPulsed}
    \end{center}
\end{figure}

A pulsed magnet for vacuum birefringence measurements and axion-like particle searches (Sec.\, \ref{Subsubsec:LSW}) was developed at the University of Tokyo \cite{Yamazaki2016}, demonstrating $9\,$T over a length of over $0.8\,$m at a $0.1\,$Hz repetition rate. The figure-of-merit for the vacuum birefringence experiments is $B^2L=54\,$T$^2$m, achieved at a repetition rate that is much higher than for other pulsed magnets used for such experiments. The system utilizes ``racetrack'' coils (Fig\,\ref{Fig_TokyoPulsed}, left) and takes full advantage of the fact that high $B^2L$ can be be achieved in a small field volume, resulting in relatively small heating and allowing for efficient cooling. The high repetition rate of the system is enabled by an energy-recycling scheme (Fig\,\ref{Fig_TokyoPulsed}, right), where the field-pulse temporal profile is bipolar and a significant fraction of the energy is thus returned to the capacitor bank at the end of the pulse. 

Operation of pulsed magnets is associated with considerable mechanical shock and acoustic disturbance presenting a significant challenge to the operation of precision experiments. What nevertheless makes such measurements possible is the finite propagation time of acoustic waves. For example, acoustic shock arrives at the optical elements in the OVAL experiment some $4\,$ms after the magnetic pulse.  

The PVLAS experiment takes advantage of several techniques to maximize the signal and isolate it from possible systematics. A 3.3\,m long Fabry-Perot cavity increases the single-pass
ellipticity by a factor of $N_{\rm pass}=2\mathcal{F}/\pi$, where $\mathcal{F}$ is the finesse of the cavity (in PVLAS, $\mathcal{F}=7\cdot 10^5$ and the circulating light power is 40 kW; the light wavelength is 1.064$\,\mu$m); a heterodyne technique is used to ensure that the observed signal is linear in the induced ellipticity (similar ``tricks'' are also used in other polarimetry experiments). PVLAS uses two permanent magnets, each producing 2.5\,T field with field length of about 0.8\,m, from AMT\&C (Moscow), which are Halbach-array dipoles with 20\,mm bore.  The Halbach-array configuration provides ``self-shielding:'' the field outside the array is small. Additional ferromagnetic shielding is installed around the sides and ends of the magnets. As a result, the fields at the outer surfaces of the cylindrical magnets is less that 0.1\,mT. The magnets rotate around their axis in order to modulate the signal. The two magnets can be independently rotated with frequencies up to 23\,Hz, which important for identifying and suppressing systematic effects. The advantages of the permanent magnets include their relatively low cost ($\approx 100\ $k\EUR for both magnets; no power supplies needed), the absence of running costs, as well as good mechanical stability. It was possible to balance the magnets with extra wights ($\approx 10\ $g) to minimize vibrations.

The standard QED vacuum birefringence ~\eqref{Eq:CMofVac Delta n} for $B=2.5\ $T is $\Delta n = 2.5\cdot 10^{-23}$, which, for the current parameters of the PVLAS experiment corresponds to induced ellipticity of $5\cdot 10^{-11}\ $rad. 

The present ellipticity sensitivity of PVLAS (which is significantly above the shot noise) is $\approx 2.5\cdot 10^{-7}\ $rad/$\sqrt{\textrm{Hz}}$ corresponding to birefringence sensitivity $\Delta n \approx 1.2\cdot 10^{-19}/\sqrt{\textrm{Hz}}$, so it would take about three years of data taking to reach statistical sensitivity corresponding to the QED expectations.

What is the way forward for the vacuum-birefringence experiments? In the case of DC magnets, from the point of view of the $B^2L$ figure-of-merit, switching to accelerator superconducting dipoles appears particularly attractive and realistic. However, rotation or sufficiently fast modulation of the field does not appear practical in this case, so the experiment would have to do with some other ways to modulate the signal. Polarization-modulation is a possible approach \cite{Zavattini2016,Zavattini2017}, but systematics-free polarimetry with a necessary sensitivity is yet to be demonstrated. An interesting idea \cite{Dobrich:2009kd,Zavattini2009,Grote2015} involves adding magnets to the arms of a gravitational-wave detector such as VIRGO, taking advantage of the existing ultrahigh-precision optical interferometers at these facilities.  

\subsubsection{Probing vacuum birefringence with high-intensity lasers}
\label{subsubsection:Probing vacuum birefringence with x-ray and high-intensity lasers}

While most of the current review is devoted to production of and experiments with static or quasi-static magnetic fields in macroscopic volumes, there are other ways to address some of the related fundamental physics using strong electromagnetic fields associated with focused ultrafast laser pulses. While we do not attempt to survey this vast research area, as one particular example, we discuss vacuum-birefringence experiments along these lines and compare their impact to the more traditional vacuum-birefringence  experiments discussed in Sec.\ \ref{subsubsection:Vacuum-birefringence tests with quasi-stationary magnetic fields}.

The field strength associated with laser pulses can exceed those of quasi-static laboratory fields by many orders of magnitude \cite{di2012extremely} (albeit over much smaller lengths and time scales). For pulsed laser beams focused into a micrometer-sized spot, these could be as high as  10$^5-10^6\ $T for magnetic fields, and, correspondingly, as high as over $10^{12}\ $V/cm for electric fields, i.e., some three orders of magnitude over the characteristic atomic field.

The principle scheme of a vacuum birefringence experiment employing a high-intensity laser field to polarize the quantum vacuum was put forward in \cite{Heinzl:2006xc}, envisioning the combination of an optical high-intensity laser pulse as pump and a bright linearly polarized x-ray pulse as probe.
As the birefringence-induced signal is inversely proportional to the wavelength of the probe and directly proportional to the number of photons available for probing, x-ray probes seem most promising, particularly given the tremendous progress in x-ray polarization purity measurements achieved in recent years \cite{Marx:2011,Marx:2013xwa}.
An experiment of this type will become possible, for example, at the Helmholtz International Beamline for Extreme Fields (HIBEF) at the European x-ray free-electron laser (XFEL).
Its potential feasibility with state-of-the-art technology was recently demonstrated \cite{Karbstein:2015xra,Schlenvoigt:2016,Karbstein:2016lby};
for related studies, see \cite{DiPiazza:2006pr,Dinu:2013gaa,Dinu:2014tsa}.
Gamma-ray probes were also proposed to detect the effect \cite{Kotkin:1996nf,Nakamiya:2015pde,Ilderton:2016khs,King:2016jnl,Bragin:2017yau}.

For further signatures of quantum vacuum nonlinearity in strong electromagnetic fields of focused high-intensity laser pulses, we refer the reader to the pertinent reviews \cite{Dittrich:2000zu,Marklund:2008gj,Dunne:2008kc,Heinzl:2008an,DiPiazza:2011tq,King2016,Karbstein:2016hlj,Inada:2017lop} and references therein.

\subsection{Searches for ultralight bosonic dark matter}
\label{Subsect: AxionSearches} 

If we assume that galactic dark matter (DM) is dominated by a single particle species, we can find the number density of the DM particles by dividing the observed DM mass density in our galaxy ($\approx 0.4\,$GeV/cm$^3$) by the mass of the particle. If the DM particles are fermions, according to the Fermi-Dirac distribution, a fraction of the particles move with respect to the galaxy with the Fermi velocity, even at zero temperature of the gas. This has an interesting and important consequence: if the DM particle's mass is less than $\sim 10\,$eV/$c^2$,   it has to be a boson. Indeed, the Fermi velocity for a gas of such light particles exceeds $\sim 10^{-3}c$, the escape velocity of the galaxy, which is incompatible with the observation that DM is bound to the galaxy. 

A gas of bosonic particle can be equivalently thought of as a field (similar to the photon/electromagnetic-field duality). According the field theory, for instance, the Klein-Gordon equation for (pseudo)scalars, the field oscillates at the Compton frequency of the bosons $mc^2/h$. Assuming that DM moves in the galaxy with ``virialized'' velocities with magnitudes $v \approx 10^{-3}c$, the bosons' total energies $\approx mc^2(1+v^2/2c^2)$ acquire a distribution corresponding to the effective Q-factor of the oscillation of $(v/c)^{-2}\approx 10^{6}$. This shows that the expected dark-matter field corresponds to a highly monochromatic radiation with a coherence time corresponding to roughly a million oscillation periods and coherence length of roughly a thousand Compton lengths of the boson.    

A variety of spin-zero (axions, axion-like particles, dilatons, etc.), spin-one (dark and hidden photons, etc.) and higher-integer-spin particles have emerged as possible DM candidates \cite{Arias2012,Baker:2013zta,Essig:2013lka,Marzola2018}. 

The axion is a pseudoscalar (odd intrinsic parity, spin-zero) particle that was hypothesized in the 1970s to explain the experimental fact that strong nuclear interactions do not violate CP invariance, the symmetry with respect to the combined operation of charge conjugation and spatial inversion. At least the violation is at a small enough level that it has evaded detection in the past six decades of dedicated searches. In recent years, axions and a more general class of axion-like particles (ALPs, which are like axions, except they do not solve the ``strong-CP problem'') have drawn renewed attention because of the realization that they could be excellent candidates for dark matter, as well as offering a solution to a number of other outstanding problems.

Axion and ALP searches can be classified into categories depending on where these particles are produced. In ``light shining through wall'' (LSW) experiments (see \cite{vanBibber1987} and references therein; Fig.\,\ref{Fig LSW concept}), axions/ALPs are created in the interaction between intense laser light and a static field of a strong magnet (this is known as ``Primakoff production''). The light photon and the static field mix to produce the scalar particles that then travel freely through a wall and are detected by converting them back to photons after they cross the wall.  While the wall is transparent for axions/ALPs, it blocks the photons completely. In ``helioscope" experiments (Fig.\,\ref{Fig Helioscope concept}),  the production of axions or ALPs is entrusted to the Sun (Helios), but detection is accomplished in the laboratory, as in LSW experiments. ``Haloscope'' experiments directly detect the DM from the galactic halo. Note that, in contrast to haloscope experiments, LSW and helioscope experiments are sensitive to ``new'' particles that may or may not be abundant in the galaxy and be part of DM.

\subsubsection{Light shining through walls}
\label{Subsubsec:LSW}
\begin{figure}[h!]
    \begin{center}
    \includegraphics[width=5.5 in]{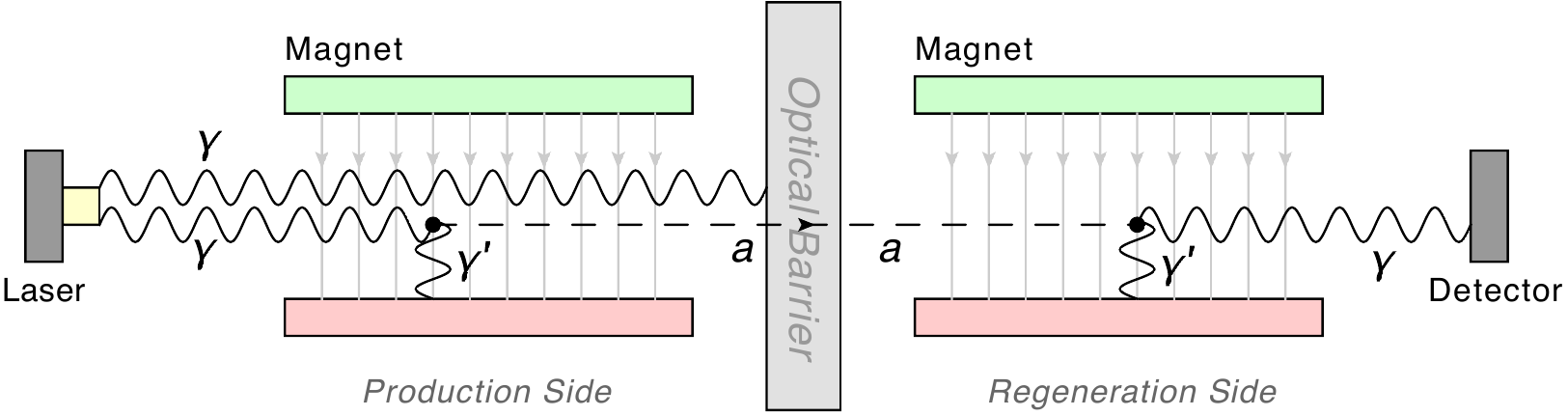}
    \caption{The concept of the light-shining-through-walls experiments. The  $\gamma^\prime$ photons are associated with the quasistatic magnetic field.}
    \label{Fig LSW concept}
    \end{center}
\end{figure}
The general idea of light-shining-through-a-wall (LSW) experiments is illustrated in Fig.\ \ref{Fig LSW concept}. On the production side of the setup, laser photons ($\gamma$) transform into axions/ALPs in the presence of the  quasistatic  field providing photons $\gamma^\prime$. The probability of such conversions is given by 
\begin{equation}
	P_{\gamma\rightarrow a} = \frac{1}{4}\big({g_{a\gamma\gamma}BL}\big)^2 F(qL),
\end{equation} 
where $g_{a\gamma\gamma}$ is the coupling constant characterizing the strength of the axion/ALP-photon coupling vertex, $q = p_{\gamma} - p_{a}$ ($p_{\gamma}$ and $p_{a}$ are, respectively,  the photon and axion/ALP momenta), $L$ is the length of the magnetic-field region. The form factor accounting for the difference between the speed of the axion/ALP and the speed of light is
\begin{equation}
F(qL) = \Bigg[               \frac{      \sin  \Big(\frac{1}{2qL}\Big)  }      {\frac{1}{2qL}}                               \Bigg]^2.
\end{equation}
An identical calculation for the probability of the conversion of the axion/ALP back to a photon in the detection region (and assuming the identical length and magnitude of the magnetic field) leads to the overall probability  
\begin{equation}
	P_{\gamma\rightarrow a \rightarrow \gamma} = \frac{1}{16}\big({g_{a\gamma\gamma}BL}\big)^4 = 6\cdot10^{-38}\Bigg( \frac{g_{a\gamma\gamma}}{10^{-10}\textrm{ GeV}^{-1}} \frac{B}{1\textrm{ T}} \frac{L}{10\textrm{ m}} \Bigg)^4,
\end{equation} 
where we assumed sufficiently light axions/ALPs, so that $qL\ll 1$ and $F(qL)\approx 1$. 

\begin{figure}[h!]
    \begin{center}
    \includegraphics[width=5.5 in]{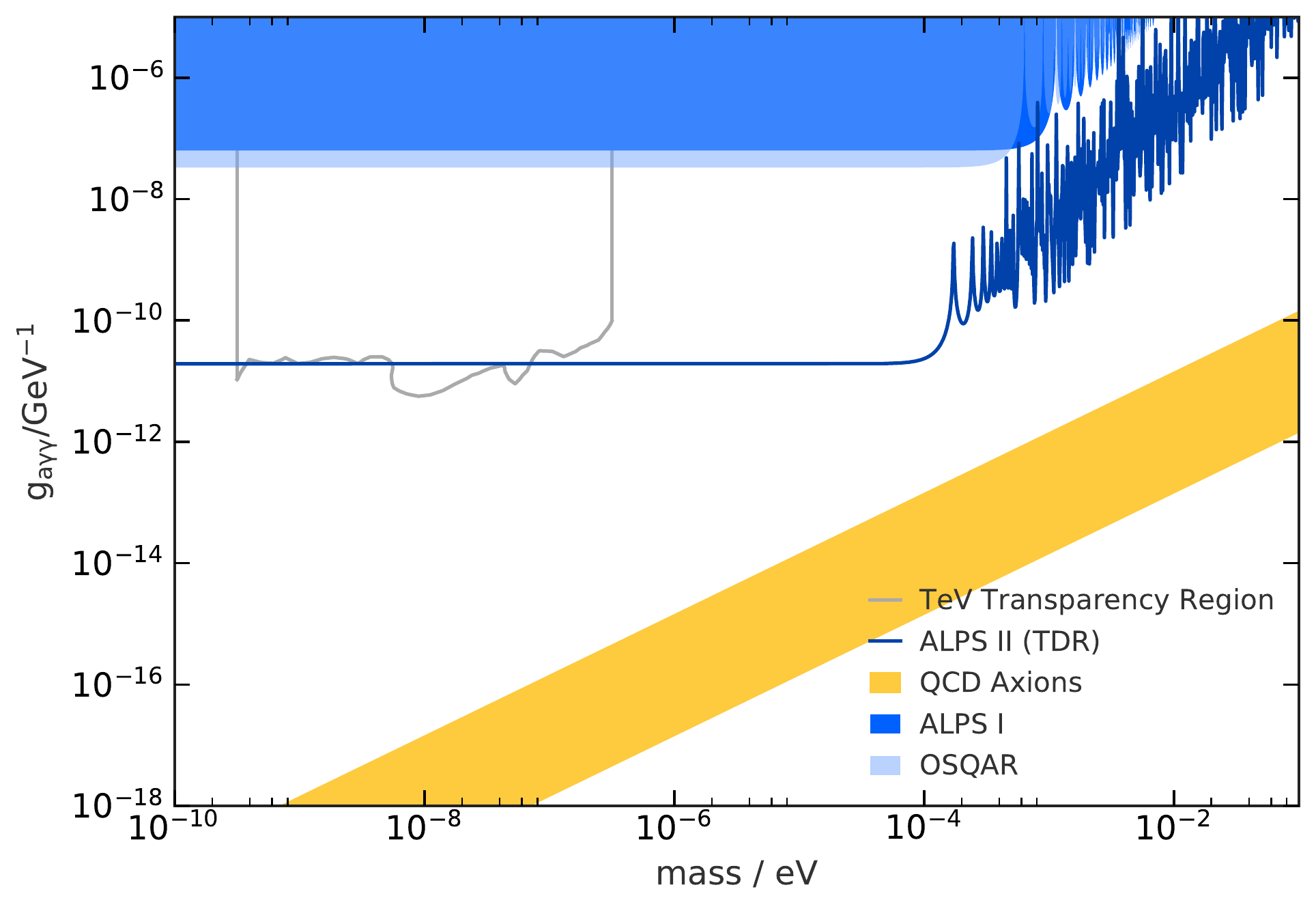}
    \caption{Overview of the current leading limits in LSW experiment (ALPS-I, OSQAR) and the projected sensitivity of the ALPS-II Experiment. The QCD-axion band as well as the region for ALPs explaining TeV transparency is indicated. ``TeV transparency'' refers to a hint for a possible range of ALP parameters from astrophysical observations that suggest excessive transparency of the universe for TeV gamma rays. The idea is that the gamma rays may ``hide'' from absorption by converting to ALPs and back in the cosmic magnetic field. TDR stands for technical design report. This overview plot was created with ALPlot (https://alplot.physik.uni-mainz.de).}
    \label{Fig LSW exclusions}
    \end{center}
\end{figure}

Various techniques can be used to significantly boost the sensitivity of LSW experiments. Using an optical cavity, one can ``trap'' the photons used to produce ALPs in the magnetic-field region thus increasing the conversion probability. Less obviously, a resonant optical cavity on the detection side gives a similar enhancement (\cite{Redondo2011} and references therein). A survey of the LSW experiments and their key parameters is given in Table~\ref{Tab: LSW experiments}, while the current sensitivity and future prospects are presented in the parameter-exclusion plot in Fig.\,\ref{Fig LSW exclusions}.
\begin{sidewaystable}[h]
\centering
\small
\begin{tabular}{cc|ccc|ccc}
\hline \hline
Experiment & Reference & Photon energy & Laser power & Power & Magnetic field & Magnetic field & $(BL)^4$ \\
& & [eV] & & buildup & strength $B$[T] & length $L$[m] & [Tm]$^4$ \\
\hline
ALPS & \cite{Ehret:2010mh} & 2.33 & 4 W & $P_{p}=300$ & 5 & 4.3 & $2\cdot 10^5$\\
\hline
BRFT & \cite{Cameron:1993mr} & 2.47 & 3 W & $P_{p}=100$ & 3.7 & 4.4 & $7\cdot 10^4$\\
\hline
BMV & \cite{Robilliard:2007bq} & 1.17 & $8\cdot10^{21} \gamma /$pulse& - & 12.3 & 0.4 & $6\cdot 10^2$\\
& & (14 pulses) & & & & \\
\hline
GammeV & \cite{Chou:2007zzc} & 2.33 & $4\cdot10^{17} \gamma/$pulse& - & 5 & 3 & $6\cdot 10^4$\\
& & (3600 pulses) & & & & \\
\hline
OSQAR & \cite{Ballou:2015cka} & 2.33 & 18.5W & - & 9 & 14.3 & $3\cdot 10^8$\\
\hline
ALPS-II & \cite{Bahre:2013ywa} & 1.16 & 30W & $P_{p}=5000$ & 5 & 100 & $6\cdot 10^{10}$\\
& & & & $P_{r}=40000$ & & & \\
\hline
LSW with X-Rays 
& \cite{Battesti:2010dm} & 50200 & 10\,mW & - & 3 & 0.150 and 0.097 & $0.017$\\
& 
     & 90700 & 0.1\,mW &\\
\hline
LSW with Pulsed Magnets 
& \cite{Inada:2016jzh}
& 9500 & 46\,mW & - 
& 8.3 T and 5.7 T  & 0.8 & $10^3$\\
and Synchrotron X Rays
& &
& &
& pulsed (duration 1ms)
& \\
\hline
\end{tabular}
\caption{Overview of experimental parameters of previous and future LSW experiments: the photon energy, the initial laser power, the power-buildup in the production and regeneration side ($P_p$ and $P_r$), as well as the magnetic field strength and length in production and regeneration sides ($B_P$, $B_R$, $L_P$, $L_R$). For all the cases, $B=B_{p}=B_{r}$ and $L=L_{p}=L_{r}$.  }
\label{Tab: LSW experiments}
\end{sidewaystable}

The choice of magnets for LSW experiments is driven by the requirement  that the magnetic field should be transverse in both the production and regeneration regions.\footnote{This follows from the fact that the amplitude of the photon-axion/ALP conversion and the inverse process are both proportional to a rotational invariant $\vec{\epsilon}\cdot\vec{B}$, where $\vec{\epsilon}$ describes the electric field of the light.} As in the case of the vacuum-birefringence experiments  (Sec.\,\ref{subsubsection:Vacuum-birefringence tests with quasi-stationary magnetic fields}), it is the length of the field that enters the figure-of-merit (Table \ref{Table:FOMs}). Long transverse-field magnets are used in accelerators, and so LSW experiments have benefited from  the use of surplus magnets from accelerators such as LHC.

Note also that the ``physics case'' for the LSW experiments is not limited to axion/ALP searches; there is, indeed, a broad range of ``new physics'' that can be explored \cite{Redondo2011}.

\clearpage
\subsubsection{Helioscopes}
\label{subsubsect:Helioscopes}

\begin{figure}[h!]
    \begin{center}
     \includegraphics[width=5.5 in]{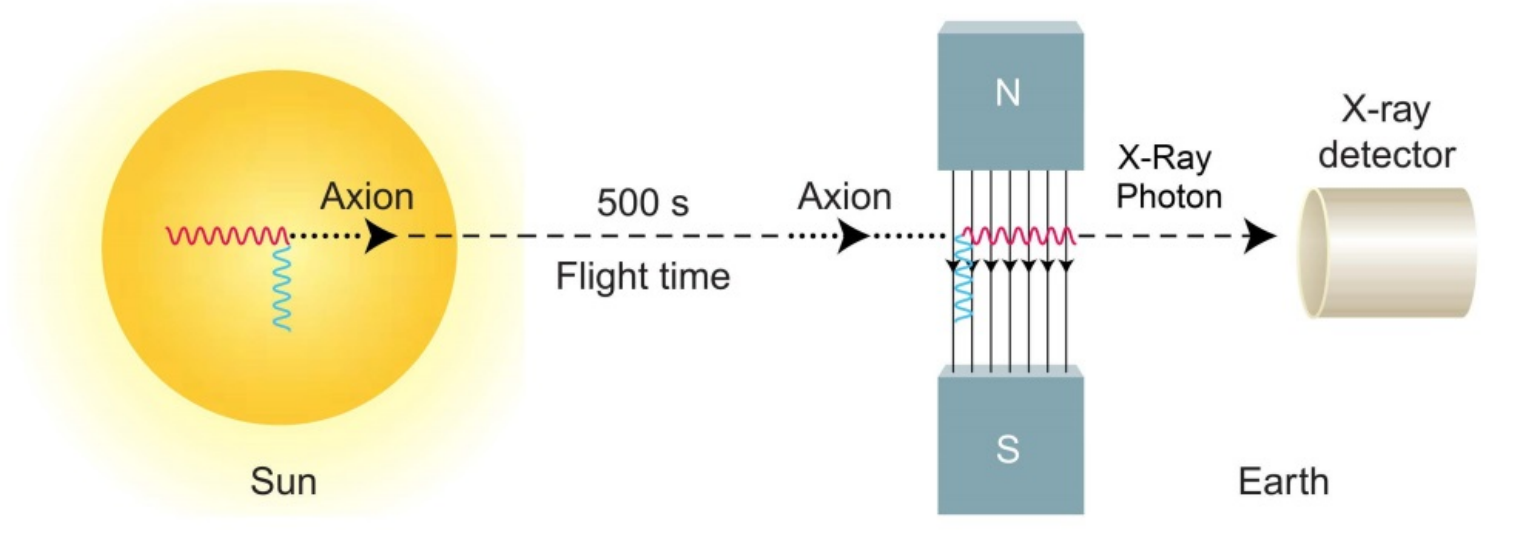}
    \caption{The axion-helioscope concept. Axions are produced in the sun and travel towards the earth. In the presence of a transverse magnetic field in the haloscope  (corresponding to the blue photon in the figure), the axions are converted into x-ray photons and detected. The energy spectrum of the x-ray photons corresponds to that of the axions produced in the sun.}
    \label{Fig Helioscope concept}
    \end{center}
\end{figure}

The general layout of axion helioscopes is illustrated in Fig.\ \ref{Fig Helioscope concept}. The figure-of-merit for a helioscope reads:
	\begin{equation}
		\textrm{FOM}_{\textrm{Helio}}\ =\ B^2L^2A = B^2V L. \label{Eq:FOM for helioscopes}
	\end{equation}
This is a simplified equation that assumes uniform transverse magnetic field of magnitude $B$ over the volume $V$ of the detector of length $L$ and area $A$. If the field varies over the volume, the FOM needs to be evaluated by appropriate volume integration taking into account that the amplitude of the axion-to-photon conversion scales as $BdL$, hence the $L^2$ dependence in Eq.\ (\ref{Eq:FOM for helioscopes}). It is important to note that this FOM is different from the FOM encountered in other experiments (Table~\ref{Table:FOMs}).

An example of a helioscope is the CERN Axion Solar Telescope (CAST) \cite{CAST2017NP}. The experiment is constructed around a 9 T LHC test magnet built in the 1990s. It has been very successful over the years, providing new significant limits on the axion-photon coupling. 

\begin{figure}[h!]
    \begin{center}
     \includegraphics[width=5.5 in]{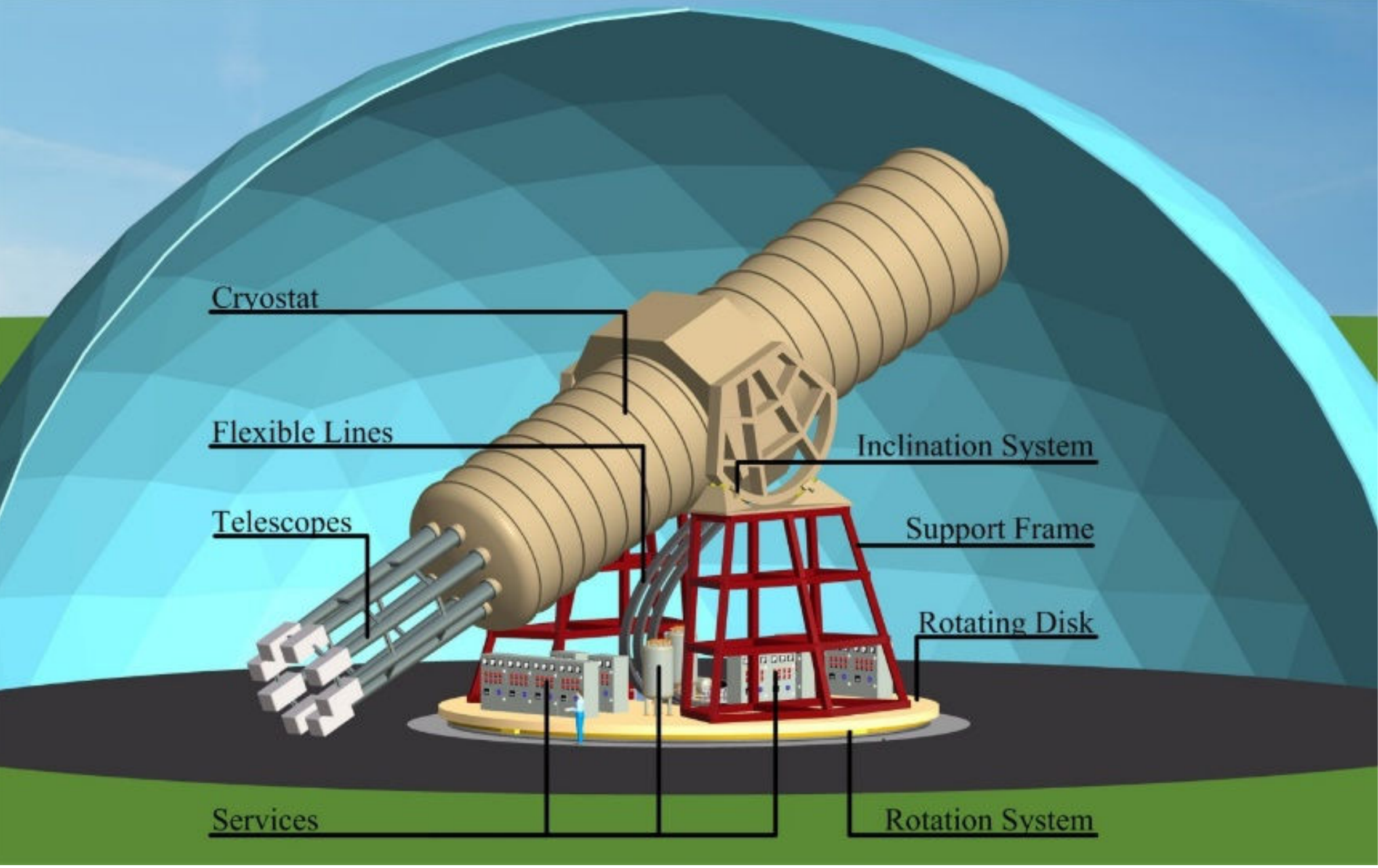}
    \caption{General view of the IAXO design whose key part is a 22\,m long eight-coil toroidal magnet enclosed in a 25~m long cryostat. Figure from \cite{IAXO2014}.}
    \label{Fig IAXO overview}
    \end{center}
\end{figure}

The currently planned International Axion Observatory (IAXO) \cite{IAXO2014,Irastorza:2015geo} is promising to reach an unprecedented level of sensitivity and discovery potential. In terms of the magnet, IAXO moves away from the geometry of accelerator dipoles and towards the detector-inspired barrel geometry, allowing for a better helioscope FOM (Eq.\ \ref{Eq:FOM for helioscopes}). IAXO design draws upon the experience of CAST, and aims at building a new large-scale magnet optimized for axion searches and extensively implementing focusing of the light produced in the solar-axion/ALP conversion and low-background photon detection techniques. The IAXO superconducting magnet  will have a toroidal multibore configuration~\cite{Shilon:2012te}. The current design considers a 22~m long, 5.2~m diameter toroid assembled from eight coils, effectively generating an average (peak) 2.5 (5.1) tesla in the eight bores of 600 mm diameter. This represents a 300 times better $B^2L^2A$ FOM compared to the CAST magnet. The toroid's stored energy will be 500~MJ. The design is inspired by the ATLAS barrel and end-cap toroids \cite{tenKate:1158687,tenKate:1169275}, the largest superconducting toroids built and presently in operation at CERN. Each of the eight magnet bores will be equipped with a detection line composed of an x-ray-optics telescope and a low-background detector. Beyond the magnet, several improvements are foreseen also in the optics and detector parameters. Figure~\ref{Fig IAXO overview} shows the conceptual design of the overall IAXO infrastructure~\cite{IAXO2014}.
 
As a first step towards IAXO, construction of a scaled-down version of the setup, BabyIAXO (Fig.\ \ref{Fig Helioscope concept}), is contemplated. The BabyIAXO magnet will be only 10~m long and will feature one bore of 60~cm in diameter. It will have a peak field of  4.1\,T, an average field of 2.5\,T, and a total stored energy of 27\,MJ. BabyIAXO will be equipped with only one set of optics and a detector, but of similar dimensions as for the final IAXO systems. BabyIAXO will therefore constitute a representative prototype for the final infrastructure. With the FOM of this demonstrator already exceeding that of CAST by an order of magnitude,  it will also provide relevant physics outcomes at an intermediate level between the current best CAST limits and the full IAXO prospects. The design and operational experience with BabyIAXO, in particular with the magnet, is expected to provide relevant feedback for the technical design of the full infrastructure and enable improvements in the ultimate FOM.
\begin{figure}[h!]
    \begin{center}
     \includegraphics[width=5.5 in]{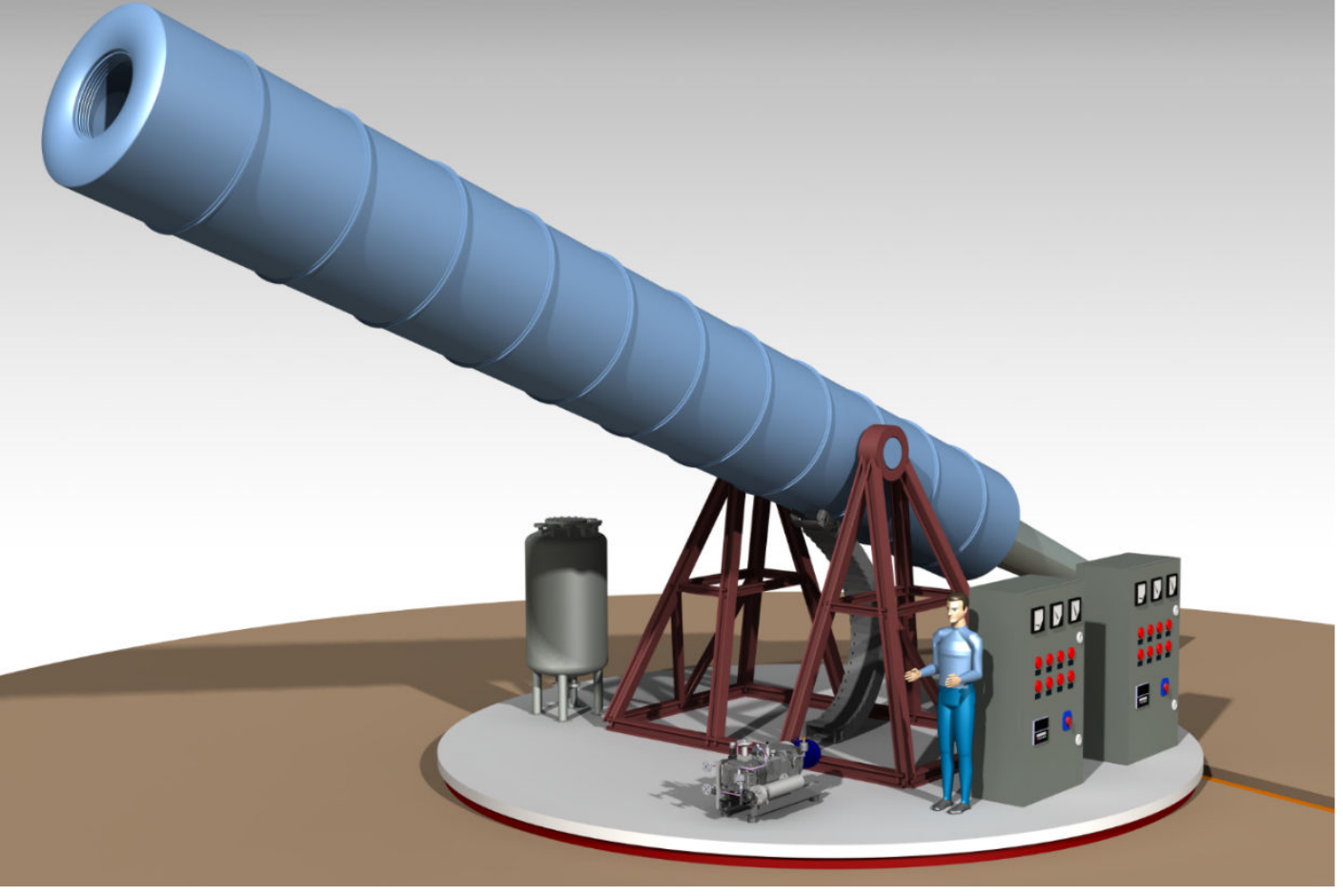}
    \caption{BabyIAXO: the proposed scaled down (there is one only one 10~m long magnet), fully functional initial-demonstration version of IAXO. Figure courtesy CERN/ATLAS Magnet Team.}
    \label{Fig BabyIAXO}
    \end{center}
\end{figure}

A list of helioscopes and their representative parameters are presented in Table \ref{tab:helioscopes}.
\begin{table}[h]
\center
\small
\begin{tabular}{cccccccc}
   \hline\hline
  Experiment & References & Status & $B$ (T) & $L$ (m) & $A$ (cm$^2$) & Focussing & $g_{a\gamma\gamma}$ (10$^{-10}$ GeV$^{-1})$ \\ \hline
  Brookhaven & \cite{Lazarus:1992ry} & past & 2.2 & 1.8 & 130 & no & 36 \\
  SUMICO & \cite{Moriyama:1998kd,Inoue:2008zp} & past & 4 & 2.5 & 18 & no & 6 \\
  CAST & \cite{Zioutas:1998cc,Zioutas:2004hi,Arik:2008mq,Arik2011,Arik2014} & ongoing & 9 & 9.3 & 30 & yes & 0.66 \\
  BabyIAXO & \cite{babyiaxo} & in design & $\sim$2.5 & 10 & 2.8$\times10^3$ & yes & 0.2 \\
  IAXO & \cite{Irastorza:2011gs,IAXO2014} & in design & $\sim$2.5 & 22 & 2.3$\times10^4$ & yes & 0.04 \\  
  \hline\hline
\end{tabular}
  \caption{Past, present and future helioscopes with the key magnet parameters. The last column represents the sensitivity achieved or expected in terms of the magnitude of the axion/ALP-photon coupling constant. The numbers for the BabyIAXO and IAXO helioscopes correspond to the design parameters considered by the collaboration.}\label{tab:helioscopes}
\end{table}

\subsubsection{Haloscopes}
\label{Subsubsec: Haloscopes}

If axions or ALPs constitute the dark matter of our galactic halo and if their mass corresponds to a Compton frequency in the microwave range, they can be detected in the laboratory via their conversion to photons in a microwave cavity permeated by a magnetic field \cite{Sikivie1983,Sikivie1985}. The signal power to be detected is
\begin{equation}
	W \propto g_{a\gamma\gamma}^2\frac{\rho_{halo}}{m_a}B^2V\cdot C\cdot Q,\label{Eq:PrimakoffPower}
\end{equation}
where $\rho_{halo}\approx 0.4\ $GeV/$c^2$cm$^{-3}$ is the density of the galactic dark matter, $m_a$ is the axion/ALP mass, $V$ is the volume of the cavity, $C\approx 0.5$ is the cavity-mode form factor and $Q \approx 10^5-10^6$ is the cavity quality factor. Resonant conversion occurs when the frequency of the cavity is close to the mass of the ALP, i.e.,
\begin{equation}
	h \nu = m_a c^2 [1 + O(\beta^2)/2],
\end{equation}
where $\beta \approx$10$^{-3}$ is the galactic virial velocity (which is an estimate of our relative velocity with respect to galactic DM) and $h$ is the Planck constant. The signal is thus highly monochromatic. However, the expected finite bandwidth of the signal limits the required quality factor for the cavity to $Q\approx 10^6$. The search for axions is performed by tuning the cavity frequency in small overlapping steps and the time integration at each scanned frequency is one of the key limiting factors. The signal-to-noise ratio can be approximated by the Dicke radiometer equation \cite{Dicke1946}
\begin{equation}
	\textrm{SNR} = \frac{W}{k_B T_{syst}}\Bigg( \frac{t}{\Delta \nu}  \Bigg)^{1/2},    
\end{equation}
with $W$ being the detection power (in the range of $10^{-23}\ $W), $k_B$---the Boltzmann constant, $T_{syst} = T + T_N$---the sum of the physical temperature $T$ and the intrinsic amplifier noise temperature $T_N$, $t$---integration time, and $\Delta \nu$---the bandwidth. SNR can be maximized by lowering $T_{syst}$. 

Note that the FOM for the magnets used in haloscopes based on axion/ALP-photon conversion is $B^2V$ [see Eq.\,\eqref{Eq:PrimakoffPower}], which is proportional to the magnetic energy stored in the field.

\begin{figure}[h!]
    \begin{center}
     \includegraphics[width=2.5 in]{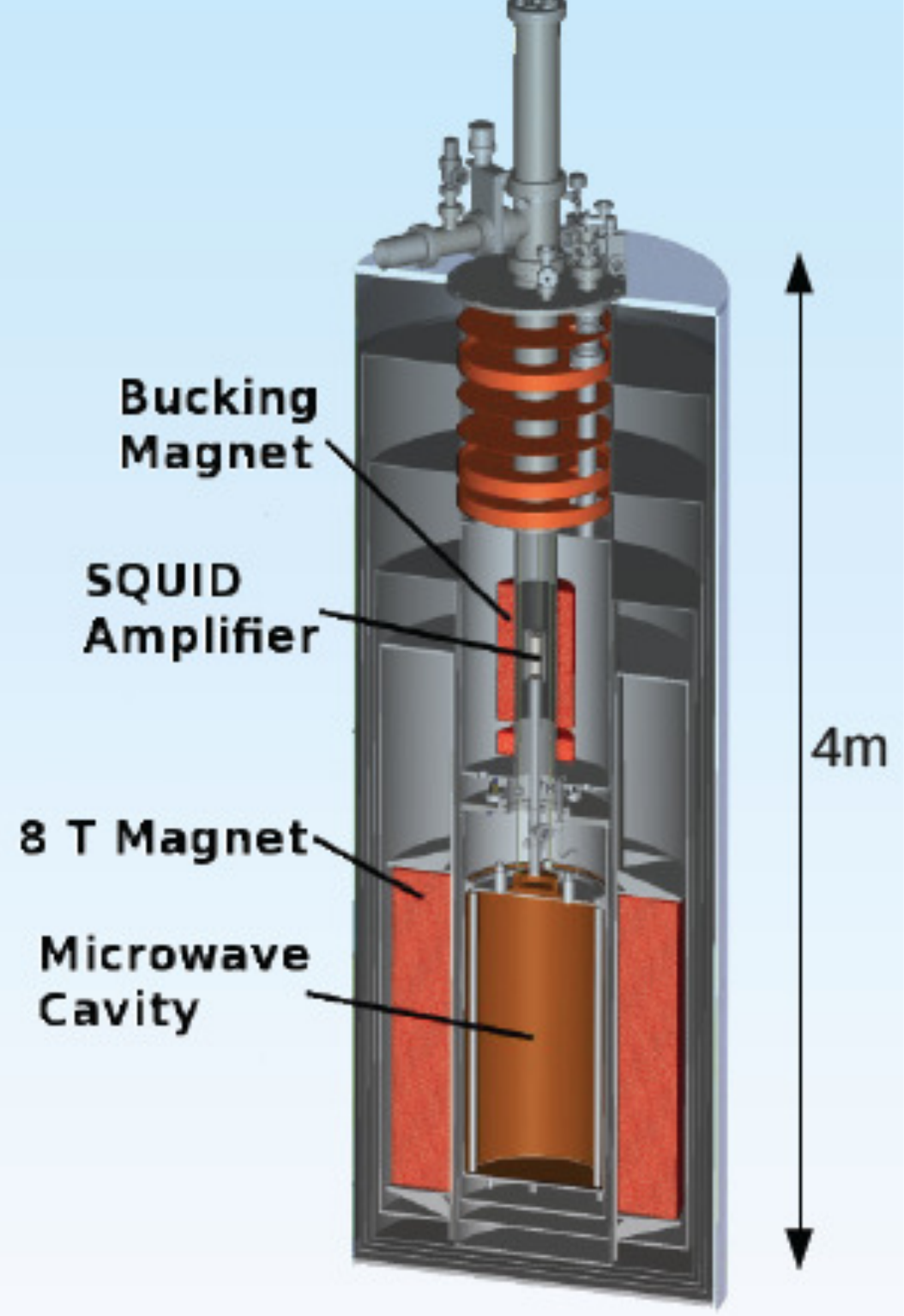}
     \includegraphics[width=1.0 in]{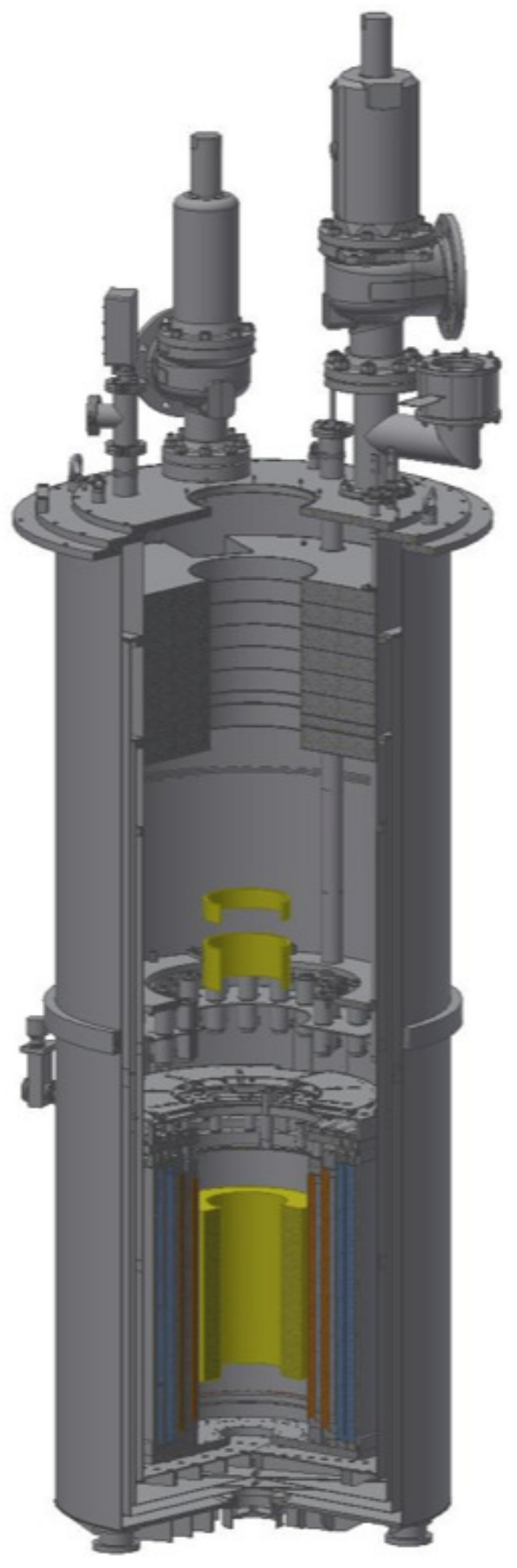}
    \caption{Left: The magnet used by the ADMX experiment. The barrel is 50\,cm in diameter and has capacity to accommodate a 220-liter cylindrical microwave cavity. The purpose of the small bucking magnet is to create a zero-field region where the field-sensitive SQUID amplifiers are placed. Right: Design of the new 24$\,$T magnet for ADMX.   The outer diameter of the magnet is about 0.9\,m. The coils are color-coded according to: yellow-HTS; 
red-Nb$_3$Sn; blue NbTi. Figure courtesy NHMFL.}
    \label{Fig ADMX magnet}
    \end{center}
\end{figure}

The ``flagship'' haloscope experiment is the Axion Dark Matter eXperiment (ADMX) at the University of Washington in Seattle. The experiment is currently running at the level of sensitivity high enough to detect QCD axions (i.e., axions capable to solve the strong-CP problem) with masses around 2.7$\,\mu$eV. The ADMX magnet was built by Wang NMR (Dublin, California), delivered in 1993. This superconducting NbTi magnet was
designed to operate at 8.5\,T in a persistent mode, however, the current switch necessary to put the magnet into the persistent mode upon charging failed on
delivery, and ADMX has used this magnet, running with current leads, at 7.6\,T, for all searches so far. An upgrade to low-noise superconducting quantum interferometer-device (SQUID) amplifiers
necessitated an additional 8\,T bucking magnet to create a field-free region about 1$\,$m above the
main solenoid (Fig.\ \ref{Fig ADMX magnet}, left). Inside the bucking coil, nested cylindrical layers of mu-metal
and niobium define a field-free region that houses a low-noise RF-SQUID amplifier and circulators.

A design for the next-generation ADMX magnet is currently in the works. It envisions a 24 T static field and a bore diameter of 16\,cm (smaller than the 50\,cm in the current magnet). Indeed, this will accommodate a smaller microwave cavity providing access to higher frequencies and, correspondingly, higher axion/ALP masses. 
The power in the photons produced from conversion of dark-matter axions in the presence of magnetic field is proportional to $B^2V$ [Eq.\,\eqref{Eq:PrimakoffPower}]. Conventional cavity resonators have geometry dictated by the boundary conditions necessary to allow the  resonances of the transverse-magnetic (TM) modes of the cavity in the magnetic field to correspond to the axion mass to be probed via $h\nu=m_ac^2$. Here $\nu$ is the frequency of the cavity $\rm{TM}$ mode used for the search, usually the $\rm{TM_{010}}$ mode. Therefore the only degree of freedom that can be altered at a given axion mass is the strength of the magnetic field. The current ADMX magnet has a $\rm{7.6\,T}$ central field, so a higher-field magnet with the proposed design field of $\rm{24\,T}$ would raise the signal-to-noise ratio by a factor of 12.6. The NHMFL design (Fig.\,\ref{Fig ADMX magnet}, right) utilizes three nested sets of windings, the outer windings being niobium-titanium, the intermediate windings being niobium-tin, and the innermost windings being YBCO tape. The choice of winding materials is guided by the quench fields in the different superconductors; closer to the bore of the magnet the windings must be quench--free at higher fields. As of February of 2018, there is no proposal submitted to any funding agency for a magnet upgrade; however such a proposal has a high priority once the first results from phase two of ADMX have been published. Publication of these results is expected in early spring of 2018.

A closely related experiment to ADMX is the Haloscope At Yale Sensitive To Axion Cold dark matter (HAYSTAC) \cite{Kenany2017,Brubaker2017} that was previously known as ADMX-HF and X3. HAYSTAC uses a 9.4 T DC magnet made by Cryomagnetics that is cooled with a pulse-tube refrigerator. Both the magnet and
the main bore (17.5\,cm in diameter, 50\,cm long) are ``dry'' in the sense that the only helium in the system is the $^3$He/$^4$He mixture in the dilution refrigerator.  A small cavity gives
sensitivity to high-mass axions (in a range complementary to that of ADMX), currently already reaching the sensitivity necessary to detect a QCD axion. A disadvantage of
a dry magnet system is that such a magnet quenches within minutes in the event of a power loss. The HAYSTAC collaboration lived through such an unfortunate experience; however, despite of the initial pessimistic assessment of the ``destroyed'' magnet, miraculously, it was possible to fully repair the system within two months.
\begin{table}[h]
\center
\small
\begin{tabular}{cccccccc}
   \hline\hline
  Experiment & References & Status & $B$ (T) & $V$ (m$^3$) & $B^2V$ (T$^2$m$^3$) & Multiple Cavities & Axion/ALP                    \\
                     &                     &            &              &                       &                                      &                            & mass range ($\mu$eV) \\ \hline
 ADMX & \cite{Rybka2017} & present & 7.4 &    &  & no & 2-3.5 \\
                          &   & future & 24 & 0.008     & 4.8 & no &                                           \\
  HAYSTAC & \cite{Brubaker2017} & present & 9.4 & 0.0015 & 0.12 & no & 23-24 \\
  ORGAN & \cite{ORGAN2017} & present & 7  & 0.003 & 0.16 &               & 60-210 \\
                &                                & future & 14 & 0.0014  & 0.27 &              & 60-210 \\
  CAPP & \cite{Petrakou2017} & commission & 12 & 0.033            & 4                  & no         & 3 - 12.5 \\ 
             & \                               & present             & 18 &  0.001         &  0.3           & no        & 15-26 \\ 
             &                                 & comission         & 25  &   0.0025     &        1.4     & no        & 12-42 \\   
             &                                 &                          &       &                    &                 & yes        & $>85$  \\   
  Grenoble & \cite{Pugnat2017NewHalo} & proposal & 9/17.5/27/      & 0.49/0.05/0.0029/                                                               & 13-0.01 			& yes/no        &  1.3--200          \\ 
                  &                                             &                & 40/43             &                             2.9$\cdot 10^{-5}$/0.45$\cdot 10^{-5}$ &        			&                    &                         \\ 
  \hline\hline
\end{tabular}
  \caption{Present and future haloscopes based on the Primakoff production of microwave photons by dark-matter axions/ALPs. In this case, $V$ refers to the volume within microwave cavities placed into magnetic field. After CAPP runs the 12\,T and the 25\,T magnets, they plan to combine them and reach 37\,T within the volume of the 25\,T magnet, with a significant boost in the sensitivity of the axion/ALP search.}\label{tab:haloscopes}
\end{table}

A new haloscope has been proposed by a French/Korean collaboration that utilizes the Grenoble hybrid magnet described in Sec.\,\ref{Subsubsect_Hybrid_Magnets} that boosts the sensitivity by both optimizing the magnet FOM and operating at low (mK-range) temperatures \cite{Pugnat2017NewHalo}. Compared to other haloscopes such as ADMX, the superconducting outsert magnet alone offers an unprecedented value of $B^2V \approx 40\,$T$^2$m$^3$ to search for axion/ALPs in the mass range of 1-3\,$\mu$eV. A further increase of $B^2V$ up to 75\,T$^2$m$^3$ is possible but would imply major changes in the structure of the Grenoble hybrid magnet as well as significant investments, which might be considered at a later stage. On the other hand, adding resistive magnets to operate at higher fields and employing small cavities will allow further unique possibilities for this new haloscope to probe axions/ALPs of larger masses, so overall the range of axion/ALP masses from $\mu$eV to hundreds of $\mu$eV can be covered with FOM of  $B^2V$ = 13-0.01 T$^2$m$^3$ (Table \ref{tab:haloscopes}).

It should be noted that magnets of a variety of geometries can be considered for use for axion/ALP searches \cite{Baker2012} and a pilot haloscope search with a toroidal magnet was recently completed \cite{Choi2017}. The significance of this is that large toroidal magnets that are built for plasma-physics research (Tokamaks) can, in principle, be used for axion searches \cite{Vallet2012}. Indeed, Tokamaks like the International Thermonuclear Experimental Reactor (ITER) can have very large $B^2V\gtrsim 3\cdot10^4\,$T$^2$m$^3$ (c.f. Table\,\ref{tab:haloscopes}).

\begin{figure}[h!]
    \begin{center}
     \includegraphics[width=5.5 in]{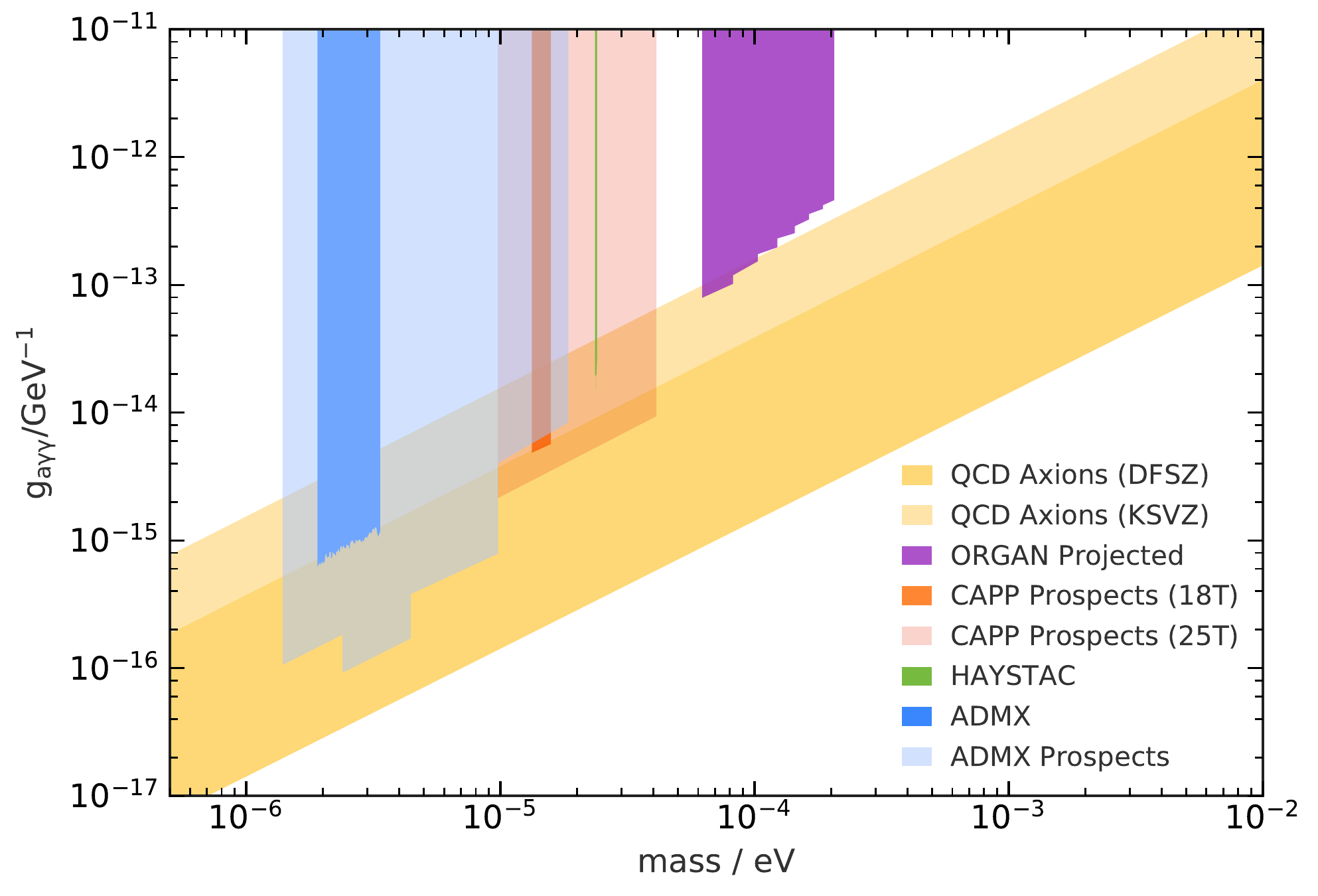}
    \caption{Current and projected limits on the axion-photon coupling constants from Primakoff-haloscope experiments under the assumption that dark matter is dominated by a single-species of axions or ALPs. This overview plot was created with ALPlot (https://alplot.physik.uni-mainz.de). }
    \label{Fig haloscope-results}
    \end{center}
\end{figure}

A summary of the Primakoff-haloscope results and projections is given in Table \ref{tab:haloscopes}  and  Fig.\,\ref{Fig haloscope-results}. Note that assuming that dark matter is dominated by a single species of axion/ALP, haloscopes have higher sensitivity to the coupling constant $g_{a\gamma\gamma}$ compared to other experiments searching for axions/ALPs in the same mass ranges.

Several new experiments and proposals address the possibility that the DM particles could be lighter than $\mu$eV. In principle, the entire range down to $\approx 10^{-22}$\,eV is ``fair game,'' limited from below by the requirement that the de Broglie wavelength of the particle not exceed the size of the galaxy.

\begin{figure}[h!]
    \begin{center}
     \includegraphics[width=5.5 in]{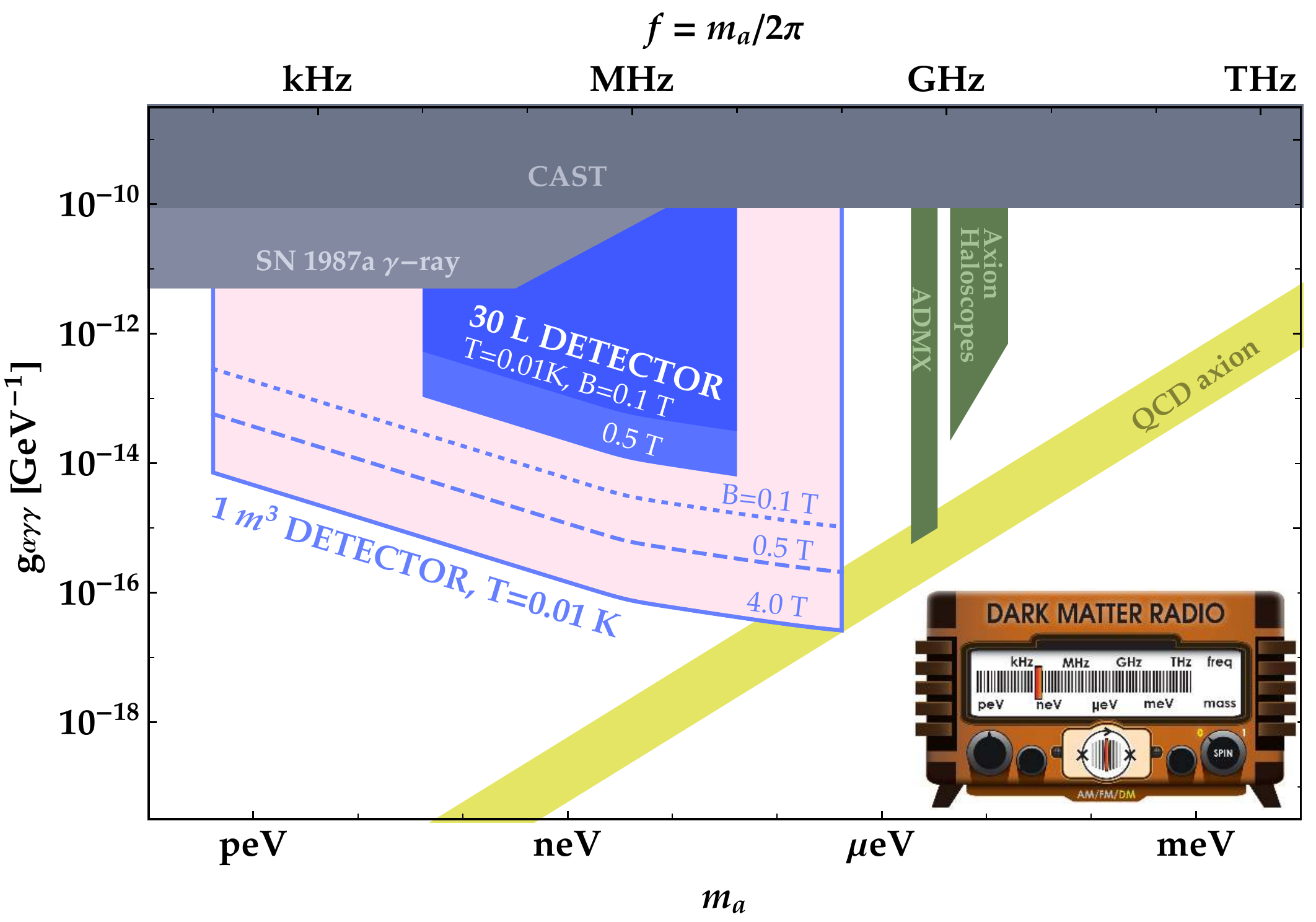}
    \caption{Projected sensitivity of various stages of the Dark Matter radio (DM radio) experiment to axion-photon coupling. The inset at the lower right shows an emblem of the experiment accurately capturing some of the essential features. Figure courtesy DM radio collaboration.}
    \label{Fig_DM_Radio}
    \end{center}
\end{figure}

The DM Radio experiment \cite{Chauhuri2015} will search for axion/ALP and hidden photon dark matter over a wide range of sub-$\mu$eV masses. The concept of DM Radio is simple: like ADMX, it is a sensitive electromagnetic resonator (a superconducting lumped-element circuit), shielded from external noise. The dark matter passes through the shield and excites the resonator when the resonant frequency is tuned to the dark-matter mass. 
DM Radio will cover many orders of magnitude in mass and coupling for the hidden photon in a relatively small and fast experiment. This broad frequency coverage is possible since DM Radio relies on a lumped element (LC) circuit. To search for axions, DM Radio needs a magnet [Fig.\,\ref{Fig_DM_Radio}]. In the presence of a magnetic field, DM radio will search for axions in a frequency range complementary to ADMX and CASPEr (see below). The DM Radio group would welcome collaboration with a facility that can provide a large magnet, with minimal fringing fields to prevent driving superconductors normal, and compatible with a high-quality-factor resonator, so that the resonant axion signal can be optimally detected. 

A similar ``lumped-circuit'' approach to dark-matter axion/ALP search is pursued by the MIT-based ABRACADABRA (A Broadband/Resonant Approach to Cosmic Axion Detection with an Amplifying B-field Ring Apparatus) experiment \cite{ABRACADABRA2016}. The ABRACADABRA collaboration envisions a three-stage program where the stages are defined by the magnet. The first stage ABRACADABRA-10~cm is a demonstrator experiment which will begin to cut into as yet unexplored ALP parameter space. It is currently preparing for its first data run with  a 1\,T toroidal magnet with inner radius of 3\,cm, maximum outer radius of 6\,cm, and maximum height of 12\,cm. ABRACADABRA-75\,cm scales the magnet to an inner radius of  25~cm, a height of 75~cm, and magnetic field of 5~T. This experiment would have some sensitivity to the QCD axion depending on the readout approach. The ultimate experiment would be a large magnet with specifications on the order of a 10\,T field with an inner radius of  4\,m, an outer radius of 8\,m, and a height of 12\,m. This may be prohibitively expensive to build, so the collaboration is exploring alternative readout strategies and geometries that may allow similar sensitivities with a more affordable magnet.

If DM consists of axions/ALPs, there exist additional possibilities to search for them based on the fact that, in addition to axion/ALP coupling to electromagnetic field (with coupling constant $g_{a\gamma\gamma}$), there are two more nongravitational coupling of axions/ALPs that are predicted by theory (see, for example, \cite{Graham2013}). One of these, corresponds to axion/ALPs interaction with two gluons instead of two photons. As a consequence, a nucleon and nonzero-spin nuclei acquire time-dependent electric-dipole moments (EDMs) whose time evolution follows that of the axion/ALP dark-matter field, oscillating at the axion/ALP Compton frequency and having a typical ``Q-factor'' of 10$^6$. It is important to note that EDMs violate both parity (P) and time-reversal invariance (T) on the account of the corresponding intrinsic properties of the axion/ALPs. The other possible axion/ALP interaction with normal matter is the so-called derivative interaction, where the gradient of the axion/ALP field (colloquially known as ``axion wind'') acts as a pseudo-magnetic field causing nuclear spin precession. In contrast to the usual magnetic field, the pseudo-magnetic field can penetrate magnetic shielding. A search for both these interactions is conducted by the Cosmic Axion Spin-Precession experiment (CASPEr) collaboration \cite{budker2014proposal,Garcon2018}, with separate setups. The search for the axion/ALP-induced oscillating EDM (CASPEr-Electric) is conducted at the Boston University, while the setup sensitive to the derivative coupling (CASPEr-Wind) is at the Helmholtz Institute in Mainz.  In both CASPEr-Electric and CASPEr-Wind a sample of polarized nuclei is immersed in a magnetic field whose magnitude is scanned. When the Larmor precession frequency becomes resonant with the axion/ALP mass, the DM field induces resonant spin-flips, and the resulting time-dependent transverse magnetization of the sample is detected with a SQUID magnetometer, so the experiments are, in effect, highly sensitive continuous-wave NMR spectrometers. Because CASPEr-Electric requires an electric field present in the sample and CASPEr-Wind does not, the two setups operate with different samples: a solid ferroelectric material containing $^{207}$Pb in the former case and liquid $^{129}$Xe in the latter. 

The requirements for magnets to be used in CASPEr are strict: they should be broadly scannable and have temporal instabilities and field inhomogeneity at better than ppm levels over the entire field range. Phase-one CASPEr-Electric is using a commercial 9\,T Cryomagnetics superconducting magnet, specified to have 0.1\% homogeneity over a 1\,cm$^3$ volume. The rated current at 9\,T is 64\,A, and the charging rate is 0.1\,A/s, therefore the magnet can be charged or discharged in ten minutes. The charging rate is an important parameter since the CASPEr-Electric experimental protocol includes sample pre-polarization in a large magnetic field, before a sweep to the target magnetic field range for the axion search. If magnetic shielding is necessary to prevent external magnetic noise, an open question is how the magnetic field profile inside the magnet will be affected by the presence of a shield. The planned magnet for CASPEr-Wind has  a spatial inhomogeneity over a 10\,mm diameter spherical volume corresponding to $<1$\,ppm NMR linewidth during a 0.2\,T sweep. The main field will be sweepable from -14.1 to 14.1\,T and the room-temperature bore will be 89\,mm in diameter.  The magnet will be equipped with a set of 4 to 8 cryoshims. Particular attention needs to be given to minimizing vibrations.

An experiment analogous to CASPEr-Wind but searching for axion/ALP coupling to electrons rather then nuclei called  QUAX (QUaerere AXion) is being developed at the University of Padova \cite{Barbieri2017}. Its principle is detection of electron spin-flips in a magnetized solid in a magnetic field induced by the pseudo-magnetic DM field. 

\subsection{Neutrino-mass measurements}
Neutrinos were assumed massless in the framework of the Standard Model, however, neutrino-oscillation experiments have shown that at least some of the masses associated with the three neutrino flavors are nonzero; moreover, the neutrino mass eigenstates are superpositions of flavor eigenstates (and vice versa). Apart from neutrino oscillations, one can glean information on neutrino masses from the analysis of cosmological data, from the neutrinoless double-beta-decay experiments, and from accurate measurements of the electron spectra in beta-decays. 

Most commonly used for mass measurements is beta-decay of tritium resulting in a $^3$He nucleus, an electron, and an electron antineutrino with a release of 18.6 keV--a small energy on the nuclear scale. Because the decay results in three particles in the final state, the spectrum of the decay electrons is continuous. The relative effect of the (anti)neutrino mass is most pronounced at the ``endpoint,'' where the electron carries maximal kinetic energy, close to the overall energy released in the decay. A major difficulty of the measurements close to the endpoint is that there are few decays resulting in highest-energy electrons. For example, only $2\cdot 10^{-13}$ of the tritium beta-decays result in electrons within $1\ $eV from the endpoint.    

The state-of-the-art beta-spectrometer Karlsruhe Trititum Neutrino Experiment (KATRIN) \cite{KATRIN2005} employs a building-size vacuum chamber and aims to achieve neutrino-mass sensitivity of better than 200$\ $meV in the next years, limited by the ability to control systematic effects. Given the fundamental importance of neutrino-mass determinations, it is highly desirable to develop alternative, perhaps, smaller-scale, experiments that are not affected by the same systematics as KATRIN. 

\begin{figure}[h!]
    \begin{center}
     \includegraphics[width=3.5 in]{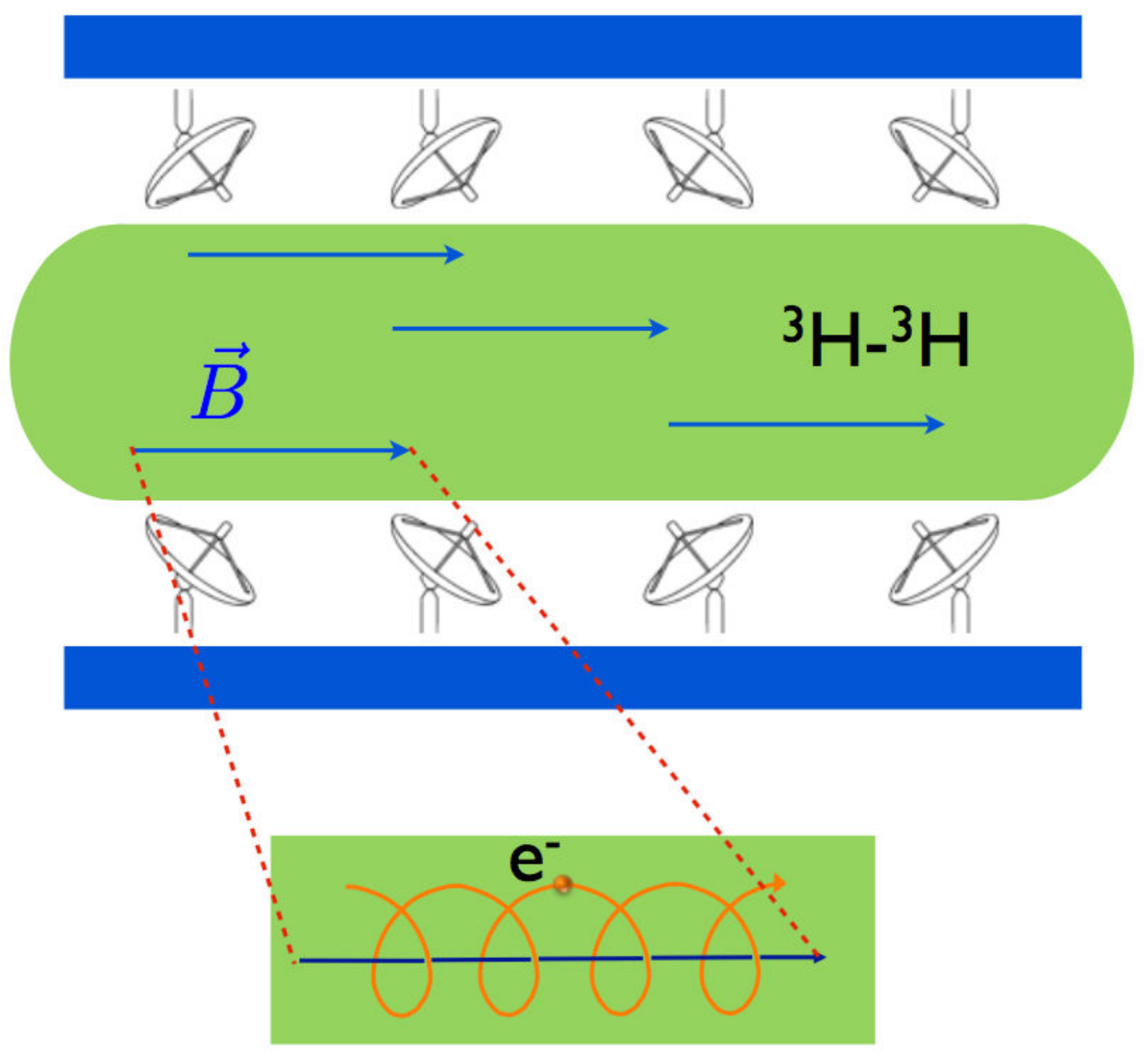}
    \caption{The concept of the neutrino-mass measurement via determination of the cyclotron frequency of electrons emitted in beta-decay of tritium \cite{Monreal2009}.}
    \label{Fig Project8 concept}
    \end{center}
\end{figure}
One such alternative approach \cite{Monreal2009} involves applying a strong magnetic field around the source and measuring the energy-dependent cyclotron frequency 
\begin{equation}
f = \frac{1}{2\pi}\cdot \frac{eBc^2}{m_ec^2 + E}
\end{equation}
of the electrons produced in beta-decay (Fig.\ \ref{Fig Project8 concept}), taking advantage of the fact that frequencies tend to be the physical quantities most amenable to precision measurements. For example $\delta E=1\ $eV of kinetic energy near the endpoint translates into a fractional cyclotron-frequency offset of $\approx \delta E/(m_ec^2)\approx 2\ $ppm, which can be readily detected if the electron is observed long enough.

In the ``Project 8'' experiment \cite{Project8_2017}, a gas volume is filled with tritium, a magnetic field is applied, and antennas are used to detect the radiation emitted by electrons undergoing the cyclotron motion. With $B=1\,$T, the cyclotron frequency is close to $28\ $GHz (the K$_\textrm{a}$ microwave band). In order to precisely measure the applied magnetic field and calibrate the apparatus, Project 8 uses the internal-conversion electrons from a $^{83m}$Kr source, which emits electrons of several discrete energies, one of them close to the energy of the electrons at the tritium endpoint.  

When it comes to detection of the radiation from single $18\,$keV electrons undergoing cyclotron motion in a $1\, $T field, the challenge lies in the low power of the emitted radiation, on the order of 1$\, $fW in total. Nevertheless, with the use of advanced microwave techniques, this feat was accomplished by the Project 8 collaboration \cite{CRES2015} in a prototype experimental setup at the University of Washington.

Sufficient observation time to establish the energy resolution is ensured by a shallow (few mT) magnetic trap in which the electrons are localized until they scatter off of gas atoms. Energy resolutions at the eV level have been achieved this way.

With the proof-of-principle of this measurement technology accomplished, two conditions must be met in a future setup for successful neutrino-mass measurement: a) accumulation of sufficient event statistics, which directly translates into a stable fiducial population of $10^{18}$ tritium atoms in the apparatus and b) excellent control of systematic effects, in particular those that affect the energy resolution. One of the dominant factors is that the final-state spectrum of molecular tritium is broadened due to the rovibrational excitations of the daughter molecule, providing a sensitivity limit of 100\,meV for molecular tritium. To achieve its design sensitivity of 40\,meV, Project 8 will therefore have to employ atomic, rather than molecular, tritium.

However, the presence of atomic tritium imposes an additional requirement. Contact with any surface catalyzes recombination of atomic tritium. To prevent recombination, tritium atoms can as well be trapped magnetically through their non-zero nuclear magnetic moment.

The task for the Project 8 magnet system is therefore threefold:
\begin{itemize}
\item	provide a highly uniform ($\Delta B/B$ at a ppm level) 1\,T solenoidal field over a $\approx 10\text{\,m}^{3}$ fiducial volume in which the electrons can emit cyclotron radiation,
\item	provide shallow mT traps in which the decay electrons are localized during observation,
\item	provide a conservative potential with a closed $|B|\approx 2\,$T contour inside all surfaces to prevent recombination of atomic tritium. The magnet system must superimpose this trapping field on the 1\,T background field without disturbing its ppm uniformity in the fiducial region.
\end{itemize}
\begin{figure}[h!]
    \begin{center}
     \includegraphics[width=6.5 in]{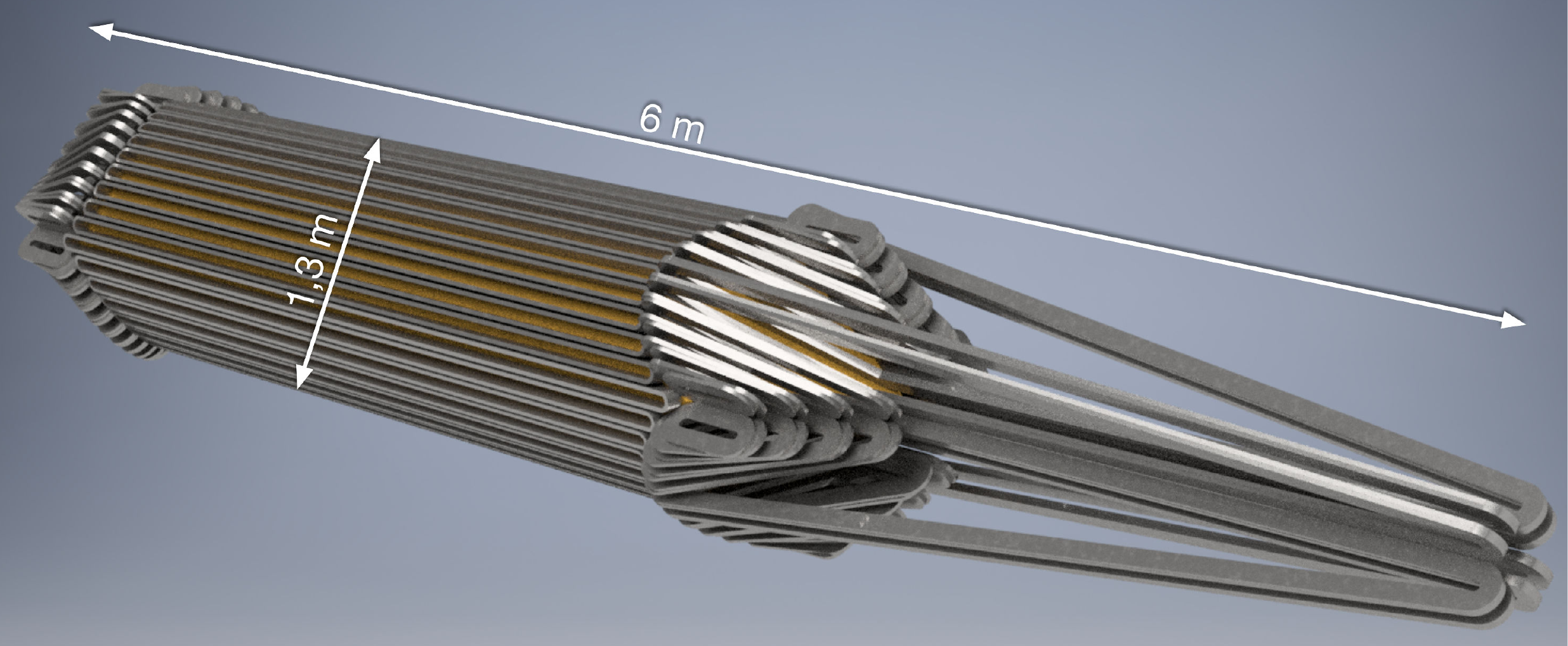}
    \caption{Design sketch of a possible Project 8 atomic-tritium trap consisting of 60 bars of $4\times 8\,$cm cross-section. The fiducial volume in which the field homogeneity is good enough for cyclotron-radiation-emission spectroscopy (CRES) is about 25\% of the physical volume.}
    \label{Fig Project 8}
    \end{center}
\end{figure}

While the design of the uniform background field and electron traps is straightforward, design of the atom trap provides several challenges. A cylindrical multipole magnet with discrete axial conductors carrying opposing currents creates a magnetic field that increases radially from its central axis. Such a Ioffe-Pritchard trap has been used to confine antihydrogen by the ALPHA collaboration. The order of the multipole in a Ioffe trap determines the radial dependence of the field, where higher orders increase the steepness of the radial increase and, therefore, the fiducial volume in which the uniformity requirement is met. However, employing multipoles of too high order, the 2\,T contour would not project inward far enough to hold atoms away from the physical surface. Additional challenges arise from closing off the cylinder in the axial direction and because the trap must be filled with atomic tritium from the outside. Figure \ref{Fig Project 8} shows a first design sketch of a possible atomic-trap configuration, in which the trap is loaded through a protruding quadrupole section, which opens up into the actual trap volume ensuring a continuous 2\,T contour inside all surfaces for a current density of $\approx 200 \text{ A}/\text{mm}^2$ per conductor.

While the Project 8 requirements call for a complex, superconducting Ioffe-Pritchard trap to store $10^{18}$ tritium atoms in a large fiducial volume with a 1\,T solenoidal field that is uniform at the ppm level, in the center of a 2\,T deep magnetic minimum, a first engineering design for a $\approx 1 \text{ m}^3$ atom-trapping demonstrator magnet is being currently undertaken by the collaboration.

\clearpage
\section{Ideas for future experiments}
\label{Sec:Ideas}

Atoms and molecules have proven to be very useful for fundamental-physics research \cite{SafronovaReview2017}. In this section, we discuss several cases where strong magnetic fields can potentially enable new breakthroughs in this area.

 \subsection{Parallel spectroscopy of complex systems at high magnetic fields: Spectroscopy 2.0}
Specific experiments require atoms or molecules with particular sets of properties. For example, atomic parity-violation effects are enhanced for heavy atoms and when opposite-nominal-parity atomic states are close in energy or, as a limiting case, are degenerate. The enhancement of the effect of the interaction under study due to the proximity of energy levels mixed by the interaction is not specific to parity violation; in perturbation theory, the mixing is inversely  proportional to the energy denominator.
While there are known ``accidental'' degeneracies, the levels do not always possess a complete set of desired properties. For example, dysprosium atoms have nearly degenerate metastable opposite-parity states of the same total electronic angular momentum that can be tuned to full degeneracy by applying magnetic fields on the order of tens of millitesla. While these states may appear ideal for parity-violation experiments \cite{Dzuba1986}, the relevant matrix elements were found to be strongly suppressed \cite{Nguyen1997}.

In designing fundamental-physics experiments, it is helpful to have a choice of systems with a range of properties in order to optimize the sensitivity to particular interactions of interest. In the case of atoms, one can turn to atomic databases \cite{NIST_ASD}, however, the spectra for atoms are not as complete as one would have hoped in the 21st century! This is particularly true for the rare-earth and actinide atoms and for highly-excited states. But ``new'' relatively low-lying levels are being found even in lighter systems such as barium \cite{Li_Ba_2004}.

The reasons for the considerable gaps in atomic tables is that dense spectra are difficult to analyze both from the experimental and theoretical points of view. 

To address these issues, a novel approach to spectroscopy of dense atomic and molecular spectra is proposed \cite{Spectroscopy2.0}. It combines three developments in spectroscopy: 1. The ability to record the spectra in a massively parallel manner using femtosecond frequency-comb techniques; 2. Availability of strong (up to over 100 T) tunable magnetic fields enabling a new dimension in spectroscopy; and finally, 3. Developments in many-body atomic and molecular theory methods, which, combined with the power of modern computers, allows for unambiguous interpretation of the multidimensional spectra with massive information content.

Here, we outline a vision for a joint effort between experimentalists and theorists to develop a modern atomic spectroscopy program, with a possibility of extension to molecular systems in the future. This program will improve the knowledge base of atomic spectra, using state-of-the-art tools developed in recent years by experimentalists and atomic theorists. The program will cover studies of atomic spectra in strong magnetic fields, providing a new, largely unexplored dimension for atomic spectroscopy.

The advantages of extending our spectroscopy program to atomic systems subject to large magnetic fields are numerous. The information provided would allow for nearly simultaneous measurement of large numbers of transition probabilities, magnetic $g$-factors of states, and the magnetic and electric dipole interaction strengths between states brought close to or through level crossings/anti-crossings. These level crossings may appear frequently at relatively modest fields (1-10 T) in heavy elements with partially filled d- and f-electron shells. Such fields can be achieved in relatively simple laboratory experiments.

Comprehensive study of the spectra of rare-earth elements (REE) in magnetic fields is additionally relevant for modern astrophysics. For example, highly anomalous abundances of REE are observed in stars with strong magnetic field ($\sim~0.1 - 1\ $T) \cite{Ryabchikova2006, Freyhammer2008}. At present, neither the origin of the magnetic field nor the abundance anomalies are well understood. Since information about conditions in cosmic objects is obtained from astronomical spectra, the reliability of derived physical parameters depends directly on the accuracy of atomic data such as transition probabilities and oscillator strengths. Although some data on the REE are available \cite{Heiter2011}, they are by far insufficient to meet all the scientific needs. This makes new laboratory measurements of current importance. 

Sufficiently complex elements have spectra with chaotic properties \cite{Flambaum1994, Dzuba2010,Viatkina2017}. The analysis techniques necessary for understanding these spectra bridge many disciplines, ranging from collisions in ultracold atoms \cite{Maier2015} to dielectronic recombination (a process where an electron is captured
by an atom and this capture is accompanied by excitation of a
initially bound electron) \cite{Flambaum2015} and the spectra of nuclei. Chaotic systems are particularly sensitive to small perturbations and can be used to study fundamental interactions and search for new physics beyond the standard model. For example, statistics of complex spectra may be used to constrain exotic T-odd, P-even interactions that indicate physics beyond the standard model \cite{Morrison2012}. A long-term goal is to continue the spectroscopy program at a high-magnetic-field facility, where magnetic fields up to 100 T or higher may be achieved. This would allow for the study of atomic spectra in environments found only in the vicinity of neutron stars, white dwarfs, or black holes \cite{Garstang1977, Miller1979}.

When the density of states reaches some critical level, the system becomes chaotic. Strong magnetic fields mix states with different total angular momenta and this effectively increases the level density. For some systems this may lead to a transition to chaotic behavior. Such ``controlled chaos'' is interesting both from the theoretical and practical points of view. It can improve our understanding of chaos in complex quantum systems, while providing new tools to study weak interactions in atoms.
\begin{figure}[h!]
    \begin{center}
     \includegraphics[width=5 in]{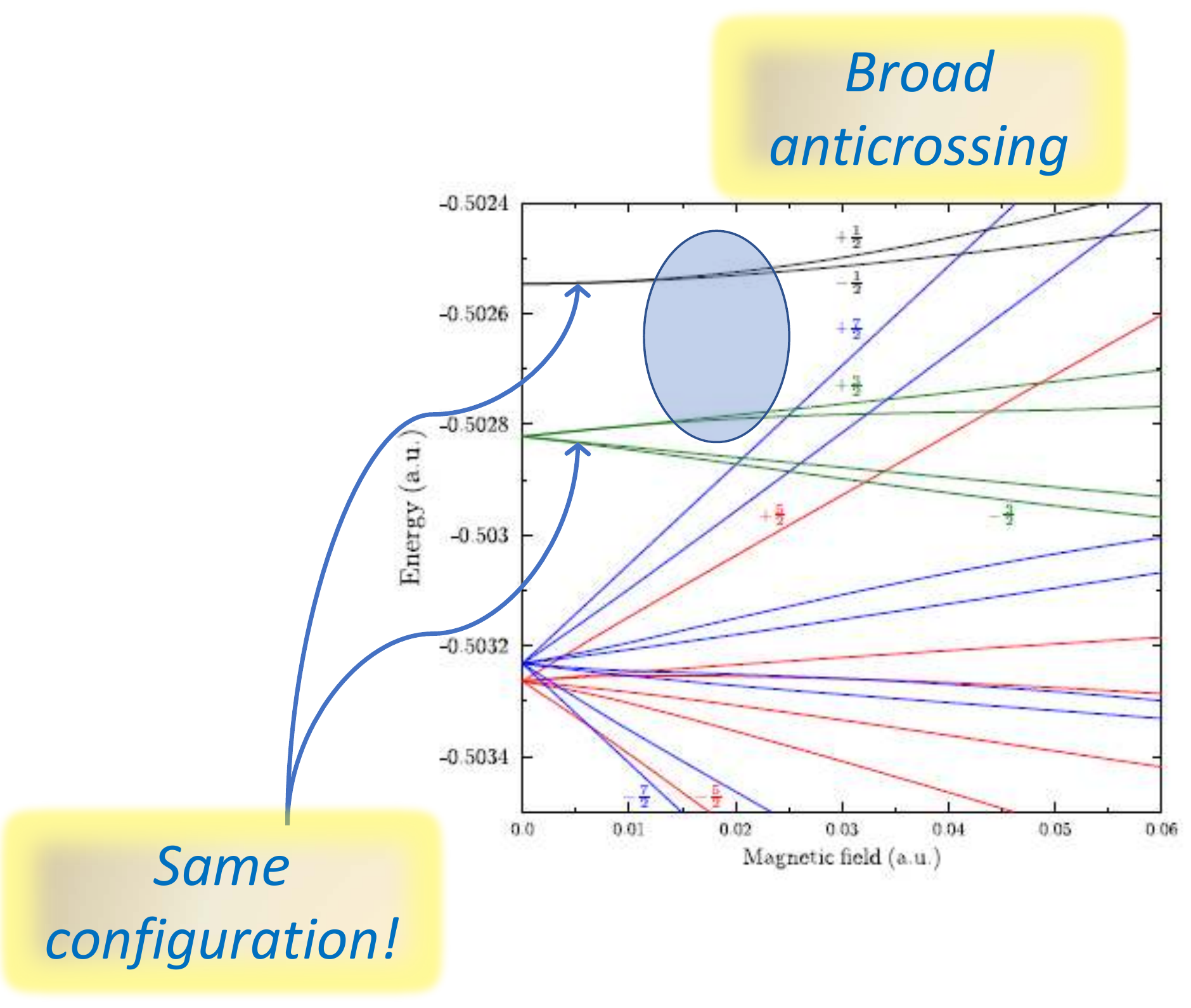}
    \caption{Selected energy levels of Ni$^+$ (Ni II) illustrating level-anticrossing behavior. Energies of electronic states and magnetic fields are given in atomic units (a.u.); an atomic unit of magnetic field is $c_1e^2/a_0\approx 1.7\cdot 10^3\ $T; $e$ is the elementary charge, $a_0$ is the Bohr radius, and $c_1$ is a constant appropriate for the units in which $e$ and $a_0$ are expressed.}
    \label{Fig BroadAnticrossing}
    \end{center}
\end{figure}

How can magnetic field help with identification of electronic states? To start with, atomic $g$-factors have been traditionally used for level identification as they depend on the angular-momentum quantum numbers, for example, $L,S$ and $J$  for the LS-coupling scheme. In dense spectra, there are level (anti)crossings as a function of magnetic field as illustrated in Fig.\ \ref{Fig BroadAnticrossing} for the case of singly ionized Ni. The magnetic-dipole (M1) operator only mixes states with the same configuration, and so the presence of anticrossings provides valuable information for level identification. This information is complementary to that from spectroscopy without magnetic field that provides information on electric-dipole (E1) couplings which are nonzero only with different, opposite-parity configurations. Furthermore, magnetic field mixes levels with different total angular momenta $J$ thus removing the usual E1-transition selection rule $\Delta J = 0,\pm 1$. Thus many transitions appear in the spectra that are forbidden in the absence of magnetic field.

In a complex system of energy eigenstates, it is often useful to discuss statistical properties of state mixing by small perturbations rather than mixing of individual pairs of states. In fact, application of magnetic field can have a profound effect on statistical properties of such mixing leading to statistical enhancement of all small perturbations. 
Studying such phenomena can improve our understanding of chaos in complex quantum systems, while providing new tools to study weak interactions in atoms.
\begin{figure}[h!]
    \begin{center}
     \includegraphics[width=5 in]{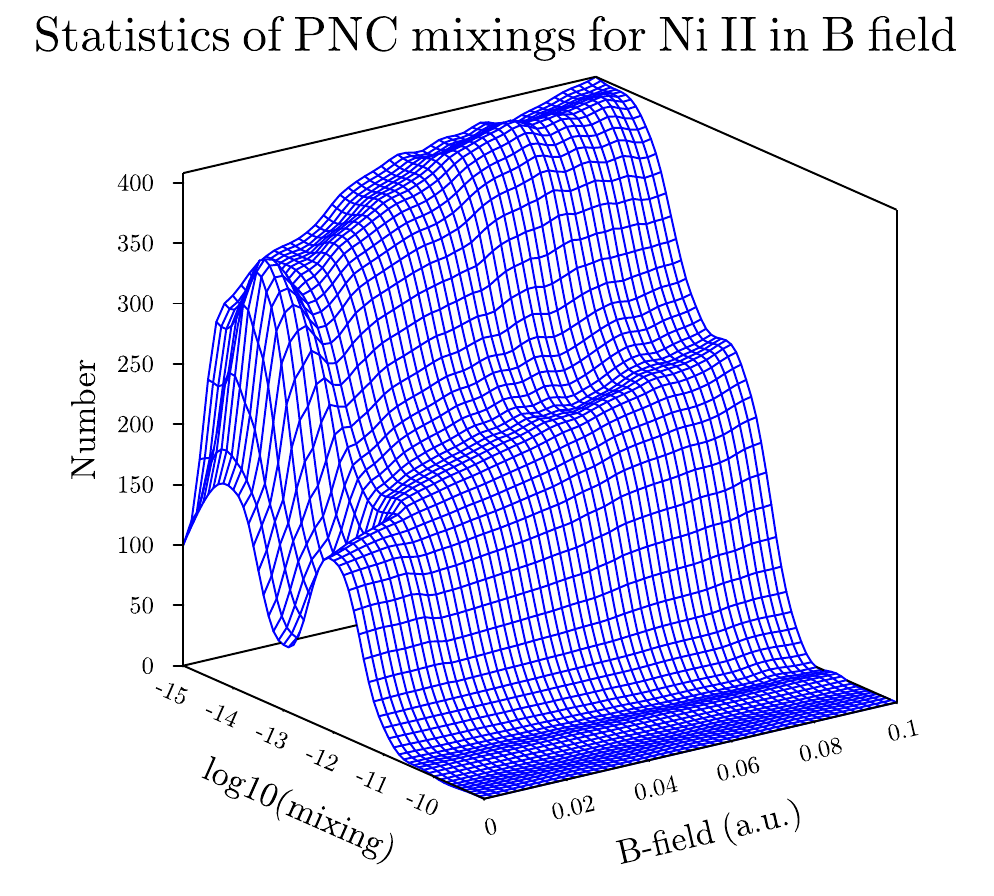}
    \caption{Calculations showing the statistics of opposite parity state mixing as a function of magnetic field. Atomic energy levels of both parities are calculated at a given magnetic field. Then, weak-interaction-induced mixings of neighboring states of opposite parity is evaluated. Taking logarithm of the absolute value of the mixing coefficients, one determines how many of the mixing values lie in a certain range, from which the plot is produced. We see that the number of mixing coefficients with magnitudes in the range between $10^{-15}$  and $10^{-12}$ grows with $B$, but there is no growth for larger mixings above $10^{-11}$.    }
    \label{Fig 3D_sumJ}
    \end{center}
\end{figure}
Figure \ref{Fig 3D_sumJ} shows a histogram for the calculated mixing of neighboring states of opposite parity for singly ionized nickel (Ni II) by a P-odd interaction \cite{Viatkina16}. We see that mixing grows with the magnetic field, but remains small. We conclude that for Ni II in the range of energies considered in this example (within approximately 30\,000 cm$^{-1}$ from the ground state) there is no transition to chaos even in magnetic fields of $\approx170\ $T. 
In Fig.\,\ref{Fig 3D_sumJ}  we see that the number of mixed states grows while the typical size of the mixing remains the same. It is predicted that at higher magnetic fields, there should be more of the larger mixing coefficients, however, it is not known yet if this will look like a phase transition or a gradual process. The average level spacing in Ni II is a few hundred cm$^{-1}$. In lanthanides and actinides, however, typical level density is much higher and transition to chaos may be possible. This can lead to much stronger enhancement of the mixing and high sensitivity to exotic fundamental interactions.

The success of the proposed spectroscopy program relies on one's ability to acquire spectra of complex system immersed in a strong (possibly, time-dependent) magnetic field in a massively parallel fashion. Advances in coherent light sources such as frequency combs provide an opportunity for basing the new broadband spectroscopy program on detection of absorption of light rather than fluorescence. In absorption, the initial electronic states of transitions can be known and controlled with high certainty. This simplifies the step of converting from an element's spectra to its electronic level structure and assigning configurations. The principle of broadband absorption spectroscopy with frequency combs was demonstrated by several groups, including the spectroscopy of molecular iodine \cite{Diddams2007}, spectroscopy of HfF$^+$ \cite{Sinclair2011}, and trace-gas detection \cite{Nugent-Glandorf2012}. In these examples absorption spectra were parallel-acquired for a wavelength range of several nanometers, with precision on the order of several MHz.
These publications demonstrate the feasibility of high resolution, broadband spectroscopy with frequency combs, albeit with limited wavelength coverage. This wavelength range was limited either by the frequency comb itself or the wavelength coverage of the dispersive imaging systems. Extending the spectral range to hundreds of nanometers will be highly desirable for the proposed program. For the light source, octave spanning frequency combs are a well-established technology, and can be acquired as complete commercial systems. 

Finally, once the spectra have been obtained as a function of magnetic field, novel theoretical techniques will be applied to analyze these data and extract the target information. An example of such theoretical tool is the recently developed software package designed to model atomic spectra with a combination of configuration-interaction and many-body perturbation-theory analysis (CI+MBPT) \cite{KPST15}. 

\subsection{Positron capture by atoms}

According to calculations \cite{Harabati2014}, about half of the atoms in the periodic table have bound states with the positron. However, such bound states have not been observed so far. Experimental observation of such bound systems, apart from being a test of atomic theory, would mark a creation of a fundamentally new class of atomic systems of particular interest for fundamental physics because they involve an antiparticle.  For example, spectroscopy of atoms containing an antiparticle can be used in searches for exotic spin-dependent interactions among the atomic constituents \cite{Ficek2018}.   

The authors of Ref.\,\cite{Harabati2014} suggested a possibility to capture
positrons to a shallow bound level using a pulse of a strong
magnetic field. The idea here is similar to the use of Feshbach resonances for molecular formation out of cold atoms, where the system can be brought into a bound state by sweeping the magnetic field. The same techniques can be used to capture electrons to an atom forming a negative ion. 

\subsection{Search for dark matter using atomic-physics methods} 

In Sec.\,\ref{Subsubsec: Haloscopes}, we discussed the CASPEr and QUAX experiments, where a spin system is brought into resonance with the Compton frequency of the DM field via tuning the energy splitting between the spin states by varying the leading magnetic field. The availability of high-field magnets may expand the range of the DM-particle masses that may be explored. Various other possibilities have also been discussed in the literature. 

Atomic and molecular transition induced by axion/ALPs were considered by Zioutas and Semertzidis \cite{ZioutasSemertzidis1988}. Axions/ALPs can induce  $1^+$ (i.e., $M1$) and $0^-$ transitions, where the number indicates the tensor rank of the transition operator and +/- indicate whether there is (-) or there is no (+) change of parity in the transition. Sikivie \cite{Sikivie_PRL_2014} considered atomic $M1$ transitions induced by the derivative coupling of axions/ALPs to nucleons and electrons. The transition probability is proportional to the square of the axion-fermion coupling. It was proposed to detect the axion/ALP-induced transitions by resonantly photoionizing excited atoms. Since the method was discussed in the context of the detection of the galactic-halo axion/ALPs, it is assumed that the atomic energy interval of the axion/ALP-driven transition is resonant with the frequency corresponding to the axion/ALP mass. 

Following Sikivie's approach, \citet{AXIOMA_2015} considered an experimental realization using  molecular oxygen at sub-kelvin temperatures, where the transition frequency is scanned by applying a strong magnetic field and the detection is done via resonant multiphoton-ionization (REMPI) spectroscopy. Another system considered in this context is optically pumped Er$^{3+}$-doped yttrium-aluminum garnet (YAG) crystals \cite{AXIOMA_YAG_2015}. 

We note that related to the discussion of the axion/ALP-induced atomic transitions is the recent work on ``cosmic parity violation''  \cite{Roberts_2014_PRD}, where $E1$ transitions between the atomic and molecular states of the same nominal parity induced by the background cosmic field that may be part of DM were considered.

\clearpage
\section{Current and forthcoming magnet technologies}
\label{Sec: Current and forthcoming magnet technologies}

\subsection{DC magnets}
\label{Sect: DC magnets}

In this section, we attempt to orient the reader in the various types of modern DC magnets that include a variety of resistive magnets, purely superconducting magnets, and hybrid magnets that combine both resistive and superconducting magnets. The development of various magnet types is driven by specific applications and available materials.

CERN is one of the world's centers for development and use of superconducting magnets. Its 14\,TeV center-of-mass energy proton-proton collider (LHC) is the largest superconducting system incorporating more than 9000 magnets.
The 8\,T dipole magnets in the LHC's 27\,km long tunnel are based on a 28-strand NbTi/Cu cable with a critical current at 1.9\,K of about 20\,kA corresponding to a maximum field of 9\,T. The feasibility of future higher-energy colliders critically depends on the magnet development. Recent years have seen significant progress towards dipoles with field strength of 16\,T.

Detector magnets are typically quite distinct from accelerator magnets, for instance, in their geometry  (``barrel'' vs. ``tube'') and volume, with detector magnets typically having orders of magnitude larger volume and stored energy. 
The world's largest detector-magnet system is currently that of the ATLAS detector at LHC \cite{ten2010atlas}. It consists of a barrel toroid, two end-cap toroids and a central solenoid, generating a 2\,T field for the inner detector and a ~1\,T for the peripheral muon detectors. The detector is 21\,m in diameter and 25\,m long, with a 8300\,m$^3$ volume containing magnetic field. The magnets operate at 4.7\,K with a current of 20.4\,kA and a peak field of 4.1\,T.  

As emphasized above, generating high magnetic fields can only be done by circulating high current densities, which are intrinsically accompanied by strong Lorentz forces. Superconductors allow circulating such currents without ohmic losses, up until a certain limiting value, imposed by the material's intrinsic properties. There is, therefore, no need for large power supplies, nor for large cooling installations, albeit at the expense of the cryogenics needed to maintain the magnet at low temperatures.  For the current workhorse material of high-field superconductivity, Nb$_3$Sn, this current corresponds to a magnetic field on the conductor of around 25\,T. The new family of high transition temperature (Tc) cuprate superconductors has critical magnetic field values above 100\,T, and therefore seems to be the dream material for the magnet engineer. However, the problem of the Lorentz forces remains, and neither the mechanical strength of high-Tc superconductors, nor their processability,  are optimal to build magnets that can handle high forces. The highest magnet field that has been generated with a high-Tc superconducting (HTS) insert is currently 46 T (Sec.\,\ref{Subsubsect_Superconducting_Magnets}). The engineering freedom using non-superconducting metals and alloys is much larger and fields above 100 T have been generated, albeit only in pulsed mode (Sec.\,\ref{Subsec:PulsedMagnets}) because of the ohmic heating problem.
Thus, if sufficient electrical power, and the corresponding cooling capacity are available (plus the budget to operate them), the magnet that generates the highest magnetic fields will be a hybrid one (Sec.\,\ref{Subsubsect_Hybrid_Magnets}), a superconducting outsert, providing several tens of tesla, in which one operates a resistive insert. Only if HTS with the same mechanical properties as the best non-superconducting conductors become available, can one hope to generate the highest possible fields with only superconductors.

\subsubsection{Conductors for DC and pulsed resistive magnets}
\label{Subsubsect: Conductors for resistive magnets}

The highest-field DC magnets in Tallahassee, Hefei, Nijmegen, Tsukuba, and Sendai use a copper-silver alloy that is $\approx 24$\,weight\% Ag. The Cu-Ag sheet used for DC resistive magnets is a bulk micro-composite consisting of a eutectic\footnote{Eutectic is a mixture of substances in specific proportions that melts and freezes at a single temperature.}  Cu-Ag phase and a proeutectic Cu phase. Inside the eutectic phase are layers of Cu and Ag with spacing $<1\,\mu$m. Inside the proeutectic Cu phase are Ag filaments $<1\,\mu$m in diameter. This is the material of choice for DC magnets operating in the 30-45\,T range as it provides an optimal combination of mechanical strength and electrical conductivity. However, for the higher-field pulsed magnets, strength becomes more important, and Cu-Nb is preferred given its higher strength/conductivity ratio.

\subsubsection{Resistive Bitter-type magnets}
\label{Subsubsect_Resistive_Magnets}

\begin{figure}[h!]
    \begin{center}
     \includegraphics[width=5.5 in]{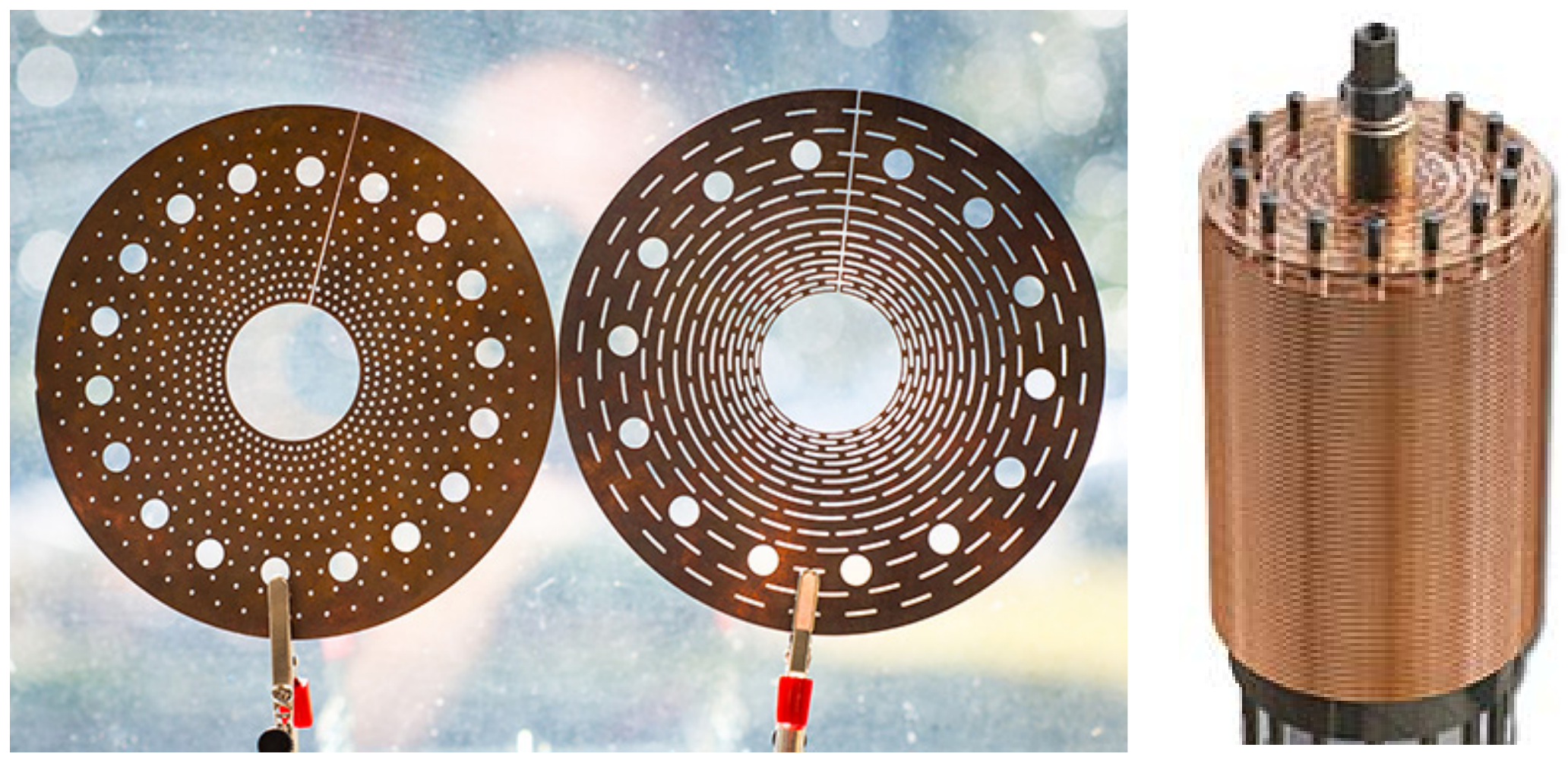}
    \caption{Left: Images of the original ``Bitter-plate'' design (round holes) and the more recent “Florida-Bitter-plate” design (staggered oval holes), used in the construction of high-field resistive electromagnets. The holes in the plates allow for the efficient flow of cooling water through the magnet, which is needed to dissipate the considerable (10s of MW) resistive heating.  The central hole in the plates is about 40\,mm in diameter. Right: these plates, with separating insulating layers, can be stacked in a helical configuration to construct a complete magnet.} \label{Fig BitterMagnetDesign}
    \end{center}
\end{figure}

Conventional superconducting electromagnets -- meaning those wound with standard low-temperature superconducting NbTi and Nb$_3$Sn wire [see Sec.\,\ref{Subsubsect_Superconducting_Magnets} for recent developments in electromagnets wound with high-temperature (high-Tc) superconducting wire] typically operate at cryogenic temperatures (1.8-5 K) and can generate magnetic fields up to about 25 T in small volumes (typically only a few cm$^3$ for fields $>22\,$T), this upper limit being imposed by the field- and temperature-dependent critical current of the superconducting wire itself.  To achieve higher magnetic fields, electromagnets must be constructed using resistive conductors, typically copper or higher-strength copper alloy (Sec.\,\ref{Subsubsect: Conductors for resistive magnets}).  However, this means that the substantial Joule heating in the coil which can amount to many megawatts of power must be continuously dissipated, otherwise the magnet will overheat and fail catastrophically. For steady-state DC electromagnets, the ability to dissipate this heat imposes a limit to the maximum achievable field. The geometry of monolithic solenoids wound with conventional wire makes it difficult to efficiently remove heat.  A breakthrough was achieved in the 1930s by Francis Bitter at the Massachusetts Institute of Technology, who used a stack of water-cooled flat copper plates or discs as the conductor instead of conventional wire.  These ``Bitter plates'' were drilled out with many round holes (Fig.\  \ref{Fig BitterMagnetDesign}) to allow cooling water to flow through and around the plates, giving increased cooling capacity \cite{https://nationalmaglab.org/about/maglab-dictionary/bitter-plate}. Using Bitter plates, eventually $28\,$T could be achieved using 20 MW of electrical power.  A further advance was the development of the “Florida-Bitter plate” in the 1990s at the NHMFL in Tallahassee; these designs used plates with staggered rows of specifically-engineered elongated holes to optimize the mechanical strength, current-carrying capacity, and cooling capacity of the magnet. Using these Florida-Bitter plates, $30\,$T resistive magnets were developed in 1995, and $35\,$T magnets were developed in 2005.  These magnets typically had 32\,mm diameter bores, were powered with a $20\,$MW power supply ($\approx 40000\,$A, $\approx 500\,$V), and were continuously cooled with chilled de-ionized water flowing at rates of order 15000 liters/minute \cite{https://nationalmaglab.org/news-events/feature-stories/making-resistive-magnets}.  Florida-Bitter-type electromagnets are now in place at many of the high-field magnet laboratories around the world, including the $38.5\,$T resistive magnet in Hefei (China) and the $37.5\,$T resistive magnet at Nijmegen (Netherlands).

\begin{figure}[h!]
    \begin{center}
     \includegraphics[width=3.5 in]{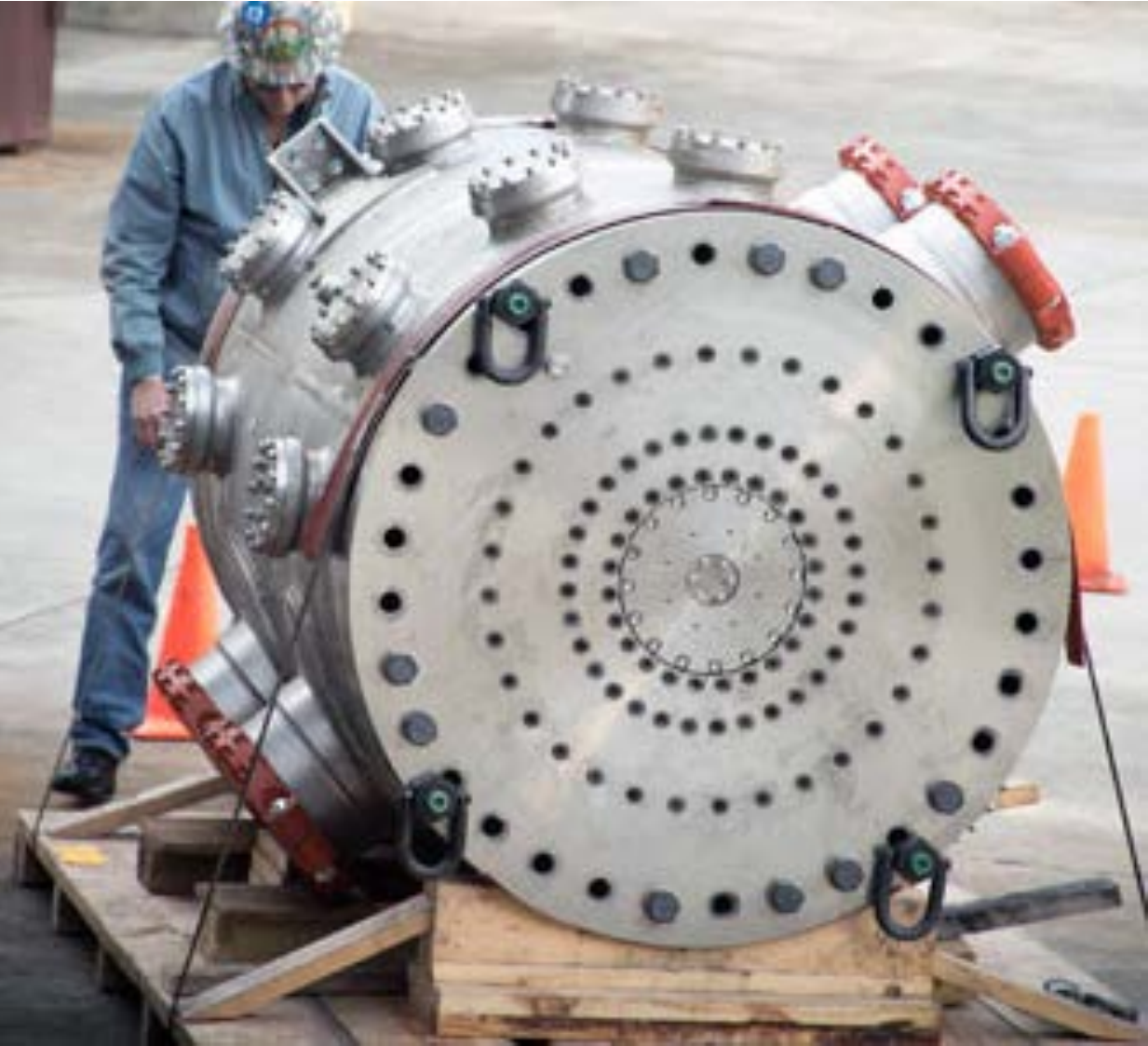}
    \caption{Photograph of the 41.4\,T resistive magnet at the NHMFL-Tallahassee (August 2017).  This magnet has a 32\,mm bore and is powered with a 32\,MW power supply.  Its design is based on Florida-Bitter plates (see text) using high-strength copper-silver-alloy conductor. Photo courtesy of the NHFML-Tallahassee.}
    \label{Fig tallahasseemagnets_resistive}
    \end{center}
\end{figure}

Driven by advances in materials science and magnet engineering, progress with Bitter-type electromagnets continues to this day. In 2017, an all-resistive $41.4\,$T magnet was demonstrated at the NHMFL in Tallahassee \cite{https://nationalmaglab.org/news-events/news/new-world-record-magnet-fulfills-superconducting-promise}, Fig.\,\ref{Fig tallahasseemagnets_resistive}, using Florida-Bitter plates of high-strength copper-silver alloy and an upgraded $32\,$MW power supply. 

While the primary scientific use of Bitter-type magnets is research in condensed-matter physics  and materials science, these magnets may also prove useful for the various types of fundamental-physics experiments discussed in this review, particularly in the cases where large field volumes are not required.

\subsubsection{Superconducting magnets}
\label{Subsubsect_Superconducting_Magnets}

\begin{figure}[h!]
    \begin{center}
     \includegraphics[width=5.5 in]{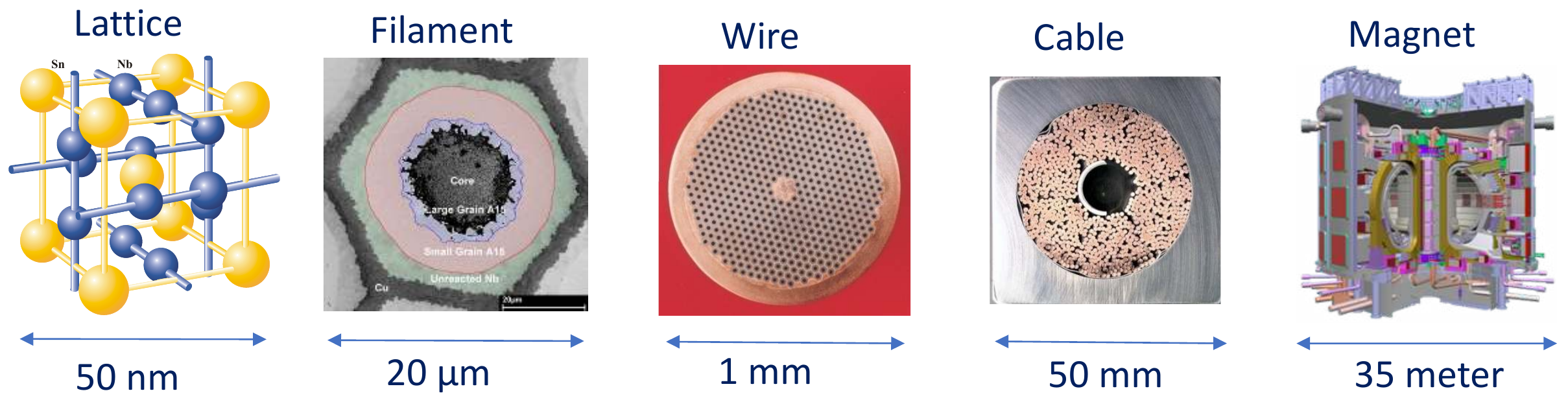}
    \caption{Designing a reliable high-performance magnet is a challenge that spans many scales, from material design at the atomic scale, all the way to engineering of a complete magnet system.}
    \label{Fig Material to Magnet}
    \end{center}
\end{figure}

Designing a modern high-performance magnet operating with multi-kA currents that needs to be safe, reliable and show no degradation over many years is a multi-scale challenge (Fig.\ \ref{Fig Material to Magnet}), starting from superconductor-material engineering at the atomic level, all the way to the complete magnet-system design. It is important to understand and control all the links in the chain.

\begin{figure}[h!]
    \begin{center}
     \includegraphics[width=6.5 in]{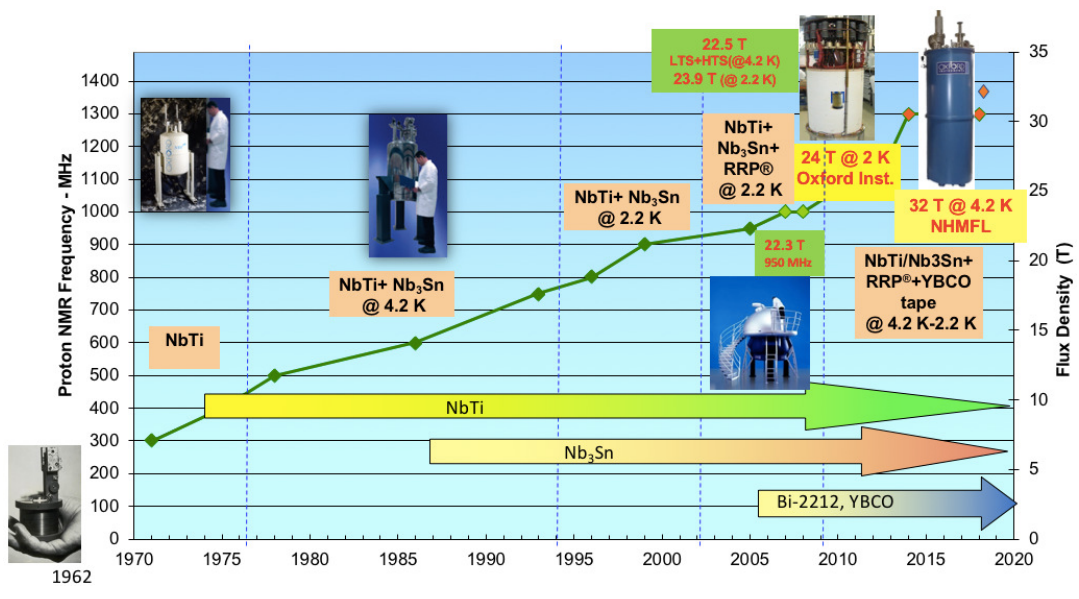}
    \caption{Solenoid magnets for NMR as a representative example of the progress in magnet technology. NHMFL: National High Magnetic Field Laboratory, Tallahassee, Florida.}
    \label{FIG_Solenoid Magnets}
    \end{center}
\end{figure}

From the discovery of superconductivity by Heike Kammerlingh Onnes in 1911, it took about 60 years to develop the first practical multitesla magnets. One of the first such magnets using NbTi wire was developed by Oxford Instruments in 1971 (Fig.\,\ref{FIG_Solenoid Magnets}). Superconducting-magnet records generally follow developments in superconductors. A representative parameter is current density. In the last 30 years, the current densities increased from 500 to 3500\,A/mm$^2$. 

High-critical-temperature (high-Tc) superconductors are key to much of the modern SC-magnet technology. The first-generation (1G) high-temperature superconducting (HTS) wires that have been commercially available since 1990s are based on the BSCCO (bismuth-strontium-calcium-copper-oxide) material. The more recent second-generation (2G) wire with better electrical and mechanical properties is based on ReBCO---rare-earth-bismuth-copper oxide. The rare-earth elements are typically gadolinium, yttrium, samarium, or neodymium.

Key properties of some superconducting materials are summarized in Table \ref{Tab SC materials}.
\begin{table}[h]
\center
\small
\begin{tabular}{cccccc}
   \hline\hline
  Material Name& Class & Critical Temperature (K) & Critical Field $B_{c2} $ & Critical Field@2.2\,K & Geometry \\ \hline
 NbTi & LTS & 9.8 & 9.5 T @ 4.2 K & 11.5 T & Multi-filamentary \\
         &         &       &                        &            & round \& rectangular wire \\ \\

 Nb$_3$Sn & LTS & 18.1 & 20 T @ 4.2 K & 23 T & Multi-filamentary \\
         &         &       &                        &            & round wire \\ \\

 MgB$_2$ & MTS & 39 & 5$-$10 T @ 4.2 K & N/A & Multi-filamentary \\
         &         &       &  1$-$3 T @ 10 K   &            & round wire \\ \\

 Bi$-$2212 & HTS & 90$-$110 & 40 T @ 4.2 K & N/A & Multi-filamentary \\
         &         &       &  10 T @ 12 K   &            & round wire \\ \\

 Bi$-$2213 & HTS & 90$-$110 & 40 T @ 4.2 K & N/A & Tape \\
         &         &       &  8 T @ 20 K   &            &  \\ 
         &         &       &  4 T @ 65 K   &            &  \\  \\

YBCO & HTS & 92$-$135 & 45 T @ 4.2 K & N/A & Tape \\
         &         &       &  12 T @ 20 K   &            &  \\ 
         &         &       &  8   T @ 65 K   &            &  \\ 

  \hline\hline
\end{tabular}
  \caption{Superconducting materials. LTS, MTS, and HTS stand for low-, medium-, and hight-temperature superconductors. N/A means that these materials as are typically not used below 4.2\,K. They can operate at lower temperatures but without particular advantage.}\label{Tab SC materials}
\end{table}

\subsubsection{Superconducting magnets based on high-temperature superconducting wire}

\begin{figure}[h!]
    \begin{center}
     \includegraphics[height=3.5 in]{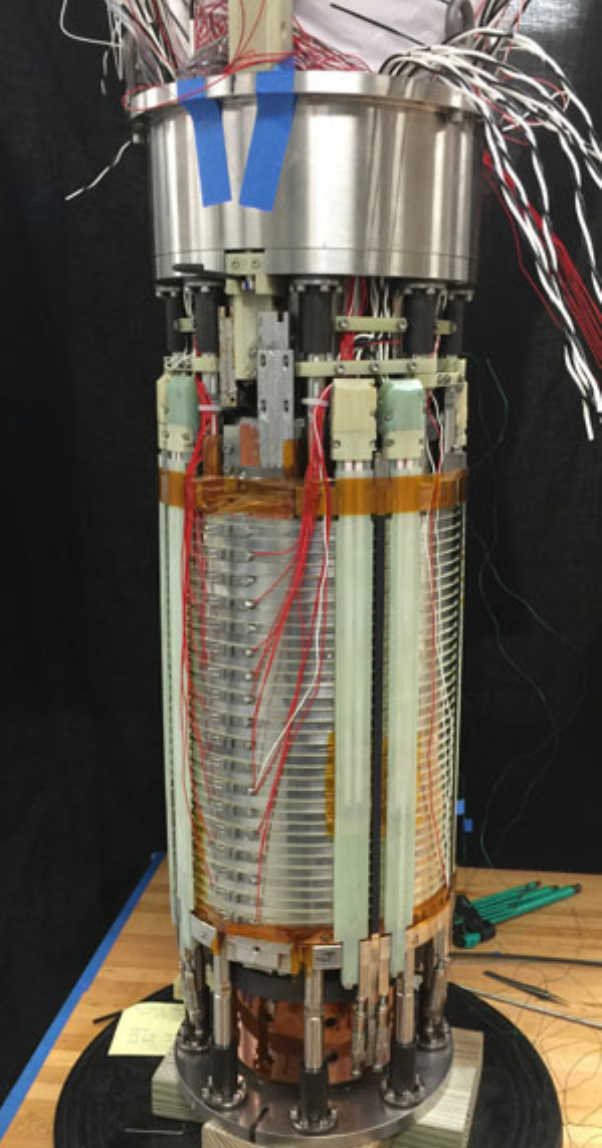}
     \includegraphics[height=3.5 in]{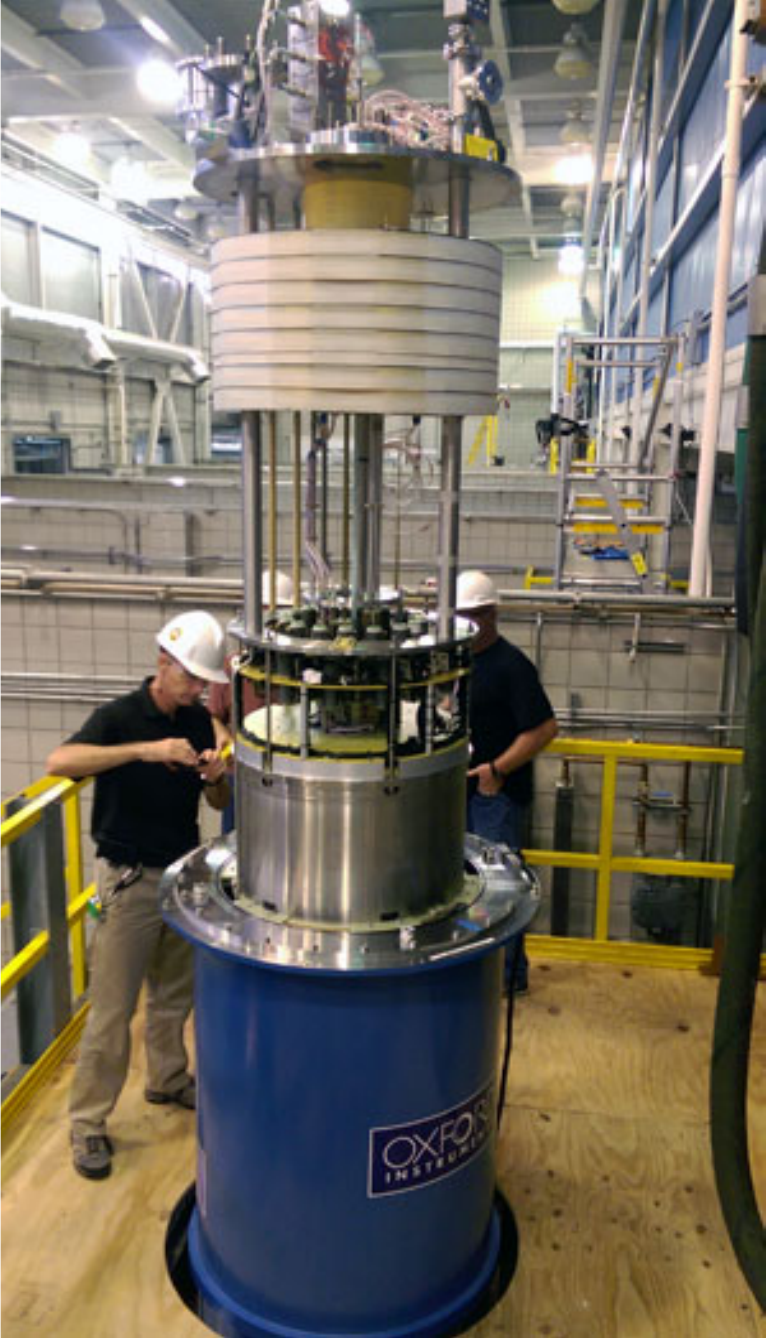}
    \caption{Left: photograph of the innermost two high-Tc-superconductor (YBCO) coils of the all-superconducting 32\,T magnet, whose successful operation was demonstrated in December 2017 at the NHMFL-Tallahassee. Total height of the pictured assembly is about 1\,m. These YBCO coils are integrated within an outer coil set made of conventional (low-Tc) superconducting wire. The LTS outsert and the cryostat were designed and developed by Oxford Instruments NanoScience. Right: magnet assembly. Photos courtesy of the NHFML-Tallahassee.}
    \label{Fig tallahasseemagnets}
    \end{center}
\end{figure}

As noted above, the field- and temperature-dependent critical current of conventional (low-Tc) superconducting wire (Nb$_3$Sn wire) limits the maximum achievable magnetic field to of order $25\,$T. However, high-temperature (high-Tc) superconductors such as YBCO can exhibit significantly higher critical currents, raising the possibility of a new generation of very high-field superconducting magnets with fields significantly exceeding $25\,$T. As with many aspects of high-field magnet development, the primary challenge lies in the materials science and engineering of the superconducting wire (or tape) itself.  Being ceramic oxides, high-Tc superconductors exhibit significantly different electric and mechanical properties from those of low-temperature superconductor materials. A breakthrough was achieved in late 2017 with the demonstration of an all-superconducting $32\,$T magnet at NHMFL-Tallahassee \cite{https://nationalmaglab.org/news-events/news/new-world-record-magnet-fulfills-superconducting-promise}; Fig.\ \ref{Fig tallahasseemagnets}. Using a combination of a low-Tc superconducting outsert (developed in partnership with Oxford Instruments) and a high-Tc superconducting insert (using YBCO conductor developed in partnership with SuperPower Inc.), this $32\,$T magnet offers a stable and homogeneous field. Given the high critical current of YBCO and other high-Tc materials, it is anticipated that superconducting magnets with considerably higher field can be developed in the coming years \cite{https://nationalmaglab.org/news-events/news/new-world-record-magnet-fulfills-superconducting-promise}. 

The feasibility of extending the field range of DC magnets all the way to 100\,T (the field range so far only accessible in the pulsed mode) was considered \cite{Iwasa2013} with an optimistic conclusion that such a magnet should be possible with present-day technology. The ``first-cut'' design is based on GdBCO superconductor arranged in a nested-coil configuration (39 coils in total), with a band of high-strength steel over each coil. The design envisions a 20-mm cold bore, a nearly 5.6-m outermost winding o.d.,
and nearly 17\,m winding height. The GdBCO conductor tape is assumed to be 12\,mm wide and just under 0.1\,mm thick, with a total length over 12\,500 km. Its magnet's inductance is 43\,kH and its magnetic energy is 122\,GJ at 2400\,A.

\begin{figure}[h!]
    \begin{center}
     \includegraphics[width=6.5 in]{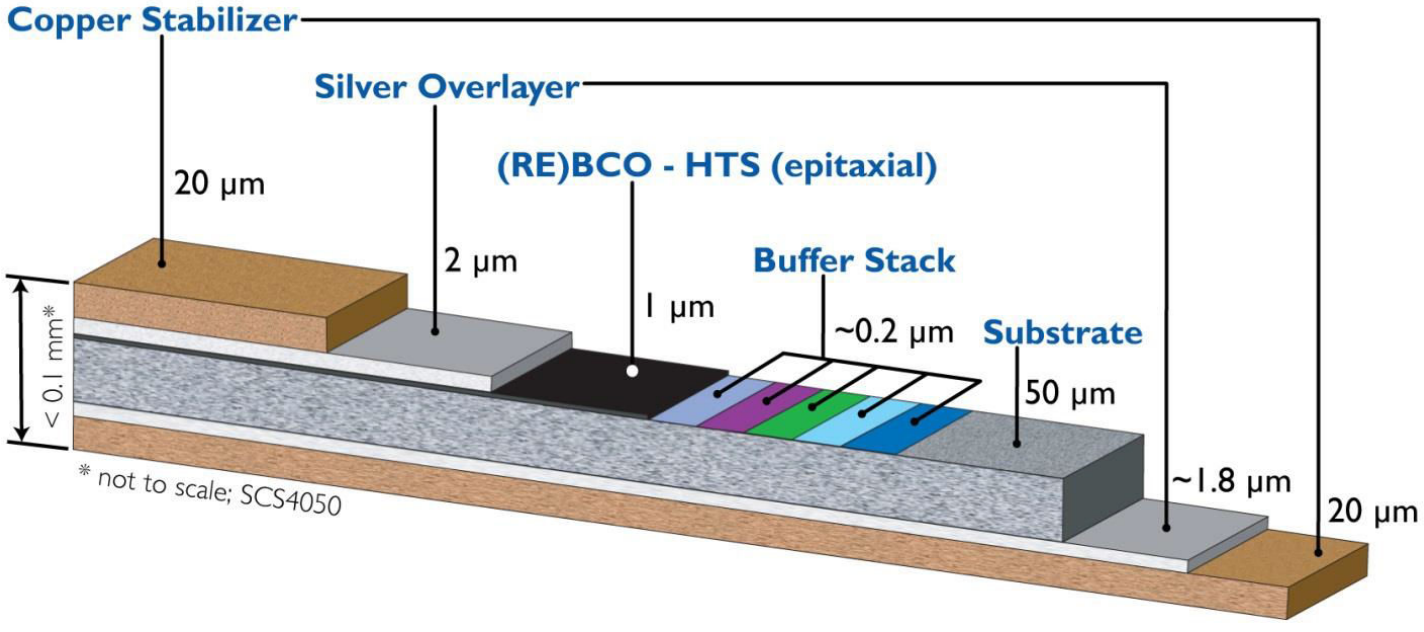}
    \caption{High-temperature superconducting cable design by SuperPower, offering high pinning capabilities to withstand the presence of a large transverse magnetic field. Figure from \cite{http://www.superpower-inc.com/content/2g-hts-wire}.}
    \label{Fig no insulation HTS}
    \end{center}
\end{figure}

High-reliability HTS cable for operation at low temperature is currently available from several companies.  A HTS cable consists of many different layers, as shown in Fig.\,\ref{Fig no insulation HTS}.
Magnets based on such cables are still in the development stage; due to the high cost of manufacturing these magnets it is imperative to have a reliable quench protection mechanism.  Even though such quench-protection systems have been developed, a more passive system could be advantageous to fully protect the magnets.  When a quench is developed, it is imperative to safely extract the large stored magnetic energy and prevent generating the equivalent heat in a small concentrated region that can destroy the coil.  An alternative to an active quench-protection system is to use no insulation between the turns in the coil; that way the current after a quench will immediately be diverted into the different turns instead of heating up the quenched spot.  No-insulation magnets were invented and applied at the Brookhaven National Laboratory \cite{Strongin1967} using low-temperature superconductors. But such coils cannot be charged fast which makes them unusable in most accelerator applications.  However, some axion/ALP searches require steady-state magnet operation, where the field stability is not critical, the magnet cycle time can be long, and the magnet field uniformity is not critical either.  

No-insulation, high temperature superconducting magnets, were first introduced in 2010 at the MIT lab of Francis Bitter \cite{NoIns1,NoIns2}.  The Center for Axion and Precision Physics Research at the Institute for Basic Science of the Republic of Korea (IBS-CAPP) already has in its laboratory two HTS, no-insulation magnets \cite{Ahn2017} made by SuNAM (\url{http://www.i-sunam.com}). One magnet with an aperture of 3.5\,cm is capable of reaching 26\,T , while the second one is capable of providing 18\,T in a 7\,cm inner diameter, over 40\,cm long magnet-bore aperture.  The 18\,T, no-insulation, HTS magnet is the first such magnet currently used in the field of axion dark matter.  In addition, IBS-CAPP is commissioning a 25\,T, 10\,cm inner diameter, 40\,cm long-solenoid, no-insulation, HTS magnet with the Brookhaven National Laboratory magnet Division, scheduled for delivery in 2019.  The 25\,T magnet will also be used to search for axion dark matter.

The no-insulation, HTS magnets exhibit superb stability in steady-state operation, while tests on quenches indicate a robust system even after multiple quenches \cite{NoIns1,NoIns2}.  Charging the magnet takes much longer than an equivalent for a magnet with insulation, with magnetic field lagging behind the current as it is distributed in all turns at the beginning.  However, the current eventually prefers to follow the superconductor instead of the copper stabilizer of Fig.\,\ref{Fig no insulation HTS}.  The B-field drifts to its final value over longer times, but the field difference does not constitute a significant parameter in the axion dark-matter search, making these magnets ideally suited for this application.

\subsubsection{Hybrid magnets}
\label{Subsubsect_Hybrid_Magnets}

DC magnetic fields up to about $45\,$T can be achieved using a hybrid approach, wherein a Bitter-type resistive electromagnet (now the `insert') is placed within the bore of a large superconducting `outsert' magnet \cite{Bird2015,Ouden2016}. Using an $11.5\,$T superconducting outsert and a $33.5\,$T resistive insert, a $45\,$T hybrid magnet (32 mm bore) was first demonstrated in 1999 at the NHMFL-Tallahassee. 

A recent advance in hybrid magnet design are the so-called ``series-connected'' hybrid magnets, in which a superconducting outsert magnet and a resistive (Florida-Bitter type) insert magnet are driven in series by the same power supply, rather than by independent power supplies as is typically the case.  Connecting the superconducting outsert and the resistive insert in series has the benefit of significantly reducing, due to the large inductance of the superconducting magnet, the high-frequency magnetic field fluctuations that are otherwise typically present in purely resistive Bitter-type magnets. The result is an extremely ``quiet'' magnetic field that is well suited to precision measurements requiring high stability (e.g., spin resonance \cite{Gan2017}). A 36$\,$T series-connected hybrid magnet with a 40 mm warm bore and $\approx 1\,$ppm uniformity over 1 cm$^3$ was recently developed and tested at the NHMFL in Tallahassee \cite{Dixon2017,https://nationalmaglab.org/news-events/news/national-maglab-racks-up-another-record}. With its high stability and excellent field homogeneity, this system is suitable for electron-spin-resonance studies at frequencies up to $1\,$THz and NMR at $\sim 1.5\,$GHz proton frequency.

\begin{figure}[h!]
    \begin{center}
     \includegraphics[width=5.5 in]{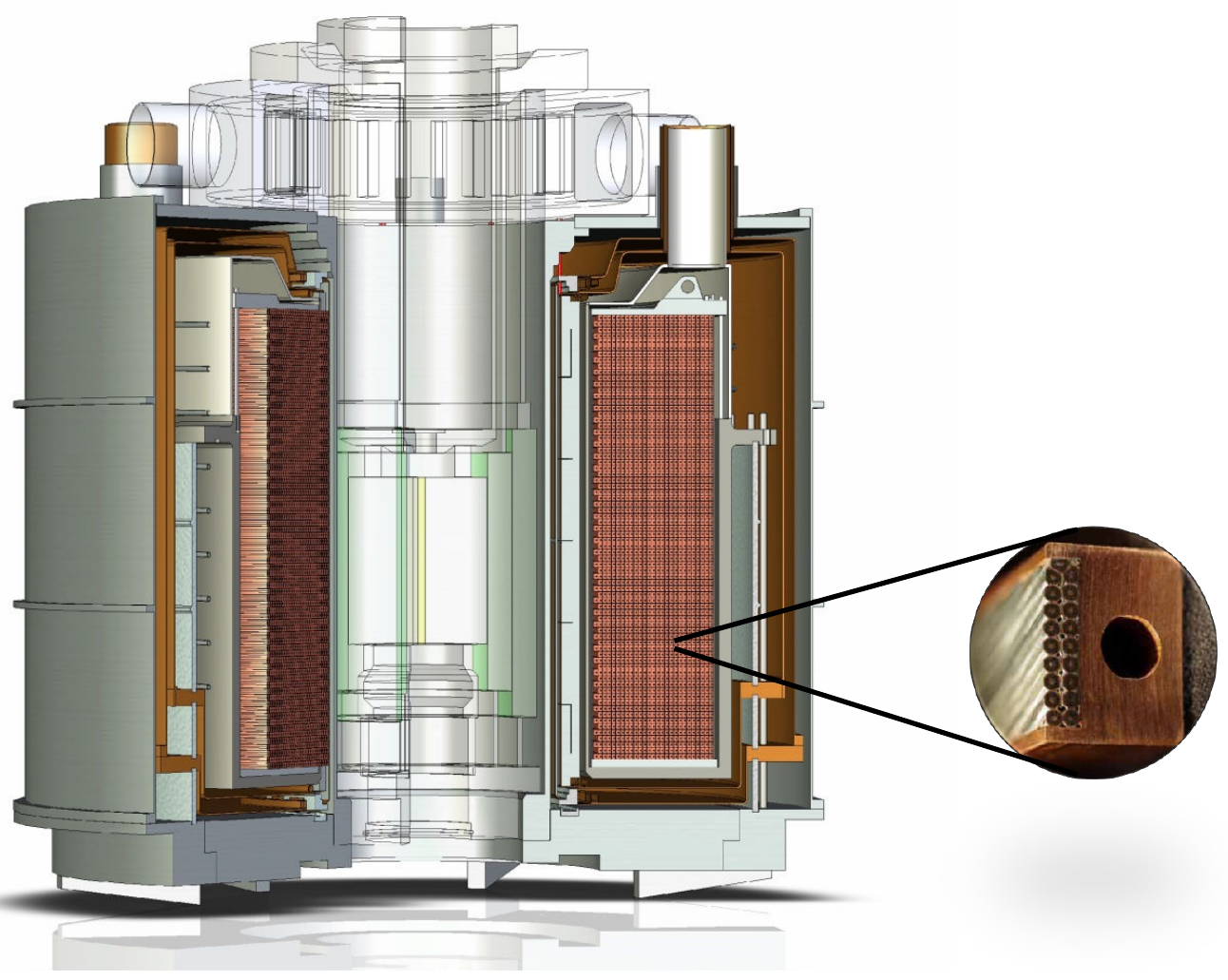}
    \caption{Three-dimensional cut-away view of the cryostat and superconducting coil of the Grenoble hybrid magnet under construction in collaboration with Saclay Nuclear Research Center (CEA-Saclay). The superconducting coil is made of the assembly of 37 interconnected double pancakes with inner-bore diameter of 1111\,mm and a total height of 1400\,mm. The total weight of the hybrid magnet is about 53 tons, including the 24 ton mass cooled down to 1.8\,K with superfluid He. The inset shows a superconductor of 17.92\,mm in height$\times$12.96\,mm in width.}
    \label{Fig Grenoble Hybrid}
    \end{center}
\end{figure}

As an example of the hybrid-magnet technology we briefly describe a system under construction at LNCMI-Grenoble (Fig.\,\ref{Fig Grenoble Hybrid}) that, in a first step, will produce a continuous magnetic field of 43\,T in a 34\,mm warm bore \cite{Pugnat2018}. There are only a few  hybrid magnets over 40 T currently operating worldwide \cite{Pugnat2014} as there remains a technological challenge due to the necessity for the superconducting coil to withstand quenches and large electromechanical forces from the warm insert, especially under fault conditions. A feature of the LNCMI system (also available for some other hybrid magnets) is that it is reconfigurable, offering the users a range of configurations from 43\,T in 34\,mm to 9\,T in 800\,mm diameter bore. This is accomplished by using different combinations of superconducting and resistive coils. The superconducting coils are based on the proven Nb-Ti/Cu material and are cooled with pressurized superfluid He to 1.8 K, producing in excess of 8.5\,T. The magnet requires a dedicated He liquefier with a production capacity of 140 l/h. The 13\,km long superconductor is made from a flat Rutherford cable (a design named after the laboratory where it was introduced) containing 19 Nb-Ti strands of 1.6 mm diameter soft-soldered onto a copper-silver hollow “stabilizer” with both ends connected to the pressurized superfluid He bath. Operation at low temperatures enhances the superconducting properties of the conductor while the use of superfluid helium allows to profit from its uniquely high thermal conductivity. As an illustration, the heat flux transported by conduction between 1.9 and 1.8 K in a 1 m long static channel of superfluid He can be up to about three orders of magnitude higher than that of a bar of oxygen-free high thermal conductivity (OFHC) copper of the same geometry.
The resistive part of the 43 T, 34 cm-diameter-bore hybrid magnet will be a combination of resistive polyhelix and Bitter coils using 24 MW of electrical power \cite{Debray2012}. By combining the Bitter insert alone with the superconducting coil, another hybrid-magnet configuration will allow to produce 17.5 T in a 375 mm diameter aperture with 12 MW.
A crucial part of the magnet design is the control and protection system. Indeed, in case of a fault, it would be necessary to safely guide and dissipate the 110 MJ of energy stored in the hybrid magnet. A upgrade of the Grenoble hybrid magnet to exceed 45\,T in the 34\,mm diameter bore will be possible thanks to the ongoing two-phase upgrade of the electrical power installation, from 24 up to 30\,MW and then up to 36\,MW.

Looking towards the future, preliminary designs for a $60\,$T hybrid magnet have been recently initiated at NHMFL \cite{Bird2015}. 

\subsection{Pulsed magnets}
\label{Subsec:PulsedMagnets}

To attain magnetic fields in excess of $\approx45\,$T (the approximate limit of hybrid DC resistive magnets today), pulsed magnets are used.  In effect, this avoids the significant ``cooling problem'' described above; in a pulsed magnet the field is ramped up to a high value and back down again before Joule heating in the copper-alloy windings destroys the magnet. High magnetic fields $>50\,$T and even exceeding $100\,$T can be achieved nondestructively for timescales typically in the range of 1-100 ms.  For pulsed magnets, the design challenge centers primarily on high-strength materials (both reinforcement materials and the conductor itself) and clever engineering design to cope with the huge magnetic pressures that exist due to the Lorentz forces associated with the high fields (50-100$\,$T) and currents ($\sim 30\,000\,$A) in the magnet.  
\begin{figure}[h!]
    \begin{center}
     \includegraphics[width=6.5 in]{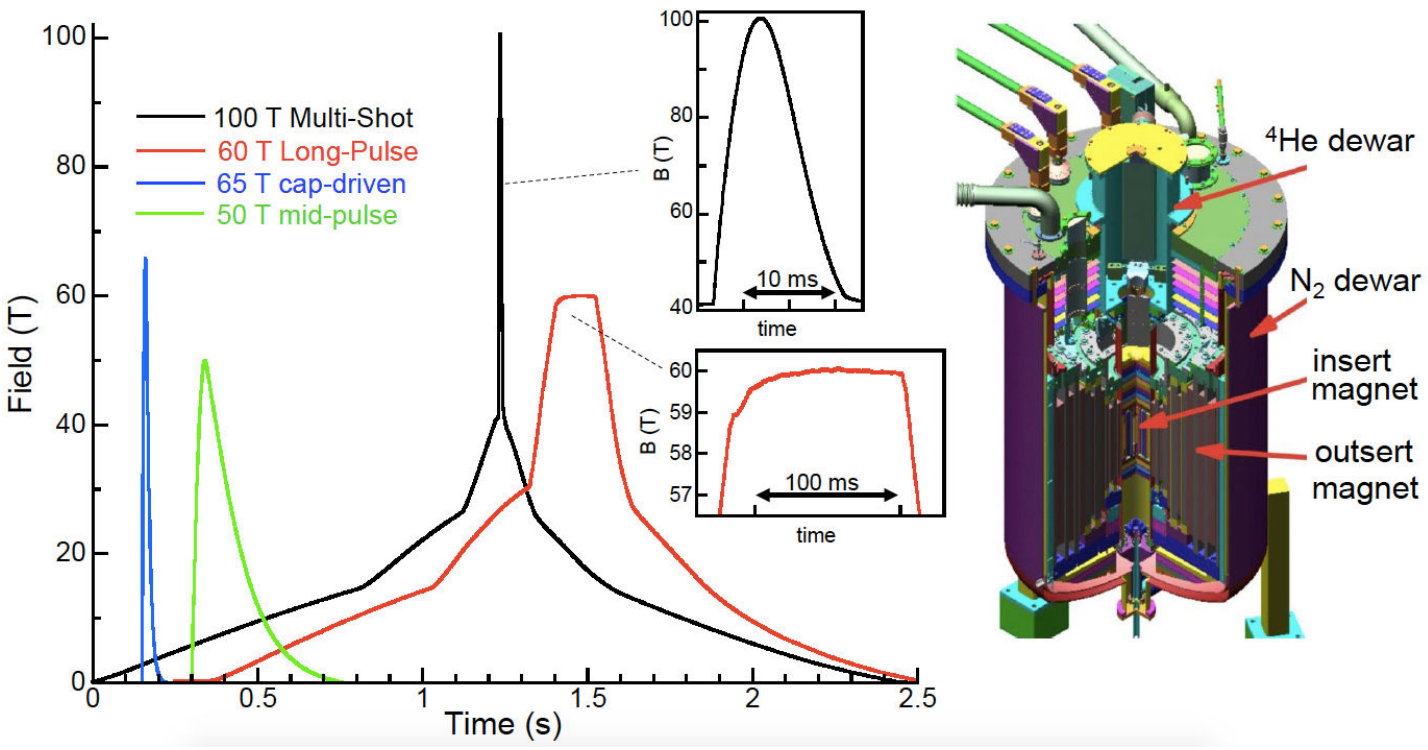}
    \caption{Measured field profiles of the various nondestructive pulsed magnets at the NHMFL-Los Alamos.  The 65 T ``standard-user'' magnets are powered with a 4 MJ capacitor bank, as is the larger-inductance 50 T ``mid-pulse'' magnet.  The 60 T Long Pulse magnet, which can maintain 60 T for 100 ms (or, for example, 50 T for 200 ms), has a user-defined waveform and is driven with a 1.4 GW motor-generator.  The 100 T Multi-Shot magnet, also depicted in the diagram on the right, has a generator-driven $\approx 40\ $T outsert magnet and a $\approx 60\ $T insert magnet that is driven with a 2.2 MJ capacitor bank; it achieved fields $>100.7\ $T in early 2012. The total height of the magnet assembly is $\approx 2\,$m.}
    \label{Fig NHMFLFieldprofiles}
    \end{center}
\end{figure}
Most often, a large capacitor bank capable of storing many megajoules of energy is used to power the magnet; however there are also other schemes employing large motor-generators (essentially, flywheels) as power sources.  For example, at the NHMFL-Los Alamos a $1.4\,$GW motor-generator is used to store up to $600\,$MJ of (rotational kinetic) energy  in a 300-ton steel shaft, which is used to drive one of the two large pulsed magnets, described briefly in the following section.

\begin{figure}[h!]
    \begin{center}
     \includegraphics[width=2.5 in]{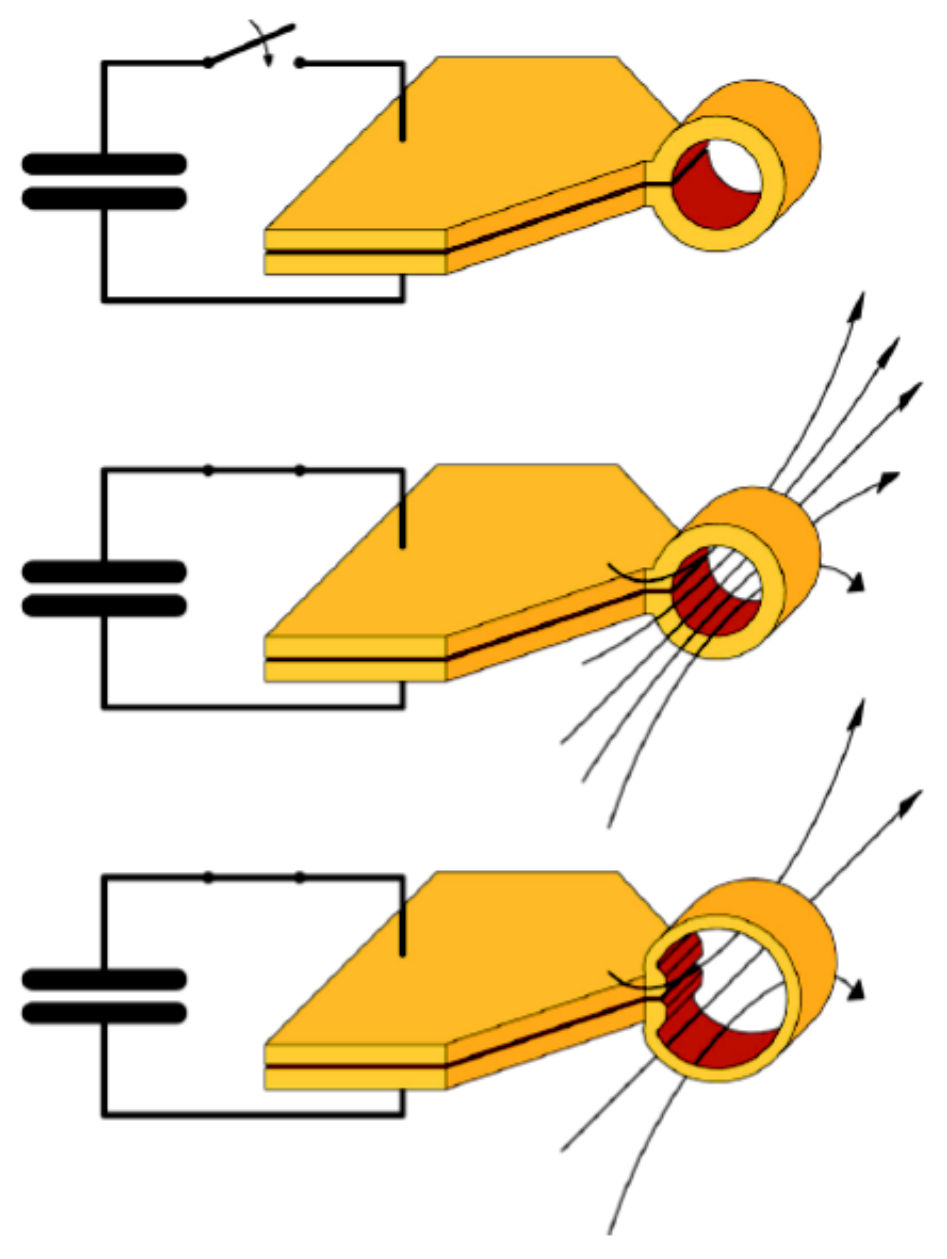}
    \caption{A schematic of a single-turn pulsed coil. The rise time of the field is determined by the capacitance and inductance of the entire circuit, which includes the capacitor bank, cabling, and the coil itself.
Once the current flows through the coil, magnetic field is generated and its pressure explodes the coil. 
The disintegration time of the coil is determined by its mass (inertia) and the applied magnetic force. 
The circuit impedance can be adjusted to make the rise time of the field shorter than the disintegration time of the coil. Typical values of the field achieved with this technique are 300$\,$T in a 5 mm diameter or 100~T in a 20~mm diameter, both with $\approx 5~\mu$s duration. Figure courtesy LNCMI \cite{LNCMI_Web}.}
    \label{Fig Megagauss Coil}
    \end{center}
\end{figure}
Magnetic fields significantly in excess of $100\,$T can be achieved for microsecond time durations in small volumes of order $1\,$cm$^3$ using copper coils, but these methods are either semi-destructive or completely destructive.  So-called ``single-turn coils'' are made from sheets of copper plate (typically 2-3$\,$mm thick) that are formed into a single loop with diameter of order 1$\,$cm (Fig.\,\ref{Fig Megagauss Coil}). A current pulse of $\approx 3\,$MA from a fast (low-inductance) capacitor bank can generate magnetic fields in the range of 150-300$\,$T for a few microseconds before the coil itself is vaporized \cite{Portugall1999,Portugall2013}. A significant advantage of the single-turn-coil technique is that the experiment in the coil (usually) remains intact; in this sense, the technique is semi-destructive. Single-turn magnet systems are currently operating at the Tokyo, Toulouse, and Los Alamos magnet laboratories.   

Significantly larger magnetic fields exceeding 1000$\,$T can be achieved on microsecond timescales using explosive or electromagnetic flux-compression techniques that are entirely destructive.  Here, a background seed field of several tesla is first introduced into a large metal ``liner'' (effectively a short copper cylinder with diameter of order 10$\,$cm).  The liner and the magnetic flux lines contained within are then rapidly compressed down to a small diameter, resulting in a corresponding significant increase in magnetic field. Flux-compression techniques using high explosives were explored in the 1990s at both Los Alamos (USA) and Sarov (formerly Arzamas-16, Russia), where magnetic fields in the range of 1000-2800$\,$T were reported \cite{Megagauss1998}.  Alternatively, electromagnetic flux-compression (EMFC) methods utilize a large single-turn primary coil placed around the liner.  When the primary coil is energized (typically with $\approx 5\,$MA of pulsed current from a capacitor bank), the nearly-equal-but-opposite secondary current induced in the liner results in forces that drive the liner inward, boosting the magnetic field. EMFC methods are currently practiced at the IMGSL (Japan), where magnetic fields exceeding  $600\,$T are routinely achieved (see Ch.\ 7 of the book \cite{Miura2008} and \cite{Takeyama2011}). Generation of fields approaching 1000~T via electromagnetic flux compression were reported recently \cite{Nakamura2018}.

\subsubsection{Pulsed magnets at NHMFL Los Alamos}
 \label{Subsubsect_NHMFLPulsedMag}

The $60\,$T Long-Pulse Magnet, which came online in 1998, can maintain a constant peak field of $60\,$T for up to $100\,$ms, in a $32\,$mm bore.  This magnet is driven with  a 1.4\,GW motor-generator as mentioned above. The overall pulse length (see the field profile in Fig.\ \ref{Fig NHMFLFieldprofiles}) is about $2\,$s.  Such a long duration at peak field in comparison with typical capacitor-driven pulsed magnets, for which time at peak field is of order milliseconds, allows for sensitive experiments that benefit from significant signal averaging or extended photon collection, such as the case in many optics and laser experiments. 

The $100\,$T Multi-Shot Magnet, which consists of a large generator-driven resistive outsert magnet ($40\,$T peak field in a large 225-mm bore) and a capacitor-driven insert magnet ($\approx 60\,$T), achieved a peak field of $100.75\,$T in 2012.  The time at peak field is of order milliseconds (see field profiles in Fig.\ \ref{Fig NHMFLFieldprofiles}), while the overall pulse duration is $\approx 2.5\,$s. 

\subsubsection{Pulsed magnets developed at LNCMI Toulouse}
 \label{Subsubsect_ToulousePulsedMag}

\begin{figure}[h!]
    \begin{center}
     \includegraphics[width=6.5 in]{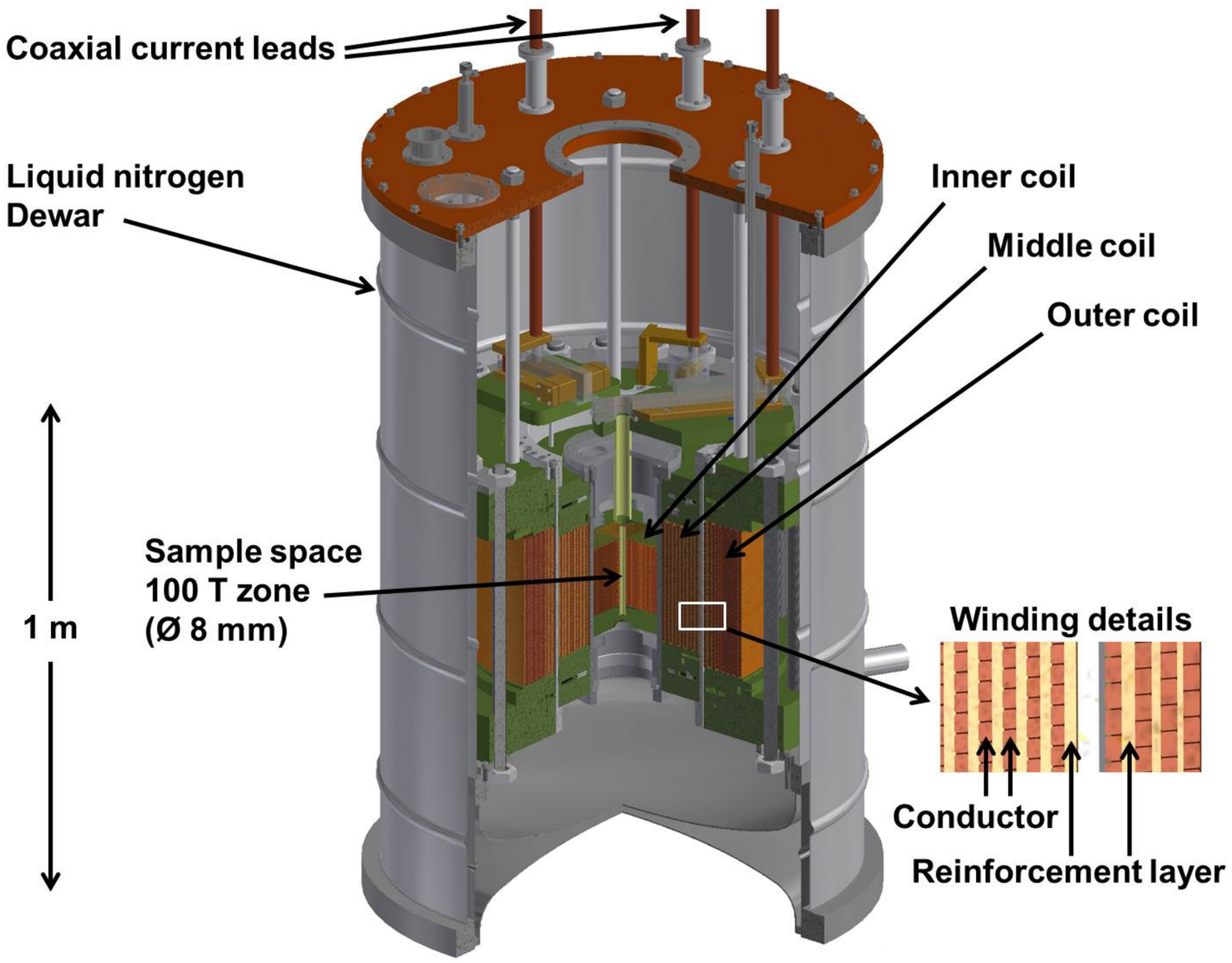}
    \caption{The LNCMI 100\,T magnet. The magnet is composed of three concentric coils powered with three independent capacitor banks. Before a pulse, the magnet is cooled with liquid nitrogen. Winding details show alternating conductor and reinforcement layers. Conductors are insulated wires made of high-strength copper alloys or micro-composites and reinforcement is made of Zylon polymer fibers. Figure courtesy LNCMI$-$Toulouse.}
    \label{Fig LNCMI 100 T magnet}
    \end{center}
\end{figure}

The pulsed magnets at LNCMI are cooled with liquid nitrogen and produce long pulses with a duration on the order of $100\ $ms at a repetition rate of 1 pulse/hour. The typical sample-space dimensions are $5-20\,$mm. The LNCMI 100\,T magnet is depicted in Fig.\,\ref{Fig LNCMI 100 T magnet}. A test of the system up to 98.8\,T is described in \cite{Beard2018}. 

\begin{figure}[h!]
    \begin{center}
     \includegraphics[width=6.5 in]{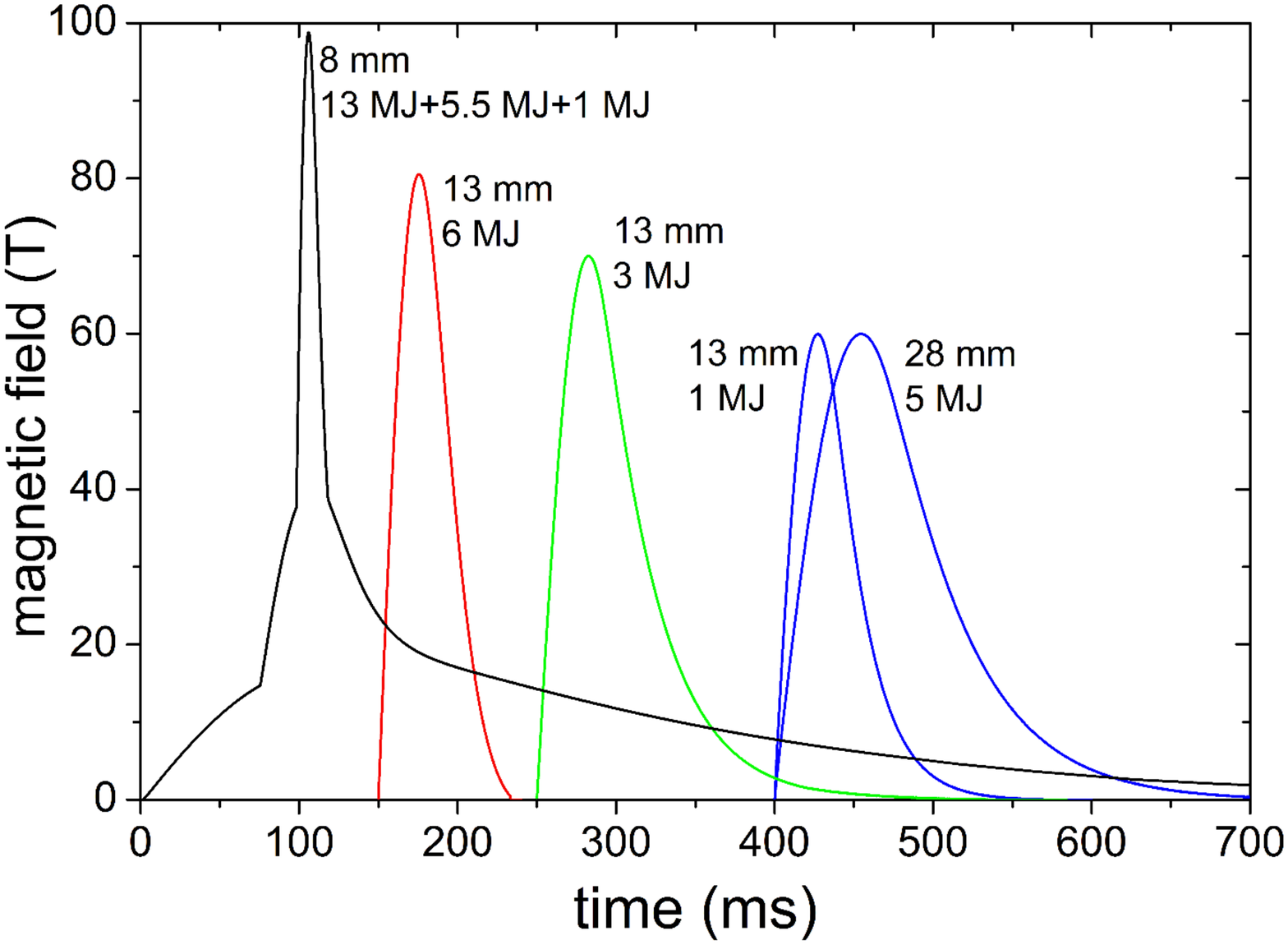}
    \caption{Temporal profiles of magnetic pulses realized by the solenoids available at LNCMI. The diameter of the free bore and the magnetic energy stored at the maximum magnetic field are indicated. The 100\,T magnet is composed of three concentric solenoids fed by three independent capacitor banks. The three different energies indicated are the three magnetic energies for the sub-coils from the outside to the inside. Figure courtesy LNCMI$-$Toulouse.}
    \label{Fig LNCMI Pulsed Mag timing}
    \end{center}
\end{figure}

Timing diagrams for several representative LNCMI pulsed magnets is presented in Fig.\,\ref{Fig LNCMI Pulsed Mag timing}.

In LNCMI pulsed magnets, the dissipated energies are in the MJ range, corresponding to peak powers in the GW range. The mechanical stresses on the coils are in excess of $3\ $GPa (30\,000 atm). These features necessitate extremely careful mechanical design and thermal management in order to ensure long lifetime of the magnet components and avoid potentially violent failures. The ability to dissipate the heat generated by a field pulse determines how frequently the magnet can be pulsed.

\begin{figure}[h!]
    \begin{center}
     \includegraphics[width=5.5 in]{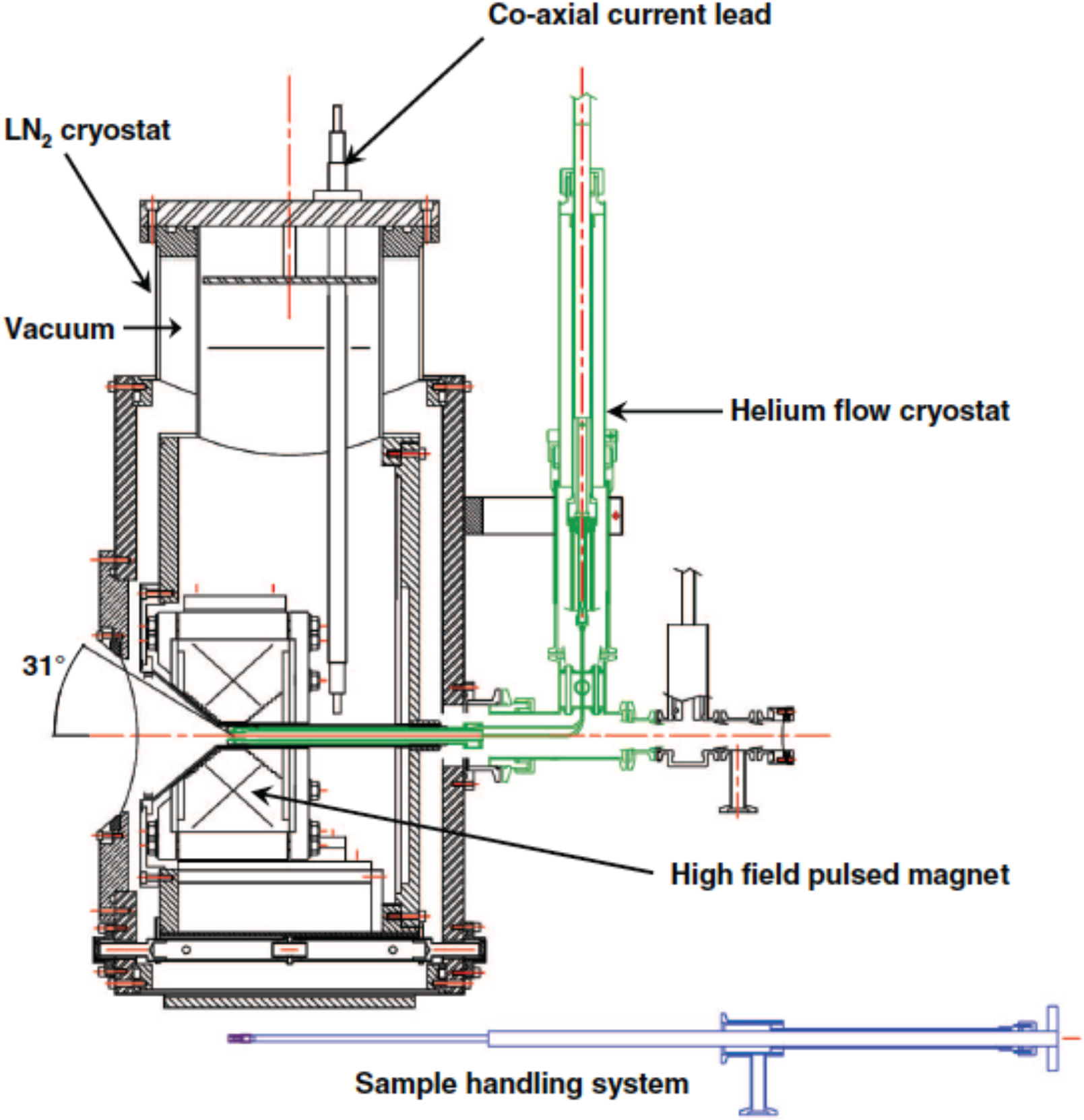}
    \caption{A 30\,T pulsed magnet for X-ray studies operating at ESRF. The solenoid coil is 16\,mm in diameter and is maintained at 77\,K. The total pulse duration is up to 60\,ms with a repetition rate of 6 pulses/hr at maximum field. The sample can be maintained at a temperature in the range of $1.5-300\ $K in a liquid-helium cryostat. Figure from \cite{Billette2012}. }
    \label{Fig 30 T magnet at ESRF}
    \end{center}
\end{figure}

In many cases, it is desirable to combine high magnetic fields with other unique facilities, necessitating the development of ``portable'' high-field magnets. An example is a high-field magnet (Fig.\,\ref{Fig 30 T magnet at ESRF}) developed by LNCMI for x-ray studies at the European Synchrotron Radiation Facility (ESRF) \cite{Billette2012}. This is a horizontal-field, 30\,T magnet system with a conical bore optimized for synchrotron-radiation powder diffraction. The magnet provides a wide-angle optical access downstream of the sample, which allows to measure many Debye rings enabling accurate crystal-structure analysis. 

Another example is a pulsed 40\,T magnet operating at the neutron-diffraction facility at the Institut Laue-Langevin (ILL).
The magnet also features a wide-angle conical access to a 8-mm diameter sample whose temperature may be adjusted within the $1.5-300\ $K range. The magnet produces 100-ms long pulses every 6 minutes corresponding to the total ``field-on'' time of about $20\,$s/day. Recently, this device was used to study magnetically ordered phases in an enigmatic material URu$_2$Si$_2$ with strongly correlated electrons \cite{Knafo2016}. In a strong magnetic field,  the material undergoes a phase transition into magnetically ordered phases. Neutron diffraction under pulsed magnetic fields
was used to identify the field-induced phases as a spin-density-wave state.

\section{Techniques for measuring high magnetic fields}
\label{Sec:B_Measurement}

Most experiments using magnets do not only require generation of a field with particular sets of parameters, the experimentalist also needs to know what the values of the parameters are. In many cases, direct measurements are required, sometimes referred to as ``magnetic-field metrology.'' We briefly discuss a selection of measurement techniques in this section.

\subsection{DC fields}

NMR is frequently the technique of choice for DC-field measurements. Commercial ``NMR precision teslameters'' cover the range of fields to $\approx 10\,$T with proton-based NMR probes (the proton gyromagnetic ratio is 42.5774806(10) MHz/T) and to $\approx 23\,$T with deuterium based probes  (the deuteron gyromagnetic ratio is 6.53569(2)\,MHz/T). Typical precision of such devices is in the several parts-per-billion (ppb) range, while relative accuracy is in the parts-per-million (ppm) range.

Significantly higher relative accuracies of better then a part in $10^{12}$ were demonstrated with NMR magnetometers based on gaseous $^3$He \cite{Nikiel2014} (absolute accuracy is limited to roughly a part in 10$^8$ by the finite accuracy of fundamental constants). In these magnetometers, nuclear spins are polarized via optical pumping. Since the atomic transitions of helium are deep in the ultraviolet, the atoms are excited into a metastable state in a discharge. There are convenient near-infrared transitions from the metastable excited state, so the electronic and nuclear spins can be polarized by a repeated absorption/de-excitation cycle. A large fraction of nuclei in the gas cell are eventually polarized where the atoms in the metastable states transfer their excitation and nuclear polarization to the ground-state atoms in a process known as ``metastability exchange.'' The efficiency of optical pumping decreases at high magnetic fields where the Zeeman energy exceeds the hyperfine-interaction energy and the hyperfine structure is ``decoupled.'' This has limited the operation range of such magnetometers to about 12\,T. The magnetometry protocol consists in measuring the frequency of the free-induction decay of the $^3$He spins after excitation with a resonant radio-frequency pulse. The coherent spin-precession times in such magnetometers are on the order of minutes, which is achieved due to averaging of magnetic-field inhomogeneities by helium atoms sampling the entire volume of the cell during the free precession and by ensuring accurate spherical shape of the cell \cite{Nikiel2014}.    

A particular challenge is metrology for the highest-field DC magnets (up to 45\,T), such as the resistive Bitter-type magnets and hybrid magnets described in Sec.\,\ref{Sect: DC magnets}. One technique, widely employed at the NHMFL-Tallahasee, is careful determination of the field-current  calibration curve ($B$ versus $I$) for each magnet using the following method: First, a measure of the magnetic field is obtained from a pickup coil in the magnet's bore. By Faraday's law of induction, the voltage induced in the pickup coil is related to the time-derivative of the magnetic field as the current is ramped from zero up to its peak value. Integrating the signal from the pickup coil therefore gives the (uncalibrated) magnetic field versus current. This curve is then calibrated by scaling it with respect to known specific fields (and therefore currents) that are separately identified with fixed-frequency NMR probes. This method provides field-measurement accuracy better than one-tenth of one percent, which is found to be superior to that obtained using four-wire Hall effect probes (owing largely to the temperature- and slightly nonlinear field-dependence of the high-field Hall effect in most materials).

\subsection{Pulsed fields}

For pulsed high-field magnets with millisecond time scales of the field evolution, a typical goal for magnetic field metrology is 0.1\% accuracy with 1 kHz bandwidth, while pulsed-coil diagnostics requires even higher bandwidth on the order of 100 kHz and higher accuracy may be required for particular applications.

There are a number of constraints typically imposed on a magnetic-field measurement in a strong-field magnet, including the need for linear response over a wide range of fields, small (typically, several mm) sensor size, construction out of nonmagnetic and mostly nonconductive materials, and the need to operate in a wide temperature range  from 400 K down to 0.1 K.

A typical implementation of the measurement device is a pick-up coil, in which the induced voltage is proportional, according to Faraday's law of induction, to the rate of the magnetic-flux change through the coil. An immediate concern for such a device is that it sensitive not only to the rate of change of magnetic induction but also to vibrations that can change the effective area of the coil.
A pick-up unit is usually constructed by winding several turns of copper enameled wire around a ceramic or low-thermal-expansion plastic bobbin. The device needs to be calibrated, carefully positioned and aligned and mounted to ensure vibration damping. In a carefully implemented device of this kind, overall relative accuracy of $\sim  2\cdot10^{-3}$ is achieved with standards laboratory equipment, while better than $10^{-4}$ was reported with metrology-grade equipment.

The pickup coils used in the pulsed magnets at the NHMFL Los Alamos are typically calibrated using the de Haas - van Alphen (dHvA) quantum oscillations in copper. The dHvA effect is a quantum mechanical phenomenon wherein the magnetization of a material oscillates as a function of magnetic field due to the Landau quantization of electron energy.  The oscillations are periodic in $1/B$, and occur at a well-known frequencies in simple cubic-lattice metals such as copper.  Counterwound copper pickup coils, typically used for measuring dHvA signals in other samples of interest, also show clear quantum oscillations due to the copper wires themselves. The actual magnetic field is calibrated by comparing the measured dHvA frequency with the known dHvA frequency for copper.  

\begin{figure}[h!]
    \begin{center}
     \includegraphics[width=5.5 in]{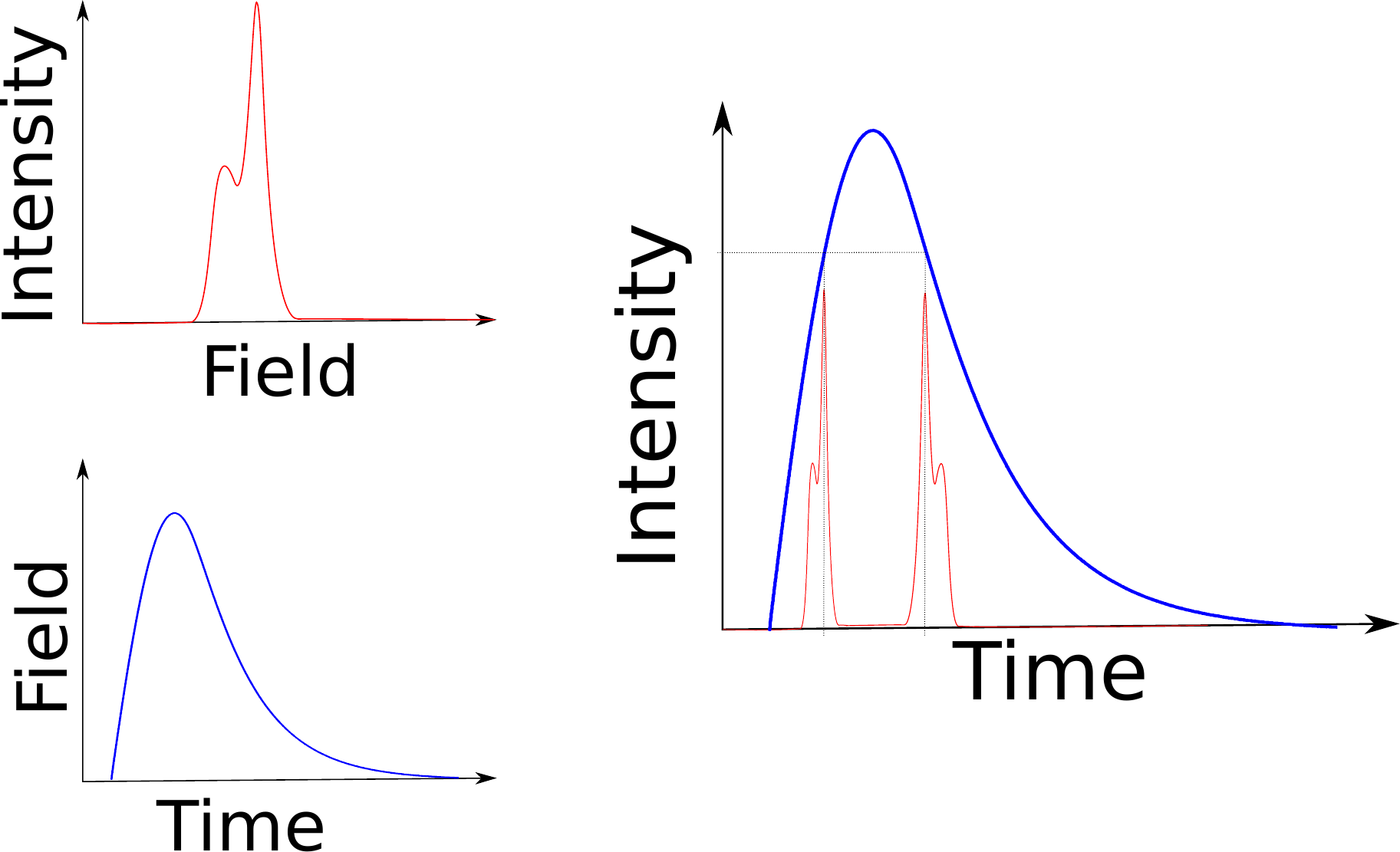}
    \caption{Resonant marking of time-variable magnetic field is based on sharp dependence of some spectroscopic signal, for instance, atomic absorption or NMR, on the applied field (upper left). During the temporal evolution of the pulsed magnetic field (lower left), the field passes through the resonant value, typically, twice, leading to the time-dependent signals marking the times the resonance is achieved (right). }
    \label{Fig_Resonant_Marking}
    \end{center}
\end{figure}
Another technique for pulsed-field metrology--resonant marking--allows to pinpoint the precise times during the magnetic-field pulse when the resonant value of the field is reached (Fig.\,\ref{Fig_Resonant_Marking}). 

A variety of nonzero-spin nuclei have been used for NMR marking. Most commonly, these are $^1$H protons and $^2$H deuterons, but $^{63}$Cu has been a choice for some of the recent work. The magnetic moment of $^{63}$Cu is about a factor of four lower than that of the proton, allowing to keep NMR frequencies in the sub-GHz range and the magnitude of the magnetic moment is known with a $10^{-5}$ accuracy. Moreover, the Knight shift of the NMR frequency due to the conduction electrons in a metal is also at the same relative level of $10^{-5}$. Additional advantages of copper include the fact that it remains solid in a broad range of temperatures and that it is a simple metal in that there are no field-induced transitions at low temperatures and high magnetic fields. All of these factors make  $^{63}$Cu a good reference material for high-field measurements, particularly those at low temperature.

The progress in RF electronics, NMR methodology and magnet technology make it now possible to obtain reasonable signal-to-noise ratios in NMR measurements using pulsed field magnets with typical pulse durations of tens of milliseconds, on certain classes of samples. These samples should have high abundances of nuclei with large gyromagnetic ratios (see Table \ref{Table:Nuclei_for_NMR}), and should have rather short nuclear relaxation times, preferably shorter, but definitely not much longer, than the field pulse duration. This is generally true for systems containing paramagnetic ions, conduction electrons or quadrupolar nuclei.
Although this technique cannot compete with NMR in DC fields in terms of sensitivity or resolution, it can give access to  NMR data in the field range above 45\,T. The current field limit of pulsed-field NMR experiments is 80\,T, and there is no reason to believe that it could not be implemented in the highest field pulsed magnets, which currently generate up to 100\,T.
\begin{table}[h]
\center
 \begin{tabular}{cccc}
 \hline\hline
 Nucleus & Relative sensitivity & Natural abundance & $f_{\textrm{NMR}}$ at 40 T (MHz) \\ 
 \hline
$^1$H & 100 & 100 & 1703.10 \\
$^7$Li & 29.3 & 100 & 661.89  \\
$^{11}$B & 16.5 & 80.4 & 546.42  \\
$^{19}$F & 83.3 & 100  & 1602.51  \\
$^{27}$Al & 20.6 & 100  & 443.77   \\
$^{31}$P  & 6.6   & 100  & 689.43   \\
$^{45}$Sc  & 30.1  & 100  & 413.71 \\
$^{51}$V    & 38.2  & 99.8 & 447.97 \\
$^{55}$Mn & 17.5  & 100  & 422.18  \\
$^{59}$Co & 27.7  & 100  & 404.10   \\
$^{65}$Cu & 11.4  & 30.9  & 483.74   \\
$^{71}$Ga & 14.2  & 60.4  & 519.39   \\
$^{93}$Nb & 48.2  & 100  & 416.85   \\
$^{115}$In & 34.7  & 95.7  & 373.19  \\
$^{121}$Sb & 23.9  & 100  & 407.56  \\
$^{141}$Pr & 29.3  & 100  & 521.49  \\
$^{151}$Eu & 17.8  & 47.8  & 423.39  \\
$^{165}$Ho & 18.1  & 100  &  363.44  \\
$^{205}$Tl & 19.2  & 70.5  & 981.56  \\
$^{209}$Bi & 13.7  & 100  & 273.68  \\
 \hline\hline
 \end{tabular}
\caption{Nuclei suitable for pulsed-field NMR. Relative sensitivity is it the product of gyromagnetic ratio and nuclear moment, arbitrarily normalized to 100 for the proton. It gives
the NMR signal strength one can expect with respect to proton NMR at the same concentration and field.}
\label{Table:Nuclei_for_NMR}
\end{table}

\begin{figure}[h!]
    \begin{center}
     \includegraphics[width=5.5 in]{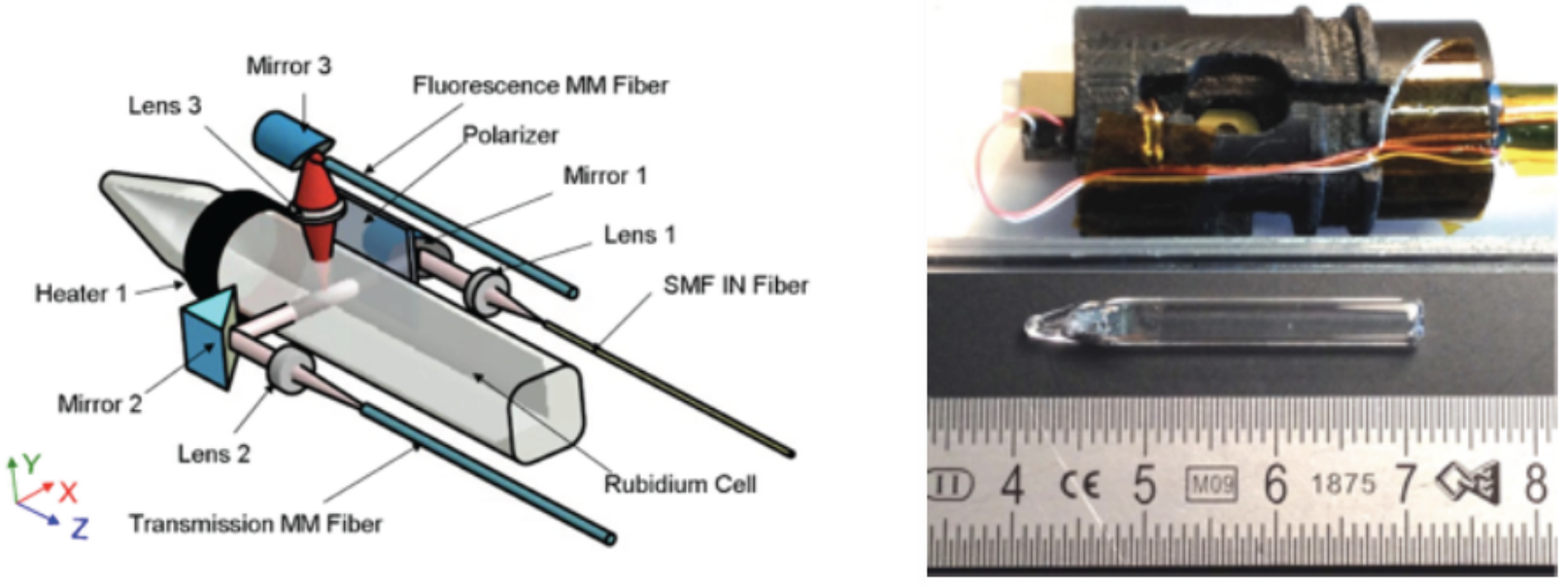}
    \caption{High-magnetic-field probe based on Rb optical spectroscopy \cite{George2017}. Left: optical measurement head; MM and SM stand for multi-mode and single-mode (optical fibers), respectively. Right: Rb vapor cell and 3D-printed holder.}
    \label{Fig_Rb_setup}
    \end{center}
\end{figure}
A technique to measure pulsed magnetic fields based on the use of rubidium in gas
phase as a metrological standard was demonstrated in Ref.\,\cite{George2017}. The idea is that the magnetic field ``tunes'' the energies of the Zeeman sublevels in the upper and the lower state and brings the transition on resonance with light of a fixed and precisely measured frequency, increasing absorption and fluorescence. The authors of Ref.\,\cite{George2017} developed an instrument based on laser inducing
transitions at about 780 nm (D2 line) in rubidium gas contained in a cell of $3\times 3\ $mm$^2$ cross
section (Fig.\,\ref{Fig_Rb_setup}).  A temperature-stabilized
fiberized probe was used to insert the cell into a high-field pulsed magnet. Transition frequencies for light linearly polarized parallel and perpendicular to the magnetic field were measured with a wavemeter. Both the light transmission and fluorescence (from a volume of $0.13\,$mm$^3$) were monitored. The sensor was operated up to fields of 58\,T and allowed precise scaling of the temporal profile of the magnetic field obtained with a pick-up-coil. In this way,  the
magnetic field strength during the entire duration of the pulse was determined 
with an accuracy of $\approx 2 \cdot10^{-4}$, better than with the calibrated pick-up coil by more then an order of magnitude. Interestingly, the absolute accuracy is not limited by the experimental method but rather by the uncertainty in the Land\'{e} $g$-factor of the 5p Rb excited state. Once the $g$-factor is known with high precision, the overall accuracy can be improved to the $10^{-5}$ level \cite{Ciampini2017}. 

In the highest-field pulsed magnets such as single-turn and flux-compression magnets, the field is often measured via magneto-optical (Faraday) rotation.
In this method, linearly polarized light propagates in a medium (e.g., glass) in the direction of the magnetic field. The polarization plane of the light rotates by an angle given by a product of the length of the sample, the strength of the magnetic and a so-called Verdet constant, which is a characteristic of the material. Major strengths of this method are that it allows for measurements that are largely immune to electromagnetic pick-up accompanying the magnetic-field pulse, the ability to measure the field evolution on fast ($\mu$s) scales, and a large dynamic range. Faraday-rotation magnetometry is used for measuring the strongest laboratory fields \cite{Nakamura2013}.

\section{Conclusions and outlook}
\label{Sec:Conclusions}

In this paper, we have provided an overview of modern technologies for generating and measuring both DC and pulsed magnetic fields and discussed a representative sampling of currently ongoing and planned fundamental-physics  experiments using high magnetic fields. There is certainly impressive progress in the field and no lack of important measurements to carry out with strong magnetic fields and a number of novel ideas and experimental concepts. We conclude with some of the highlights and and outlook of forthcoming developments in high magnetic field technologies.

\begin{itemize} 

\item Both DC and pulsed resistive magnets have been working reliably for decades. Fields continue to rise via more sophisticated engineering and larger power supplies. Magnets with geometries adapted to specific experimental requirements are possible and exist, for example, at NHMFL- Tallahassee  (25\,T split) and Berlin (25\,T conical) for scattering experiments, as well as at LNMCI-Toulouse and in Tokyo, where these magnets are used for vacuum-birefringence experiments, see Sec.\,\ref{Subsect:VB}.

\item	High-Tc superconducting magnets producing DC fields up to 32\,T already exist and will become generally available in the next decade but will require significant investments and dedicated operating environment. Lower field values but with geometries adapted to specific experimental requirements (related to the figures-of-merit such as those listed in Table\,\ref{Table:FOMs})  are feasible but require considerable engineering effort.

\item The NHMFL all-superconducting magnet that reached 32\,T has a thoroughly tested quench-protection system that allows this to be the first SC magnet for physics experiments to operate above 24\,T. This 8\,T leap in field (33\%) is a harbinger of a revolution in SC magnets for a variety of fields of science. To put this in perspective, the previous 8\,T increase (from 16\,T to 24\,T) took 42 years (1975 – 2017, Fig.\,\ref{FIG_Solenoid Magnets}). The step from 24\,T to 32\,T took about a year (24\,T became available in January 2017).

\item There is a ``quiet revolution'' going on in the field of superconducting magnets, owing to recent advances and development of high-Tc superconductors made from, for example, YBCO. Whereas the critical current of conventional low-Tc superconducting wire (NbTi, Nb$_3$Sn) limits the maximum achievable field to about 23.5\,T, high-Tc materials such as YBCO remain superconducting even in magnetic fields exceeding 45\,T (see Table \ref{Tab SC materials}).  Already, recent developments described above have boosted the peak field in a YBCO-based all-superconducting magnet to 32\,T at NHMFL- Tallahassee. As better and stronger wires and tapes are made from these high-Tc materials, the maximum fields that are achievable with all-superconducting magnets will increase correspondingly.   Similarly, the physical size and design complexity of the magnets can also increase, which is important for many magnets designed for fundamental science, where a combination of field strength and magnet size (length, volume) are key parameters (Table\,\ref{Table:FOMs}). 

\item Next-generation SC magnet technology should allow both higher fields than before as well as to allow experiments previously conducted in resistive magnets to move to SC magnets where there will be less field ripple, less acoustic noise, lower operating costs, and more time available.

\item Significant advances in superconducting magnets for a variety of applications have been demonstrated by commercial companies using low-temperature superconducting materials. This includes the high-field MRI magnets (9.4\,T with 900\,mm bore) as well the high-field NMR and research magnets up to 23\,T, which are narrow-bore systems using NbTi and Nb$_3$Sn materials. Recently, a new class of compact high-field, wide-bore magnets developed by commercial companies  is  facilitating all-superconducting magnets for over 30\,T using LTS and HTS materials. The 32\,T all superconducting magnet at NHMFL used a 17\,T HTS insert developed by NHMFL which was integrated into a 15\,T/250\,mm LTS outsert developed by Oxford Instruments.

\item	Hybrid magnets producing 45\,T will become more available in the coming years. The next-generation of hybrid systems that are currently discussed are expected to come online around 2030. 

 \item Pulsed magnets producing 100\,T with a few millisecond pulse duration are already here (in Los Alamos since 2012) and will become more available during the next years. Lower field values but with geometries adapted to specific experimental requirements are quite feasible but require engineering efforts.

\end{itemize}

High magnetic fields  have enabled important fundamental-physics discoveries. One example involving small-scale condensed-matter physics is the discovery of the Quantum Hall Effect \cite{Klitzing1986} by Klaus von Klitzing, at the time of discovery working with a high-field magnet at LNCMI, for which he was awarded the 1985 Nobel Prize in Physics. We hope that this review has convinced the reader that other fundamental-physics discoveries will likely be forthcoming due to the remarkable progress of high-magnetic-field technologies and the abundance of ideas for new experiments. But there is also a reason for us to be humble: in spite of the impressive progress in strong-field generation, even the strongest fields we can imagine producing in macroscopic volumes in the laboratory still come many orders of magnitude short of the strongest fields found elsewhere in the universe, for example, in neutron stars where magnetic fields are expected to be as high as $10^{11}$\,T.

\section{Glossary}
Some of the acronyms used in this review are presented in Table \ref{Table:abbreviations}.
\begin{table}[h]
\caption{Abbreviations and their meanings.}
\medskip 
\center
\begin{tabular}{ll}
\hline \hline
Abbreviation~~ & Meaning \\
\hline
\rule{0ex}{2.6ex} AMO & atomic, molecular and optical physics \\
\rule{0ex}{2.6ex} ADMX & axion dark matter experiment (UW, Seattle) \\
\rule{0ex}{2.6ex} ALPs & axion-like particles \\
\rule{0ex}{2.6ex} BSCCO & bismuth-strontium-calcium-copper-oxide (HTS) \\
\rule{0ex}{2.6ex} CAST & CERN axion solar telescope \\
\rule{0ex}{2.6ex} CAPP & Center for Axion Precision Physics (Korea) \\
\rule{0ex}{2.6ex} CULTASK & CAPP's ultralow-temperature axion search in Korea \\
\rule{0ex}{2.6ex} CERN & European Organization for Nuclear Research \\
\rule{0ex}{2.6ex} CHMFL & High Magnetic Field Laboratory of the Chinese Academy of Sciences \\
\rule{0ex}{2.6ex} DM & dark matter \\
\rule{0ex}{2.6ex} EMFC & electromagnetic flux compression \\
\rule{0ex}{2.6ex} ESRF & European Synchrotron Radiation Facility \\
\rule{0ex}{2.6ex} FOM & figure of merit \\
\rule{0ex}{2.6ex} HLD & Hochfeld-Magnetlabor Dresden \\
\rule{0ex}{2.6ex} HTS & high-temperature superconductor \\
\rule{0ex}{2.6ex} IAXO & International axion observatory \\
\rule{0ex}{2.6ex} ILL & Institut Laue--Langevin (Grenoble)\\
\rule{0ex}{2.6ex} LHC & Large Hadron Collider at CERN \\
\rule{0ex}{2.6ex} LNCMI & Laboratoire National des Champs Magn\'{e}tiques Intenses \\
\rule{0ex}{2.6ex} LTS & low-temperature superconductor \\
\rule{0ex}{2.6ex} NHMFL & US National High Magnetic Field Laboratory \\
\rule{0ex}{2.6ex} NLED & nonlinear electrodynamics \\
\rule{0ex}{2.6ex} SC & superconductor, superconducting \\
\rule{0ex}{2.6ex} QCD & quantum chromodynamics \\
\rule{0ex}{2.6ex} QED & quantum electrodynamics \\
\rule{0ex}{2.6ex} TML & Tsukuba Magnet Laboratory \\
\rule{0ex}{2.6ex} YBCO & yttrium barium copper oxide (HTS) \\
\rule{0ex}{2.6ex} XFEL & X-ray Free-Electron Laser \\

\hline \hline
\end{tabular}
\label{Table:abbreviations}
\end{table}

\section*{Acknowledgment}  
The HIMAFUN project was supported by the European Magnetic Field Laboratory (EMFL), Laboratoire National des Champs Magn\'{e}tiques Intenses (LNCMI) and the Helmholtz Institute Mainz.
Contributions to this work were supported in part by IBS-R017-D1 of the Republic of Korea. DB acknowledges support from the European Research Council (ERC) under the European Unions Horizon 2020
research and innovation programme (grant agreement No
695405), from the Simons and Heising-Simons Foundations, and from the DFG Reinhart Koselleck
project. The authors are grateful to Mark Bird, John Blanchard, Greg Boebinger, Michael Hartman, Werner Heil, Alec Lindman, Oliver Portugall, Leslie Rosenberg, Maria Simanovskaia, Alexander Sushkov, Michael Tobar, Karl van Bibber, Lindley Winslow, and Max Zolotorev for valuable input.

\bibliographystyle{model1-num-names}
\bibliography{HIMAFUNbib.bib}

\end{document}